# A general theory to estimate Information transfer in nonlinear systems

Carlos A. Pires (*), (1), David Docquier, (2), Stéphane Vannitsem, (2)

(1) Universidade de Lisboa, Faculdade de Ciências, Instituto Dom Luiz, Lisboa, Portugal.
(2) Royal Meteorological Institute of Belgium, Brussels, Belgium.

∗ Corresponding author: E-mail address: clpires@fc.ul.pt

Address: Carlos A. L. Pires. Faculdade de Ciências, Edifício C8, DEGGE, Campo Grande 16, 1749-016 Lisboa, Portugal

**Abstract**

A general theory for computing information transfers in nonlinear stochastic systems driven by deterministic forcings and additive and/or multiplicative noises, is presented, satisfying indistinctly closed, periodic or pdf vanishing boundary conditions of the state-space. It extends the Liang-Kleeman (LK) framework of causality inference to nonlinear cases based on information transfer across system variables (Liang, 2016. Information flow and causality as rigorous notions ab initio. *Phys. Rev. E*, 94: 052201. DOI: 10.1103/PhysRevE.94.052201). We present an effective method of computing formulas of the rates of Shannon entropy transfer (RETs) between selected causal and consequential variables, the 'Causal Sensitivity Method' (CSM), relying on the estimation from data of conditional expectations of the system forcings and their derivatives. Those expectations are approximated by nonlinear differentiable regressions, leading to a much easier and more robust way of computing RETs than the 'brute-force' approach calling for numerical integrals over the state-space and the knowledge of the multivariate probability density function of the system. The CSM is furthermore fully adapted to the case where no model equations are available, starting with a nonlinear model fitting from data of the consequential variables, with the subsequent application of CSM to the fitted model. RETs are decomposed into deterministic and stochastic components, being compensated by the self generation of entropy. Moreover, RETs are decomposed into sums of single one-to-one RETs plus synergetic terms (of pure nonlinear nature) accounting for the joint causal effect of groups of variables. State-dependent (or specific) RET formulas are also introduced, puting in evidence where in state-space the entropy transfers and local synergies are more relevant. A comparison of the RETs estimations is performed between: 1) the 'brute-force', expensive (taken as benchmark), probability-density-based approach (AN), 2) the CSM-based approach with and/or without model fitting, and the multivariate linear (ML) approach, in the context of two different models: (i) a model derived from a potential function and (ii) the classical chaotic Lorenz system, both forced by additive and/or multiplicative noises. The analysis demonstrates that the CSM estimations are robust, cheaper, and less data-demanding than the AN-reference values in the different experiments, providing evidence of the possibilities and generalizations offered by the method (e.g. causality diagnostics between subspaces) and opening new perspectives on real-world applications.

## 1. Introduction

Causal inference has a long history in a wide range of fields as discussed for instance in the overview of Pearl [1] from medicine to social sciences. The nice example put forward by Pearl [1] is the question on the influence of symptoms on the actual disease of a patient. Finding the directionality of the influence from the cause to the consequence is therefore key in answering such type of question, building in that way a graph of influence from one process to another.

One famous approach to analyze such types of problems is the Granger's causality which consists in evaluating whether one variable X, provides additional predictive power to the forecast of another variable Y, implying that X is a cause of Y [2]. This approach has been applied in many different contexts from economy to atmospheric variability [3-4]. Since these early developments, other methods have been designed, in particular in the context of information theory. In the 1970s, this question was addressed by the development of the directed information theory [5,6]. Since then, many information-theoretic techniques have been designed and compared, e.g. Schreiber [7] and Hlavackova-Schindler et al. [8] . Other approaches based on dynamical systems theory or connected directed networks have been also developed in parallel and illustrated in different applications [9,10].

One important development in the conjunction of dynamical systems theory and information theory has been initiated by Liang and Kleeman [11]. They developed a theory of the rate of information transfer between specific variables starting from the fundamental dynamical equations describing the evolution of the Shannon entropy. The theory was then considerably expanded in a series of papers by Liang [12-18], culminating on one side to analytical expressions of the rate of information transfer in multivariate systems, and on the other side, to specific simple expressions that can be used when dealing pragmatically with time series. The latter developments were done by assuming that the system underlying the dynamics of the observed data could be approximated by a multivariate linear stochastic system with additive noise [18], i.e. a multivariate Ornstein-Uhlenbeck process. These tools have, in particular, considerably been used in the context of climate science [19-25].

This linear approximation, however, is restrictive as some important nonlinear interactions could be missed. In the present work, the extension of the estimation to nonlinear dynamical systems will be addressed. The main objective is therefore to find a closed formula, to evaluate the rate of entropy transfer (RET) – equivalent to the rate of information transfer in Liang-Kleeman's (LK)'s theory - in nonlinear systems forced by additive or multiplicative noise that can be robustly estimated from data (e.g. time-series or time-evolving ensembles), under closed and/or periodic or pdf vanishing boundary conditions. The main idea is to apply the hereafter called 'Causal Sensitivity Method' (CSM) relying on the computation of conditional expectations (and their derivatives or sensitivities) of the deterministic and stochastic terms of the model, for a selected consequential variable.

This applies also to the computation of the Shannon entropy budget terms, bridging them with the RETs. This raises the important concept of synergetic causality, coming from the joint effects of groups of variables, not accounted for in single variable-to-variable RETs. Formulas of RETs have been obtained for closed or periodic boundary conditions in the state-space. Open conditions in the state-space call for additional boundary terms in the entropy budget.

Comparison and validation of the different approaches to estimating the RETs is performed, namely:

a) The analytical-numerical (AN) method where the model stochastic equations are known and the probability density function (pdf), (analytical or estimated from data) is used with 'brute-force' in the integral formula of RET in state-space.

b) The CSM approach developed in the context of this study, based on conditional expectations (obtained from nonlinear regressions) and their derivatives with respect to the selected consequential variable. Here the model equations are supposed to be known.

c) The model fitting (MF), from data, of both the deterministic term and of the diffusivity coefficients, with subsequent application of the CSM to the fitted model. Only the evolution equation of the selected consequential variable is necessary.

d) The multivariate linear (ML) approach proposed in Liang [18], supposing a multivariate linear model fitting forced by additive noise.

The analysis is performed in the context of two different models: a potential model following Liang [16,17] and the famous Lorenz model [26], where the advantages of the CSM and MF approaches are exemplified.

Sections 2 and 3 are devoted to the fundamental theory of the entropy budget and the rates of entropy transfer respectively. Section 4 explains the numerical aspects of the different techniques. Section 5 describes the results obtained with the potential model for which one performs a sensitivity analysis to the presence of linear, and nonlinear terms and additive or multiplicative noises, as well as with respect to the time-series length. In Section 6, the approach is also illustrated in the context of the Lorenz model, with the specific purpose of providing robust examples of estimations based on data only, and with a particular interest of trying to obtain good RET estimations from quite short time-series, simulating real situations of low data availability. Finally in Sec. 7, the main results are summarized and discussed, together with potential extensions of this type of analysis to more complex and higher dimension systems. Three appendices are added with theoretical proofs, numerical technicalities and a list of symbols and achronyms.

## 2. Theory of the entropy budget

Let us consider a $D$-dimensional stochastic continuous system described by the state vector $\mathbf{X} = (X_1, \ldots, X_D)^T$ (where $T$ stands for transpose) and a parameter vector $\boldsymbol{\theta}$, which evolves by Euler-Bernstein equations:

$$X_i(t + dt) = X_i(t) + F_i(\mathbf{X}, \boldsymbol{\theta}, t)dt + \sum_{k=1}^{D_n} B_{i,k}(\mathbf{X}, \boldsymbol{\theta}, t)dW_k, i = 1, \ldots, D \tag{1}$$

where $F_i(\mathbf{X}, \boldsymbol{\theta}, t)$ are the deterministic terms, a component of the $D$-dimensional differentiable vector $\mathbf{F}$. Stochastic forcings are introduced where $dW_k \sim \sqrt{dt}\ N(0,1)$ are independent Wiener processes and $B_{i,k}$ are noise-diffusion coefficients, merged in the differentiable matrix $\mathbf{B}$, from which one builts the positive-definite matrix of pdf diffusivities $\mathbf{G} \equiv \mathbf{B}\mathbf{B}^T$ with elements $g_{i,j} \equiv \sum_{k=1}^{D_n} B_{i,k} B_{j,k}$.

We will hereafter denote by $X_{\sim i} \equiv (X_1, \ldots, X_{i-1}, X_{i+1}, \ldots, X_D)^T$ the set of variables different from $X_i$ or its complementary vector, which will be used to discuss the effect of $X_{\sim i}$, taken as a set of causal variables on $X_i$ taken as a consequential variable.

In this work, we adopt the approach introduced originally by Liang [11-18,27], relying upon the concept of Shannon entropy $H_{X_i} = E(-\log \rho_i)$ of $X_i$ and its rate of change $\frac{dH_{X_i}}{dt}$. In the formula, $\rho_i$ denotes the probability density function (pdf) of $X_i$ and $E$ stands for expectation operator over the state-space. When the rate is positive (negative), there is a loss (gain) of information and $X_i$ becomes less (more) precise. For Gaussian pdfs, $\frac{dH_{X_i}}{dt}$ is the the relative rate of variation of the variance $var(X_i)$ per time unit (e.g. $\frac{dH_{X_i}}{dt} = 2$ means that variance doubles at every unit of time).

In that framework, one aims to decompose the rate of entropy change (REC) $\frac{dH_{X_i}}{dt}$ as:

$$\frac{dH_{X_i}}{dt} = T(X_{\sim i} \to X_i) + \frac{dH_{X_i,self}}{dt} \tag{2}$$

where the first term comes from the influence of $X_{\sim i}$ on $X_i$ or the rate of entropy transfer (RET) and the second term is the self entropy generation (SEG), obtained in a process where $X_{\sim i}$ is 'frozen in time', i.e. REC=SEG+RET. This problem was solved for two-dimensional systems (Theorem 5.1. of Liang [27]) leading to the following rate of entropy transfer:

$$T(X_2 \to X_1) = T(X_{\sim 1} \to X_1) = -E\left(\frac{1}{\rho_1}\frac{\partial F_1 \rho_1}{\partial X_1}\right) + \frac{1}{2}E\left(\frac{1}{\rho_1}\frac{\partial^2 g_{1,1}\rho_1}{\partial X_1^2}\right) \tag{3}$$

where $X_{\sim 1} = X_2$ is the complementary of vector of $X_1$ in two-dimensional systems. That RET expression can be generalized to any dimension $D \geq 3$ in the form:

$$T(X_{\sim i} \to X_i) = -E\left(\frac{1}{\rho_i}\frac{\partial F_i \rho_i}{\partial X_i}\right) + \frac{1}{2}E\left(\frac{1}{\rho_i}\frac{\partial^2 g_{i,i}\rho_i}{\partial X_i^2}\right) = T(X_{\sim i} \to X_i)_F + T(X_{\sim i} \to X_i)_g, \tag{4}$$

where $T(X_{\sim i} \to X_i)_F$ and $T(X_{\sim i} \to X_i)_g$ are respectively the determinist RET (D-RET), subindexed with $F$, depending on $F_i$ and the stochastic RET (S-RET), subindexed with $g$, depending on $g_{i,i}$. The same subindexes will be taken along the manuscript for other quantities. The proof of (4) (in Appendix A1) comes from the expression of the Frobenius-Perron operator governing the time evolution of the pdf of $X_i$ in the stochastic process where $X_{\sim i}$ is frozen in the period $[t, t + dt]$, by using a similar procedure to that of Propositions VI.2, VI.3 and VI.4 of Liang in 2016 [16]. Since, the full state-space is spanned by $X_i$ and $X_{\sim i}$, there are no 'indirect' transfers, via excluded variables.

The rate of variation of $H_{X_i}$ (2) is obtained from the Fokker-Plank equation, governing the time evolution of the pdf $\rho_\mathbf{X}$ of the system:

$$\frac{\partial \rho_\mathbf{X}}{\partial t} = -\sum_{j=1}^{D}\frac{\partial F_j \rho_\mathbf{X}}{\partial X_j} + \frac{1}{2}\sum_{j,k=1}^{D}\frac{\partial^2 g_{j,k}\rho_\mathbf{X}}{\partial X_j \partial X_k}, \tag{5}$$

by multiplying by $-(1 + \log \rho_i)$ and integrating over the state-space domain. We further indistinctly assume one of the boundary conditions in the state-space: 1) vanishing $\rho_\mathbf{X}$ when $\|\mathbf{X}\| \to \infty$ or at the border of the pdf support set, 2) closed and non-diffusive boundary conditions i.e $\rho_\mathbf{X} F_i = \sum_{j,k=1}^{D}\frac{\partial g_{j,k}\rho_\mathbf{X}}{\partial X_k} = 0$ at borders perpendicular to the $X_i$ direction, 3) periodic boundary conditions at the support set border.

Under those quite generic boundary conditions, we obtain the rate of entropy change decomposed into the deterministic part (D-REC), depending on $F_i$ and the stochastic part (S-REC), depending on $g_{i,i}$ as:

$$\frac{dH_{X_i}}{dt} = \left(\frac{dH_{X_i}}{dt}\right)_F + \left(\frac{dH_{X_i}}{dt}\right)_g, \tag{6a}$$

$$\left(\frac{dH_{X_i}}{dt}\right)_F = E\left(\frac{\partial F_i}{\partial X_i}\right) + E\left[F_i \frac{\partial \log \rho_{\sim i|i}}{\partial X_i}\right], \tag{6b}$$

$$\left(\frac{dH_{X_i}}{dt}\right)_g = -\frac{1}{2}E\left[g_{i,i}\frac{d^2 \log \rho_i}{dX_i^2}\right], \tag{6c}$$

where $\rho_{\sim i|i}$ is the conditional pdf of $X_{\sim i}$ given $X_i$.

When any of the above considered boundary conditions do not hold, then extra terms must be added to (6b,6c). For instance, for open boundary conditions, we must add terms $\oint_{\partial S} \rho_{\mathbf{X}}[-\log(\rho_i)]\,\mathbf{F}\cdot\mathbf{n}\,dl$ and $\oint_{\partial S} \rho_{\mathbf{X}}[-\log(\rho_i)]\frac{1}{2}\nabla(\nabla\cdot\mathbf{G})\cdot\mathbf{n}\,dl$, respectively to $\left(\frac{dH_{X_i}}{dt}\right)_F$ and $\left(\frac{dH_{X_i}}{dt}\right)_g$, representing the externally driven inward (deterministic and diffusive) fluxes of entropy $H_{X_i}$ along the border $\partial S$ of the pdf support $S$ in the state-space. In the expressions, $\mathbf{n}$ is the inward unit vector, the dot represents inner product, $\nabla$ is the gradient operator and dl is the infinitesimal element of $\partial S$. Those terms appear, for instance, when a non-null $\mathbf{F}$ is imposed by some external control at the system boundary $\partial S$ as it in certain thermodynamical open systems. An example of that is the case of a mass of moist air subjected to liquid-evaporation and vapour-condensation and taking the mass $M_v$ of vapour as a dynamical state variable, which is lower-bounded by zero (dry conditions) and upper-bounded by the saturation value. Therefore, the sum of boundary terms contributing to the budget of Shannon entropy (uncertainty) is $-\rho_{dry}\log(\rho_{dry})\dot{M}_{v,dry} + \rho_{sat}\log(\rho_{sat})\dot{M}_{v,sat}$ where $\rho_{dry}, \rho_{sat}$ are the pdfs at dry and saturated states respectively, $\dot{M}_{v,dry}$ is the rate of liquid-evaporation at dry conditions and $\dot{M}_{v,sat}$ is the rate of vapour-condensation at saturated conditions. That shows that the importation/exportation of entropy through at the bounds of state-variables can affect the causality. Despite their great interest, the effect of boundary fluxes in the state-space is out of the scope of this study and will henceforth not be considered.

Therefore, under closed or periodic boundaries, the S-REC $\left(\frac{dH_{X_i}}{dt}\right)_g$ is further decomposed as:

$$\left(\frac{dH_{X_i}}{dt}\right)_g = \left(\frac{dH_{X_i}}{dt}\right)_{g,a} + \left(\frac{dH_{X_i}}{dt}\right)_{g,m} \tag{7a}$$

$$\left(\frac{dH_{X_i}}{dt}\right)_{g,a} = \frac{1}{2}E\left[g_{i,i}\left(\frac{d\log\rho_i}{dX_i}\right)^2\right] \geq 0 \tag{7b}$$

$$\left(\frac{dH_{X_i}}{dt}\right)_{g,m} = -\frac{1}{2}E\left[\frac{g_{i,i}}{\rho_i}\frac{d^2\rho_i}{dX_i^2}\right] \tag{7c}$$

where (7b) and (7c) are, respectively, the additive-noise part of S-REC and the multiplicative-noise part of S-REC.

(7b) is always non-negative, being an entropy source coming from the dynamical noise. The term (7c) is null whenever $g_{i,i}$ is constant (i.e. under an additive noise). A non-vanishing (7c) value calls for a state dependent or multiplicative noise.

After taking the difference between (6a) and (4), we obtain the SEG decomposed in its deterministic and stochastic parts:

$$\frac{dH_{X_i,self}}{dt} = \frac{dH_{X_i}}{dt} - T(X_{\sim i}\to X_i) = \left(\frac{dH_{X_i,self}}{dt}\right)_F + \left(\frac{dH_{X_i,self}}{dt}\right)_g \tag{8}$$

When we looks for the expressions of RET and SEG, is useful to decompose $F_i$ and $g_{i,i}$ as:

$$F_i(X_i, X_{\sim i}) = \hat{F}_i(X_i) + F_{i,\sim i}(X_i, X_{\sim i}) \tag{9a}$$

$$g_{i,i}(X_i, X_{\sim i}) = \hat{g}_{i,i,\sim i}(X_i) + g_{i,i,\sim i}(X_i, X_{\sim i}) \tag{9b}$$

where $\hat{F}_i, \hat{g}_{i,i,\sim i}$ are exclusive functions of $X_i$ and $F_{i,\sim i}, g_{i,i,\sim i}$ are the causal-depending parts of $F_i, g_{i,i}$ since they contain a cross dependency (from the point of view of $X_i$) on $X_{\sim i}$.

## 2.1. Deterministic part of RET and SEG

For the considered boundary conditions and after, some integral manipulations (e.g. integration by parts), we obtain the D-RET and D-SEG written respectively as:

**Theorem 1**

$$T(X_{\sim i} \to X_i)_F = E\left[F_i \frac{\partial \log \rho_{\sim i|i}}{\partial X_i}\right] = E\left[F_{i,\sim i} \frac{\partial \log \rho_{\sim i|i}}{\partial X_i}\right], \tag{10a}$$

$$\left(\frac{dH_{X_i,self}}{dt}\right)_F = E\left(\frac{\partial F_i}{\partial X_i}\right) \tag{10b}$$

See proof of (10a) in Appendix A2. (10b) is the difference between (6b) and (10a). From (10a,b), it is clear that D-RET (10a) and D-SEG (10b) are linear in terms of $F_i$, i.e. the D-RET and D-SEG of a linear combination of deterministic terms and diffusivities is the respective linear combination of D-RETs and D-SEGs.

The term (10a), can be interpreted by considering the surprise function $-\log \rho_i = -\log \rho + \log \rho_{\sim i|i}$ whose averaged value is $H_{X_i}$. The rate of variation accompanying the state-space evolution is $\frac{\partial(-\log \rho_i)}{\partial t} + \sum_{k=1}^{D} F_k \frac{\partial(-\log \rho_i)}{\partial X_k}$. The 'advective' term due to $F_i$ in the sum is $F_i \frac{\partial(-\log \rho_i)}{\partial X_i} = F_i \frac{\partial(-\log \rho)}{\partial X_i} + F_i \frac{\partial \log \rho_{\sim i|i}}{\partial X_i}$ and thus $F_i \frac{\partial \log \rho_{\sim i|i}}{\partial X_i}$ in (10a) contributes to augment entropy following the state-space trajectory. Furthermore, only the parcel of $F_i$ with cross-dependency on $X_{\sim i}$ (i.e. $F_{i,\sim i}$) contributes to the D-RET.

The entropy generation (10b) is the contribution of $F_i$ for the average state-space speed divergence. When all the parcels of $\text{div}(\mathbf{F}) \equiv \sum_{k=1}^{D} \frac{\partial F_k}{\partial X_k}$ are constant and negative, that leads to the contraction of state-space volumes and to the decreasing of entropy.

Therefore, the D-RET (10a) can be further written as (proven in Appendix A3):

$$T(X_{\sim i} \to X_i)_F = E\left[\frac{d}{dX_i} E(F_{i,\sim i}|X_i)\right] - E\left(\frac{\partial F_{i,\sim i}}{\partial X_i}\right) \tag{11}$$

The first term of the r.h.s. of (11) is the average derivative with respect to $X_i$ (or sensitivity) of the conditional expectancy $E(F_{i,\sim i}|X_i)$ of the causal-depending term $F_{i,\sim i}$ for a given $X_i$. Moreover, note that (11) is zero with $F_{i,\sim i}$ changed into $\hat{F}_i$. A non-null inference of causality (11) means that causes, (through $F_{i,\sim i}$), are acting in a different way on different consequence values $X_i$. For this reason, the application of (11)-like formulas will be part of the hereafter referred 'Causal Sensitivity Method' (CSM).

Now, recalling that the full expectation decomposes as $E(\ldots) = E_i[E_{\sim i}(\ldots|X_i)]$, we can write (11) as an the average of the specific D-RET, i.e:

$$T(X_{\sim i} \to X_i)_F = E(\mathcal{F}_{i,\sim i,F}) \tag{12a}$$

$$\mathcal{F}_{i,\sim i,F}(X_i) \equiv \frac{d}{dX_i} E_{\sim i}(F_{i,\sim i}|X_i) - E_{\sim i}\left(\frac{\partial F_{i,\sim i}}{\partial X_i}\Big|X_i\right). \tag{12b}$$

The function (12a) provides information about the states where the entropy transfer is positive, or negative, and also about its intensity.

A very important point, is the fact that the RET formula (10a) depends explicitly on the pdf (obtained from integration of the Fokker-Planck equation or estimation from data), whereas the alternative formula (11) depends only on conditional expectancies like $E(F_{i,\sim i}|X_i)$. This is a key advantage with respect to (10a) because the estimation of (10a) calls for a 'brute-force' and computationally expensive integration over the state-space, growing exponentially with dimension $D$ whereas (11) represents a much cheaper approach. There, $E_{\sim i}(F_{i,\sim i}|X_i)$ can be estimated from some linear or non-linear fitting of $F_{i,\sim i}$ as a function of $X_i$ by using an ensemble of realizations of the pdf of the state-vector $\mathbf{X}$, where the pdf can be either a transient ensemble or a long enough time-series if $\rho_\mathbf{X}$ is ergodic. Note in (11) that $E(F_{i,\sim i}|X_i) = E_{\sim i}(F_{i,\sim i}|X_i)$ in which $E_{\sim i}$ is the expectation over the variables in $X_{\sim i}$.

A corollary of (11) comes out:

**Corollary of Theorem 1**

If $F_{i,\sim i} = f_1(X_i) f_2(X_{\sim i})$, i.e. the deterministic term is multiplicative-separable in terms of factors depending on $X_i$ and $X_{\sim i}$ respectively, then:

$$T(X_{\sim i} \to X_i)_F = E_i \left[ f_1 \frac{d}{dX_i} E_{\sim i}(f_2|X_i) \right] \tag{13}$$

The proof is immediate since, using (11), we get $E_i \left[ \frac{d}{dX_i} E_{\sim i}(F_{i,\sim i}|X_i) \right] - E \left( \frac{\partial F_{i,\sim i}}{\partial X_i} \right) = E_i \left\{ \frac{d}{dX_i} [f_1 E_{\sim i}(f_2|X_i)] \right\} - E_i \left[ E_{\sim i} \left( f_2 \frac{df_1}{dX_i} | X_i \right) \right] = E_i \left[ f_1 \frac{d}{dX_i} E_{\sim i}(f_2|X_i) \right]$. The above separation is quite common (e.g. $F_i$ given by a multivariate polynomial of the variables). From (13), we easily verify that, If $X_{\sim i}$ is statistically independent from $X_i$, then $E_{\sim i}(f_2|X_i)$ is independent of $X_i$, its derivative is null and hence (13) vanishes. Therefore, a non-vanishing D-RET (13) comes from the statistical dependency on the consequential variable $X_i$ of the causal-dependent factor $f_2(X_{\sim i})$.

Let us give a concrete example of (13) in dimension $D = 3$, with $i = 1, \sim i = 2,3$ and $F_{i,\sim i} = F_{1,\sim 1} = X_1 X_2^2 + X_2 X_3$, which leads to $T(X_{\sim 1} \to X_1)_F = E \left[ X_1 \frac{d}{dX_1} E(X_2^2|X_1) \right] + E \left[ \frac{d}{dX_1} E(X_2 X_3|X_1) \right]$. That needs the estimation of two conditional expectations $E(X_2^2|X_1)$ and $E(X_2 X_3|X_1)$ and their derivatives with respect to $X_1$. Thanks to the variable's separability, the CSM becomes a very expeditious and elegant way of computing RETs.

**2.2. Stochastic part of RET and SEG**

The S-REC (6c) depends on the diffusivity $g_{i,i}$ and how it depends on $X_i$ and $X_{\sim i}$. The additive-noise part of S-REC (7b) is retained in the S-SEG, i.e.:

$$\left( \frac{dH_{X_i,self}}{dt} \right)_{g,a} = \left( \frac{dH_{X_i}}{dt} \right)_{g,a} = \frac{1}{2} E_i \left[ E(g_{i,i}|X_i) \left( \frac{d \log \rho_i}{dX_i} \right)^2 \right] \geq 0, \tag{14}$$

which contributes to increase the entropy. In a random walk process (14) is the sole entropy budget term, leading to a variance increasing with a constant time rate $g_{i,i}/2$.

In what concerns the multiplicative-noise term (7c), it decomposes as:

$$\left( \frac{dH_{X_i}}{dt} \right)_{g,m} = \left( \frac{dH_{X_i,self}}{dt} \right)_{g,m} + T(X_{\sim i} \to X_i)_g \tag{15}$$

whose expressions are given by:

**Theorem 2**

$$\left( \frac{dH_{X_i,self}}{dt} \right)_{g,m} = E_i \left( -\frac{d^2 E(g_{i,i}|X_i)}{dX_i^2} \right) \tag{16a}$$

$$T(X_{\sim i} \to X_i)_g = \frac{1}{2} E \left( \frac{1}{\rho_i} \frac{\partial^2 g_{i,i} \rho_i}{\partial X_i^2} \right) = \frac{1}{2} E \left[ \frac{1}{\rho_i} \frac{\partial^2 g_{i,i \sim i} \rho_i}{\partial X_i^2} \right]. \tag{16b}$$

We must note here, that the S-SEG terms (14, 16a) depend uniquely on the conditional expectation of the diffusivity $E(g_{i,i}|X_i)$, which is an exclusive function of $X_i$, independently of any cross-dependency. The term (16a) is proportional to the concavity of $E(g_{i,i}|X_i)$, through the second derivative in (16a) (see proof in Appendix A4). The non-vanishing of the S-RET (16b) calls for a cross-dependency of the diffusivity $g_{i,i}$ on $X_{\sim i}$. In fact, if $g_{i,i}$ is constant or owns a unique self-dependency on $X_i$, then (16b) is zero.

After, some integral manipulations, the S-RET (16b) get a CSM-like expression:

$$T(X_{\sim i} \to X_i)_g = \frac{1}{2} E\left(\frac{\partial^2 g_{i,i\sim i}}{\partial X_i^2}\right) + \frac{1}{2} E\left[\frac{d^2}{dX_i^2} E(g_{i,i\sim i}|X_i)\right] - E\left[\frac{d}{dX_i} E\left(\frac{\partial g_{i,i\sim i}}{\partial X_i}\bigg|X_i\right)\right] \tag{17}$$

which depends on the causal sensitivities (first and second derivatives) of the conditional diffusivity $E(g_{i,i\sim i}|X_i)$ with respect to $X_i$.

The proof of (17) comes in the Appendix A5. Like (11), the formula (17) has no explicit dependency on the pdf, hence its calculation is rather cheaper.

The specific contribution for the S-RET (17) like for the D-RET in (12b) is:

$$\mathcal{F}_{i,\sim i,g}(X_i) \equiv \frac{1}{2} E\left(\frac{\partial^2 g_{i,i\sim i}}{\partial X_i^2}\bigg|X_i\right) + \frac{1}{2} E \frac{d^2}{dX_i^2}(g_{i,i\sim i}|X_i) - \frac{d}{dX_i} E\left(\frac{\partial g_{i,i\sim i}}{\partial X_i}\bigg|X_i\right) \tag{18}$$

From (17) we infer the corollary below:

**Corollary of Theorem 2**

If $g_{i,i,\sim i} = f_3(X_i) f_4(X_{\sim i})$, i.e. the diffusivity is multiplicatively-separable in terms of factors depending on $X_i$ and $X_{\sim i}$ respectively, then

$$T(X_{\sim i} \to X_i)_g = \frac{1}{2} E_i \left[f_3 \frac{d^2}{dX_i^2} E_{\sim i}(f_4|X_i)\right] \tag{19}$$

The proof looks like that of the Corollary of Theorem 1, playing with second derivatives and separating the factors $f_3, f_4$ in the integrals spanning $X_i$ and $X_{\sim i}$ respectively in the expectation operator. The formula (19), like (13) provides a very simple form to compute the S-RET.

**2.3. Stationary balance of entropy**

If the pdf $\rho_X$ is stationary, for instance when the system (1) is autonomous and reaches the ergodic pdf, then the entropy $H_{X_i}$ evolves towards a stationary value and all the terms of $\frac{dH_{X_i}}{dt}$ must balance, i.e. their sum is zero as:

$$\frac{dH_{X_i}}{dt} = \left(\frac{dH_{X_i,self}}{dt}\right)_F + T(X_{\sim i} \to X_i)_F + \left(\frac{dH_{X_i,self}}{dt}\right)_g + T(X_{\sim i} \to X_i)_g = 0 \tag{20}$$

The SEG contributions can be estimated from model equations, through the expressions of $F_i, g_{i,i}$ and also from the marginal pdf $\rho_i$, estimated from data. The RET terms can be obtained by the CSM expressions (11,17) from data-estimated conditional expectations on $X_i$ and their derivatives. That is important to check the correct estimation of every term. Under additive noise conditions, the D-RET is given by:

$$T(X_{\sim i} \to X_i)_F = -\left(\frac{dH_{X_i,self}}{dt}\right)_F - \left(\frac{dH_{X_i,self}}{dt}\right)_{g,a} \tag{21}$$

Furthermore, In the case of a noise-free dissipative model that moves to the balance: $T(X_{\sim i} \to X_i)_F = -\left(\frac{dH_{X_i,self}}{dt}\right)_F > 0$.

## 3. Rate of entropy transfer due to a single variable

### 3.1. Integral expressions of the RET

Let us now consider the RETs, $T(X_j \to X_i)$, $j \neq i$, between single variables as in [16]. The integrated influence of $X_{\sim i}$ upon the consequential variable $X_i$, through $T(X_{\sim i} \to X_i)$ cannot in general be split as the sum of single RETs, $T(X_j \to X_i)$, $j \neq i$, except when variables in $X_{\sim i}$ are conditionally independent with respect to $X_i$. We will come back to this point at the end of this section.

Liang [11-18] has established formulas for the deterministic and stochastic, single contributions of $T(X_j \to X_i)$, $j \neq i$. We introduce here alternative formulas to those of Liang, looking like (11-17) and relying on the CSM approach.

We start with a result (proven in Appendix A6).

**Theorem 3**

The rate of entropy transfer $T(X_j \to X_i)$ towards the consequential variable $X_i$ due the influence of the causal variable $X_j$, $j \neq i$, is decomposed into a D-RET and S-RET as:

$$T(X_j \to X_i) \equiv T(X_j \to X_i)_F + T(X_j \to X_i)_g, \qquad (22a)$$

$$T(X_j \to X_i)_F = -\int_{\mathbb{R}^D} \rho_{j|i} \frac{\partial(F_i \rho_{\sim j})}{\partial X_i} d\mathbf{X} = E\left[F_i \frac{\rho_{j|i}}{\rho_{j|\sim j}} \frac{\partial \log(\rho_{j|i})}{\partial X_i}\right], \qquad (22b)$$

$$T(X_j \to X_i)_g = \frac{1}{2}\int_{\mathbb{R}^D} \rho_{j|i} \frac{\partial^2(g_{i,i}\rho_{\sim j})}{\partial X_i^2} d\mathbf{X} = E\left[\left(-\frac{1}{2\rho_{\sim j}}\frac{\partial(g_{i,i}\rho_{\sim j})}{\partial X_i}\right)\frac{\rho_{j|i}}{\rho_{j|\sim j}}\frac{\partial \log(\rho_{j|i})}{\partial X_i}\right], \qquad (22c)$$

where $\rho_{\sim j}$ is the pdf of the set of variables different from $X_j$ and $\rho_{j|i}, \rho_{j|\sim j}$ are conditional pdfs. Liang [16] has obtained $T(X_j \to X_i)$ as the difference between the rate $\frac{dH_{X_i}}{dt}$ and the of rate of change of $H_{X_i}$ in a process in which $X_j$ is frozen. The middle terms of (22b,c) are from Liang [16], and the rightmost ones are the new relationships based on the conditional pdfs. For the two-dimensional case $\frac{\rho_{j|i}}{\rho_{j|\sim j}} = 1$ and $T(X_j \to X_i) = T(X_{\sim i} \to X_i)$. Moreover, if variables $X_j \neq X_i$ are statistically independent from the remaining variables $X_k \neq X_i, X_j$, then $\frac{\rho_{j|i}}{\rho_{j|\sim j}} = 1$, which is equivalent to the two-dimensional case spanned uniquely by the causal and consequential variables $X_j, X_i$.

By summing (22b) and (22c), we get a compact formula:

$$T(X_j \to X_i) = E\left[R_i \frac{\rho_{j|i}}{\rho_{j|\sim j}}\frac{\partial \log(\rho_{j|i})}{\partial X_i}\right] = E\left[R_{i,j}\frac{\rho_{j|i}}{\rho_{j|\sim j}}\frac{\partial \log(\rho_{j|i})}{\partial X_i}\right] \qquad (23a)$$

$$R_i = F_i - \frac{1}{2\rho_{\sim j}}\frac{\partial(g_{i,i}\rho_{\sim j})}{\partial X_i} = F_i - g_{i,i}\frac{1}{2}\frac{\partial \log(\rho_{\sim j})}{\partial X_i} - \frac{1}{2}\frac{\partial g_{i,i}}{\partial X_i}. \qquad (23b)$$

where $R_i$ is a generalized speed in state-space composed by the deterministic speed ($F_i$) and stochastic components, this one, poiting towards lower pdf values and lower diffusivities (i.e. pointing against the gradient). Note in (23a) that $\frac{\partial \log(\rho_{j|i})}{\partial X_i} = \frac{\partial \log cop(i,j)}{\partial X_i}$ where $cop(X_i, X_j) \equiv \frac{\rho_{i,j}}{\rho_i \rho_j}$ is the copula function, expressing the form how variables are interrelated, independently of the marginal pdfs. The average value of $\log[cop(X_i, X_j)]$ is the mutual information between $X_i$ and $X_j$. We can shown that (23a) appears with opposite sign in the budget equation of the copula entropy $-I(X_j, X_i) \equiv -E[\log(\frac{\rho_{i,j}}{\rho_i \rho_j})]$, which is the opposite of the mutual information, and thus the RET is an indirect entropy transfer, via the copula entropy.

The use of $D$-dimensional integrals for estimating the RETs like (23a), calls for the estimation of the pdf of the system. This can be difficult due to data restrictions and sampling, even for moderately low dimensions. Moreover, the state-space discretization and bounding required for the numerical computation of the integrals represents an additional source error. This will be further illustrated in Secs. 5 and 6 when computing RETs for two different models.

There are alternative integral formulas for the RETs containing less sources of errors, which relies on the sensitivities (derivatives) of $F_i, g_{ii}$ with respect to $X_j$, agreeing with the Theorem below, proven in Appendix A7:

**Theorem 4**

$$T(X_j \to X_i) = E\left[\frac{1}{\rho_{j|\sim j}} \frac{\partial Pr_{ex}(j|i)}{\partial X_i} \frac{\partial R_i}{\partial X_j}\right] \quad (24)$$

where we use the conditional exceedance probability of $X_j$ given $X_i$: $Pr_{ex}(j|i) \equiv \text{Prob}(u_j \geq X_j | X_i) = \int_{X_j}^{+\infty} \rho_{j|i}(u_j | X_i) du_j = 1 - Pr(j|i)$. Its derivative in (24) is $\frac{\partial Pr_{ex}(j|i)}{\partial X_i} = \int_{X_j}^{+\infty} \frac{\partial \rho_{j|i}(u_j | X_i)}{\partial X_i} du_j = \int_{X_j}^{+\infty} \rho_{j|i} \frac{\partial \log \rho_{j|i}}{\partial X_i} du_j$ which vanishes if $X_j$ and $X_i$ are statistically independent.

The sensitivity of the generalized speed $R_i$ (23b) with respect to the causal variable $X_j$ appearing in (24) is:

$$\frac{\partial R_i}{\partial X_j} = \frac{\partial F_i}{\partial X_j} - \frac{1}{2} \frac{\partial g_{i,i}}{\partial X_j} \frac{\partial \log(\rho_{\sim j})}{\partial X_i} - \frac{1}{2} \frac{\partial^2 g_{i,i}}{\partial X_j \partial X_i} \quad . \quad (25)$$

Splitting the first, second and third terms leads to corresponding formulas for the D-RET $T(X_j \to X_i)_F$ and S-RET $T(X_j \to X_i)_g$ as:

$$T(X_j \to X_i)_F = E\left[\frac{1}{\rho_{j|\sim j}} \frac{\partial Pr_{ex}(j|i)}{\partial X_i} \frac{\partial F_i}{\partial X_j}\right] \quad (26a)$$

$$T(X_j \to X_i)_g = E\left[\frac{1}{\rho_{j|\sim j}} \frac{\partial Pr_{ex}(j|i)}{\partial X_i} \left(-\frac{1}{2} \frac{\partial g_{i,i}}{\partial X_j} \frac{\partial \log(\rho_{\sim j})}{\partial X_i} - \frac{1}{2} \frac{\partial^2 g_{i,i}}{\partial X_j \partial X_i}\right)\right] \quad (26b)$$

Let us interpret heuristically the sign of $T(X_j \to X_i)$ in (24). A positive value of $T(X_j \to X_i)$ is favoured by $\frac{\partial Pr_{ex}(j|i)}{\partial X_i}$ and $\frac{\partial R_i}{\partial X_j}$ having the same sign. In particular, $\frac{\partial Pr_{ex}(j|i)}{\partial X_i} > 0$ means that for a state-space displacement $\delta X_i > 0$, the tail (or exceedance) probability $\text{Prob}(u_j \geq X_j | X_i)$ is larger, and thus values greater than $X_j$ are more likely, i.e. $\delta X_i > 0$ drives $\delta X_j > 0$. Furthermore, if $\frac{\partial R_i}{\partial X_j} > 0$, that induces a variation $\frac{\delta R_i}{\delta X_i} = \frac{\partial R_i}{\partial X_j} \frac{\delta X_j}{\delta X_i} > 0$ and therefore contributes positively to the divergence of the generalized speed and in analogy with (10b) to an increase of $H_{X_i}$.

Given the dependence on the sensitivity term $\frac{\partial R_i}{\partial X_j}$ in (25), we have the following corollary:

**Corollary of Theorem 4**

Only the additive terms of $R_i$ in (8), depending explicitly on $X_j$ contribute to $T(X_j \to X_i)$ and therefore $R_i$ can be substituted in (8) by

$$R_{i,j} \equiv F_{i,j} - \frac{1}{2\rho_{\sim j}} \frac{\partial(g_{i,i,j}\rho_{\sim j})}{\partial X_i} = F_{i,j} - g_{i,i,j}\frac{1}{2}\frac{\partial \log(\rho_{\sim j})}{\partial X_i} - \frac{1}{2}\frac{\partial g_{i,i,j}}{\partial X_i} \tag{27}$$

where $F_{i,j}$, $g_{i,i,j}$ (like 9a,b) is the sum of the additive terms of $F_i, g_{i,i}$ that depend on $X_j$, i.e. satisfying $\frac{\partial F_i}{\partial X_j} = \frac{\partial F_{i,j}}{\partial X_j}$ and $\frac{\partial g_{i,i}}{\partial X_j} = \frac{\partial g_{i,i,j}}{\partial X_j}$. For instance for D=3, if $F_1 = X_1 X_2 + X_2^2 + X_1 X_3$, then $F_{1,2} = X_1 X_2 + X_2^2$ and $F_{1,3} = X_1 X_3$.

The difference $R_i - R_{i,j}$ includes all the terms not depending on $X_j$, in particular the diffusion speed, under additive noise, i.e. the term $-g_{i,i}\frac{1}{2}\frac{\partial \log(\rho_{\sim j})}{\partial X_i}$ in (27). In the case of additive or self-dependent multiplicative noise, we have $R_{i,j} = F_{i,j}$. The restriction to terms in $R_{i,j}$ in (27) simplifies considerably the computation of (24) since many terms can be discarded. It also shows that all the terms without dependence on $X_j$ have only an implicit impact on the RET via the shape of the pdf.

### 3.2. Expressions of the RETs based on the Causal Sensitivity Method (CSM)

Let us now focus on how (26a,b) can be modified in order to enable a simple estimation based on conditional expectations, by using the CSM already used to reach (11-17),

In order to simplify the notation, we split the variables into three kinds: the causal variable $X_i$, the consequential variable $X_j$ (the target) and the remaining outer variables within the state vector, denoted by $X_k$ where $k \equiv \sim(j,i)$ stand for variables different from the pair $(X_j, X_i)$. The outer space, spanned by $X_k$ can be extended from the null set up to any variable physically possible or in practice those which have plausible connections to $X_j, X_i$ in a certain context and so $X_k$ spans the contextual variables, like third-part agents interacting with $X_j, X_i$.

That outer space also contains all the possible factors that can lead from $X_j$ to $X_i$, by any sort of causal mechanism. The average of the deterministic term and diffusivity along the outer space, for a fixed value of the consequential variable, takes a very important role in the causality diagnostics like the RETs. The RET values can even depend crucially on the extent of the outer space of variables.

Let us denote the overbar quantities as

$$\bar{F}_{i,j} \equiv E_k(F_{i,j}|X_i) = E_{\sim(i,j)}(F_{i,j}|X_i) \tag{28a}$$

$$\bar{g}_{i,i,j} \equiv E_k(g_{i,i,j}|X_i) = E_{\sim(i,j)}(g_{i,i,j}|X_i) \tag{28b}$$

where

$$E_k(\ldots|X_i) \equiv \int_{X_k} \rho_{k|i}(\ldots) dX_k = \int_{X_{\sim(i,j)}} \rho_{\sim(i,j)|i}(\ldots) dX_{\sim(i,j)} \tag{28c}$$

is the conditional expectation to $X_i$, over the outer space of variables $k = \sim(j,i)$. Those quantities are bivariate functions of $(X_i, X_j)$. In particular, in the two-dimensional isolated case (cause, consequence) or in the two-dimensional equivalent case where the outer variables are totally independent of $X_i$, we have $\bar{F}_{i,j} = F_{i,j}$ and $\bar{g}_{i,i,j} = g_{i,i,j}$.

For instance, in fluid dynamics, the Reynold turbulent stresses, parametrized as a function of large-scale variables $X_i, X_j$ is a conditional expectation of the form $\bar{F}_{i,j}$.

Let us also denote by $E_i(\ldots)$ and $E_j(\ldots)$ the expectations over the causal and consequential space respectively and by $E_{i,j}(\ldots) = E_i[E_j(\ldots)|X_i]$ the joint average.

Now, we have the CSM equivalent theorem (proven in Appendices A8,A9) :

**Theorem 5**

$$T(X_j \to X_i)_F = E_i\left[\frac{d}{dX_i} E_j(\bar{F}_{i,j}|X_i)\right] - E_{i,j}\left(\frac{\partial \bar{F}_{i,j}}{\partial X_i}\right) \tag{29a}$$

$$T(X_j \to X_i)_g = \frac{1}{2} E_{i,j}\left(\frac{\partial^2 \bar{g}_{i,i,j}}{\partial X_i^2}\right) + \frac{1}{2} E_i\left[\frac{d^2}{dX_i^2} E_j(\bar{g}_{i,i,j}|X_i)\right] - E_i\left[\frac{d}{dX_i} E_j\left(\frac{\partial \bar{g}_{i,i,j}}{\partial X_i}\bigg|X_i\right)\right] \tag{29b}$$

where the expectation's variables are sub-indexed in the expectation operator, for ease of interpretation.

(29a,b) show that the RETs depend on conditional consequential averages $E_j(\ldots|X_i)$ of the deterministic terms and diffusivities (averaged over the outer space) and their first and second derivatives with respect to the $X_i$. The terms in (29a,b) do not vanish in general because the above used operators: $E_j(\ldots|X_i)$ and $\frac{d(\ldots)}{dX_i}$ do not commute.

The formulas (29a,b) are linear in $F_{i,j}$, $g_{i,i,j}$, and thus the single RETs coming from a linear combination of deterministic terms or diffusivities is the linear combination of single RETs produced by each term.

The RET is the average value of the specific rate of entropy transfer $\mathcal{F}_j(X_i)$, in analogy with (12,18) i.e.

$$T(X_j \to X_i) = E_i[\mathcal{F}_{i,j}(X_i)], \tag{30}$$

to which contribute the corresponding determinist and stochastic components:

$$\mathcal{F}_{i,j}(X_i) = \mathcal{F}_{i,j,F}(X_i) + \mathcal{F}_{i,j,g}(X_i) \tag{31a}$$

$$\mathcal{F}_{i,j,F}(X_i) = \frac{d}{dX_i} E_j(\bar{F}_{i,j}|X_i) - E_j\left(\frac{d\bar{F}_{i,j}}{dX_i}\bigg|X_i\right) \tag{31b}$$

$$\mathcal{F}_{i,j,g}(X_i) = -\frac{d}{dX_i} E_j\left(\frac{\partial \bar{g}_{i,i,j}}{\partial X_i}\bigg|X_i\right) + \frac{1}{2}\frac{d^2}{dX_i^2} E_j(\bar{g}_{i,i,j}|X_i) + \frac{1}{2} E_j\left(\frac{\partial^2 \bar{g}_{i,i,j}}{\partial X_i^2}\bigg|X_i\right) \tag{31c}$$

We must note that the value of $\mathcal{F}_{i,j}(X_i)$ is constant for Gaussian pdfs associated to linear systems but can vary for non-Gaussian pdfs, yielding to contributions to the rate of entropy that can fluctuate from positive to negative values depending on $X_i$.

The formulas (29a,b) have the major advantage as compared with (26a,b) that its computation does not explicitly require the pdf of the system. In practice, the terms $\bar{F}_{i,j} = E_k(F_{i,j}|X_i)$, $\bar{g}_{i,i,j} = E_k(g_{i,i,j}|X_i)$ and other terms like $E_j(\bar{F}_{i,j}|X_i)$ and $E_j\left(\frac{\partial \bar{F}_{i,j}}{\partial X_i}\bigg|X_i\right)$ are conditional expectations with respect to $X_i$ and can be estimated by nonlinear regression using a certain set of basis functions (e.g. monomials) or by parametrizing the joint pdfs by Bayesian methods, followed by an estimation of the average. The regression can be obtained from an ensemble of realizations of state-space trajectories - like time-series – exploring the ergodic pdf, or a time-evolving transient ensemble where the regression coefficients vary on time.

Theorem 5 has important corollaries allowing for simple applications to quite general systems, linear, or nonlinear with additive or multiplicative noises with or without cross dependencies. They also provide relevant information about the rates of entropy transfer in models where the governing equations are known.

**Corollary 1 of Theorem 5**

If $F_{i,j} = A_{ij}X_j = \bar{F}_{i,j}$ where $A_{ij}$ is constant, i.e. the deterministic term is linear with respect to the causal variable $X_j$, then

$$T(X_j \to X_i)_F = A_{ij} E_i \left[ \frac{d}{dX_i} E_j(X_j|X_i) \right] \tag{32}$$

The systems verifying that condition can even depend nonlinearly both on $X_i$ and variables $X_k, k \neq i,j$. The proof is straightghforward, since $\frac{\partial \bar{F}_{i,j}}{\partial X_i} = 0$ and $E_j(\bar{F}_{i,j}|X_i) = A_{ij}E_j(X_j|X_i)$. Then the RET comes as $E_i\left[\frac{d}{dX_i}E_j(\bar{F}_{i,j}|X_i)\right] = A_{ij}E_i\left[\frac{d}{dX_i}E_j(X_j|X_i)\right]$ providing the result. The result also applies in particular to linear systems giving rise to:

**Corollary 2 of Theorem 5**

For a linear system of dimension $D$:

$$\frac{d\mathbf{X}}{dt} = \mathbf{AX} + \mathbf{B}\frac{d\mathbf{W}}{dt} \tag{33}$$

where $\mathbf{A}$ is a constant matrix and noise is additive, i.e $\mathbf{B}$ is a constant matrix, the ergodic pdf $\rho_\mathbf{X}$, is Gaussian with zero average and covariance matrix $\mathbf{C_{x,x}}$ with entries $C_{ij}$ and the rate of entropy transfer is:

$$T(X_j \to X_i) = \frac{C_{ij}}{C_{ii}} A_{ij} = \frac{C_{ij}}{C_{ii}} \frac{1}{\det(\mathbf{C_{x,x}})} \sum_{l=1}^{D} C_{d(i),l} \Delta_{lj}, \tag{34}$$

where the $(i,j)$ entry of $\mathbf{A}$ is $A_{ij} = \frac{\partial F_i}{\partial X_j}$; $C_{d(i),l}$ is the covariance between $\frac{dX_i}{dt}$ and $X_l$ $(l=1,..,D)$ and $\Delta_{lj}$ is the $(l,j)$ cofactor-entry of $\mathbf{C_{x,x}}$. In the ergodic Gaussian (null average) case, we have $E_j(X_j|X_i) = \frac{C_{ij}}{C_{ii}}X_i$ and thus $\frac{d}{dX_i}E_j(X_j|X_i) = \frac{C_{ij}}{C_{ii}}$. After using (32), we get the middle term of result (34), recovering the result of Eq. (14) of Liang [18].

The last equality is obtained by taking the matrix right hand side product of $\frac{d\mathbf{X}}{dt}$ with $\mathbf{X}^T$ in (33), followed by the expectation operator. Recalling that $E\left(\frac{d\mathbf{w}}{dt}\mathbf{X}^T\right) = \mathbf{0}$, we obtain the covariance matrix between time derivatives of the state vector and $\mathbf{X}$ as $\mathbf{C}_{\frac{d\mathbf{X}}{dt},\mathbf{x}} = E\left(\frac{d\mathbf{X}}{dt}\mathbf{X}^T\right) = \mathbf{A}E(\mathbf{XX}^T) = \mathbf{AC_{x,x}}$. Then $\mathbf{A} = \mathbf{C}_{\frac{d\mathbf{X}}{dt},\mathbf{x}} \mathbf{C_{x,x}}^{-1}$. Then, the result follows from the fact that $(\mathbf{C_{x,x}}^{-1})_{lj} = \frac{1}{\det(\mathbf{C_{x,x}})}\Delta_{lj}$.

In the particular case of a two-dimensional linear system driven by additive noise, we have, after computing the cofactors, the Liang [15] result:

$$T(X_j \to X_i) = \frac{C_{ii}C_{ij}C_{j,d(i)} - C_{ij}^2 C_{i,d(i)}}{C_{ii}^2 C_{jj} - C_{ii}C_{ij}^2} \tag{35}$$

A very important corollary of Theorem 5, based on the CSM is the case where the deterministic terms of $F_{i,j}$ and diffusivity terms of $g_{i,j}$ have a separate product dependence between the causal variable and the remaining variables, leading the following result (proven in Appendices A10,A11):

**Corollary 3 of Theorem 5**

For the separable case $F_{i,j} = f_1(X_j)f_2(X_i, X_k)$ and $g_{i,i,j} = f_3(X_j)f_4(X_i, X_k)$, where $f_1, f_2, f_3, f_4$ are certain continuous differentiable functions, the RET simplifies as:

$$T(X_j \to X_i)_F = E_i\left[E_k(f_2|X_i)\frac{d}{dX_i}E_j(f_1|X_i)\right] \tag{36a}$$

$$T(X_j \to X_i)_g = \frac{1}{2}E_i\left[E_k(f_4|X_i)\frac{d^2}{dX_i^2}E_j(f_3|X_i)\right] \tag{36b}$$

Those values correspond to averages of specific RETs: $\mathcal{F}_{i,j,F}(X_i) = E_k(f_2|X_i)\frac{d}{dX_i}E_j(f_1|X_i)$ and $\mathcal{F}_{i,j,g}(X_i) = \frac{1}{2}E_k(f_4|X_i)\frac{d^2}{dX_i^2}E_j(f_3|X_i)$ respectively for (36a) and (36b). The result can be extended as follows when separability works for $f_2$ and $f_4$:

**Corollary 4 of Theorem 5**

In the particular case $F_{i,j} = f_1(X_j)f_{2a}(X_i)f_{2b}(X_k)$, then:

$$T(X_j \to X_i)_F = E_i\left[f_{2a}(X_i)\, E_k(f_{2b}|X_i)\frac{d}{dX_i}E_j(f_1|X_i)\right] \tag{37a}$$

In a similar way, for $g_{i,i,j} = f_3(X_j)f_{4a}(X_i)f_{4b}(X_k)$, we get:

$$T(X_j \to X_i)_g = \frac{1}{2}E_i\left[f_{4a}(X_i)\, E_k(f_{4b}|X_i)\frac{d^2}{dX_i^2}E_j(f_3|X_i)\right] \tag{37b}$$

The above closed formulas should be applied to each term of the deterministic part and diffusivity. These relations constitute the practical basis of the CSM for the single rates of entropy transfer.

Let us give a concrete example in a 3D system to compute $T(X_2 \to X_1)$ where

$$F_1 = \exp(X_1)\sin(X_2) + X_1 X_3 \tag{38a}$$

$$g_{1,1} = (X_3 - X_1 X_2)^2 = X_3^2 + X_1^2 X_2^2 - 2X_1 X_2 \geq 0 \tag{38b}$$

The first step is to obtain $F_{1,2} = \exp(X_1)\sin(X_2)$ and $g_{1,1,2} = X_1^2 X_2^2 - 2X_1 X_2$ to which we apply (37a,b), obtaining:

$$T(X_2 \to X_1)_F = E_1\left\{\exp(X_1)\frac{dE_2[\sin(X_2)|X_1]}{dX_1}\right\} \tag{39a}$$

$$T(X_2 \to X_1)_g = \frac{1}{2}E_1\left[X_1^2 \frac{d^2 E_2(X_2^2|X_1)}{dX_1^2} - 2X_1 \frac{d^2 E_2(X_2|X_1)}{dX_1^2}\right] \tag{39b}$$

### 3.3. Causal synergies and rates of entropy transfer

At the beginning of Sec. 3.1, the problem of relating the global RET $T(X_{\sim i} \to X_i)$ with the RET between pairs of variables was posed. In order to figure out the link between these different quantities, we consider the simplest situation of dimension $D = 3$ (without loss of generality) and a multiplicatively separable deterministic term $F_1(X_1, X_2, X_3) = f_1(X_1)f_2(X_2)f_3(X_3)$. Then, applying (13), the RET $T(X_{\sim 1} \to X_1)$ is:

$$T(X_{\sim 1} \to X_1)_F = T(X_{2,3} \to X_1)_F = E_1\left[f_1 \frac{d}{dX_1}E_{2,3}(f_2 f_3|X_1)\right] \tag{40}$$

Now, let us decompose the averaged conditional product as $E_{2,3}(f_2 f_3|X_i) = E_2(f_2|X_i)E_3(f_3|X_1) + cov(f_2, f_3|X_1)$, i.e. the product of conditional expectations plus the conditional covariance. Next, after taking the derivative and applying the single RET expression (37a), we get:

$$T(X_{\sim 1} \to X_1)_F = T(X_{2,3} \to X_1)_F = T(X_{2,3} \to X_1)_{F,sing} + +T(X_{2,3} \to X_1)_{F,syn} \tag{41a}$$

$$T(X_{2,3} \to X_1)_{F,sing} = T(X_2 \to X_1)_F + T(X_3 \to X_1)_F \tag{41b}$$

$$T(X_{2,3} \to X_1)_{F,syn} = E_1\left[f_1 \frac{d}{dX_1} cov(f_2, f_3|X_1)\right] \tag{41c}$$

which shows that, the joint RET is the sum of single RETs (41b) plus a synergetic term (41c) related to the conditional covariance between the functions of the complementary variables $X_2, X_3$. The same procedure can be replicated for multiple variables putting into evidence the effect of synergies of couples, triplets and other groups of variables which are accounted for in multivariate forms of the mutual information, like the Interaction Information for triplets [28].

The sum of single RETs equals the global RET when all the non-consequential variables are conditionally independent with respect to the consequential variable.

This led us to present the concept of synergetic rate of entropy transfer as the difference between the global RET and the sum of single RETs and the corresponding deterministic and stochastic contributions:

$$T(X_{\sim i} \to X_i)_{syn} = T(X_{\sim i} \to X_i) - \sum_{j \neq i} T(X_j \to X_i) = T(X_{\sim i} \to X_i)_{F,syn} + T(X_{\sim i} \to X_i)_{g,syn} \tag{42a}$$

$$T(X_{\sim i} \to X_i)_{sing} = \sum_{j \neq i} T(X_j \to X_i) = T(X_{\sim i} \to X_i)_{F,sing} + T(X_{\sim i} \to X_i)_{g,sing} \tag{42b}$$

For linear systems the deterministic synergetic RET vanishes, though this is not a sufficient condition of linearity.

### 3.4 Rate of entropy transfer between transformed variables

We address in this section the problem of how variable changes can affect the values of $T(X_j \to X_i)$. Ideally, in a naive view, any independent variable changes should not influence the RET because the causality only depends on the link between variables and not on how variables or processes are represented, in particular the physical units.

The next theorem checks in a rigorous manner, the effect of one-to-one diffeomorphism variable-changes in the rate of entropy transfer. For that, let us consider new variables $\hat{X}_i, \hat{X}_j, \hat{X}_k$ defined in a union of open intervals (eventually not connected), which are functions respectively of $X_i, X_j, X_k$, where $k \equiv \sim(j,i)$ stands for variables different from $j, i$ and where, without loss of generality, all the jacobians are strictly positive, i.e. $\mathfrak{J}_i \equiv \frac{d\hat{X}_i}{dX_i} > 0, \mathfrak{J}_j \equiv \frac{d\hat{X}_j}{dX_j} > 0, \mathfrak{J}_k \equiv \frac{d\hat{X}_k}{dX_k} > 0$. Then, we have the following result proven in Appendix A12.

**Theorem 6**

The rate of entropy transfer (RET) between one-to-one diffeomorphism variable changes is:

$$T(\hat{X}_j \to \hat{X}_i) = T(X_j \to X_i) + T(X_j \to X_i)_{tr} \tag{43a}$$

$$T(X_j \to X_i)_{tr} = -\frac{1}{2} E\left[\frac{1}{\rho_{j|\sim j}} \frac{\partial Pr_{ex}(j|i)}{\partial X_i} \frac{\partial g_{i,i}}{\partial X_j} \frac{d \log \mathfrak{J}_i}{dX_i}\right] = E_i\left\{\frac{d \log \mathfrak{J}_i}{dX_i}\left[\frac{d}{dX_i} E_j(\bar{g}_{i,i,j}|X_i) - E_j\left(\frac{\partial \bar{g}_{i,i,j}}{\partial X_i}|X_i\right)\right]\right\} \tag{43b}$$

The result (43b) shows that the RET correction $T(X_j \to X_i)_{tr}$ is a unique function of the diffusion term, being the deterministic part of $T(X_j \to X_i)$ kept invariant for any one-to-one diffeomorphisms of variables. Only the stochastic RET can change, being invariant when any of the factors in (43b) vanish, i.e.: the noise is additive or multiplicative without cross dependencies, i.e. $\frac{\partial g_{i,i}}{\partial X_j} = 0$ or when the variable transformation $X_i \to \hat{X}_i$ is affine (translation and scaling) leading to $\frac{\partial \log \Im_i}{\partial X_i} = 0$. This guarantees the RET invariance under quite generic conditions. This is a revision of Remark 1 of Theorem III.3 of Liang [17], showing that in certain conditions, it is possible to make invertible changes of the consequential variable, keeping the value of the entropy transfer. The variables $\hat{X}_k$ where $k \equiv \sim(j,i)$ can additionally be an invertible mixture of $X_k$, since $\rho_{j|\sim j}$ is preserved by that transformation. The r.h.s. of (43b) shows that the correction, due to the variable change is obtained by conditional expectations of the diffusivity. The result (43b) can be important for the RET estimation, since certain variable changes can make the pdf simpler, as for instance Gaussian anamorphoses of the single variables [29] where $\hat{X}_i, \hat{X}_j, \hat{X}_k$ are such that the marginal pdfs follow a standard Gaussian pdf, though the multivariate pdf is not necessarily Gaussian.

When the variable changes are not injective, the problem is more complicated. For instance, for $X_i \to \hat{X}_i = X_i^2$ we have the entropy $H(X_i^2) = H(X_i) + E(\log|2X_i|) - H(X_i|X_i^2)$, where $H(X_i|X_i^2)$ is the entropy of inverse images of the transformation, which vanishes in the case of injective variable changes. This problem is not addressed here.

**4. Numerical approach to the computation of RETs**

The estimation of the rates of entropy transfer (RETs) is a more complicated task than the estimation of the single, multivariate and conditional entropies, for which numerous estimators exist in the literature as seen in the review [30]. The main difficulty comes from the RET dependency, not only on the pdfs and conditional pdfs (like the entropy) but also on the partial derivatives of pdfs (10a, 23a). The later can be difficult to evaluate, especially if pdfs display abrupt changes or their support set has a fractionary Hausedorff dimension as in strange chaotic attractors, leading to non-differentiable pdfs (albeit some pdf projections like marginal pdfs can be differentiable). This difficulty is alleviated if enough (in intensity) stochastic forcing noise is added to the evolution equations of deterministic chaotic systems, thus producing non-vanishing pdf values at the fractal gaps of the strange attractors. Therefore, the estimation of pdf derivatives by finite differences call for the prescription, both of the full pdf and of conditional pdfs along a regular grid of the state-space. This constraint call for the estimation of RET integral formulas using integrals over the pdf, as used by Liang in [16,17], which will be considered as a benchmark to compare with the CSM approach.

Therefore, in testing the methods to evaluate the rate of entropy transfer, $T(X_j \to X_i), (i,j = 1, \dots D, i \neq j)$, four different approaches are considered: the Analytical-Numerical (AN) approach which consists in using the direct estimations of the probabilities and integrals over the pdfs, the Causal Sensitivity Method (CSM), the Model Fitting (MF) followed by application of CSM and finaly the Multivariate Linear (ML) approach of Liang [18], which supposes a linear model fitting and Gaussian distributed noises. The four methods will be applied into two low-order models of dimension $D = 3$ whose equations are known: (1) a model derived from a potential function [16,17], with some generalizations to include multiplicative noise, linearity and non-linearity, all weighted by appropriate parameters and (2) the classical Lorenz chaotic model [26], driven by additive or multiplicative noise, weighted by controlled parameters. The details of the four referred approaches are described below. Sections 5 and 6 present model results.

**4.1. Analytical-Numerical (AN) approach**

In the first approach, the so called '*Analytical-Numerical*' (AN), we evaluate the integrals implicitly in (26a,b). An analytical or a data-estimated multivariate (3D) pdf as well as the analytical values of the sensitivities to the causal variable: $\frac{\partial F_i}{\partial X_j}, \frac{\partial g_{i,i}}{\partial X_j}, \frac{\partial^2 g_{i,i}}{\partial X_j \partial X_i}$ are used in the centroids $(X_{i\prime}, X_{j\prime}, X_{k\prime})$ of equal-volume voxels (3D cells) with sides $\Delta_i, \Delta_j, \Delta_k$, respectively in the variables $X_i, X_j, X_k, (k \neq i, j)$. The indexed voxel $(i', j', k') \in \{1, \ldots, N_i\} \times \{1, \ldots, N_j\} \times \{1, \ldots, N_k\}$ belongs to a regular coarse-grained grid-mesh of the parallelepipedic subset $[L_i, U_i] \times [L_j, U_j] \times [L_k, U_k]$ of the state-space in which the bulk of the multivariate pdf lies. In the forthcoming discretized expressions, we use *plica-indices* for the respective variable (e.g. $i'$ is the index discretizing variable $X_i$). This method was used in Liang [16,17] but using integrals of (22b,c).

Recall that in three dimensions, there is only one outer variable $k \neq i, j$. The probability of each voxel $(i', j', k')$ is denoted as $P(i', j', k') = \tilde{\rho}_\mathbf{X}(X_{i\prime}, X_{j\prime}, X_{k\prime})/Z$, where the normalizing constant is $Z = \sum_{i'=1}^{N_i} \sum_{j'=1}^{N_j} \sum_{k'=1}^{N_k} \tilde{\rho}_\mathbf{X}(X_{i\prime}, X_{j\prime}, X_{k\prime})$, such that probabilities sum one altogether and $\tilde{\rho}_\mathbf{X}$ is prescribed or estimated. The pdf is estimated as $\tilde{\rho}_\mathbf{X}/(\Delta_i \Delta_j \Delta_k)$.

The one- or two-indexed probabilities are sums over the remaining indices, e.g. $P(i', j') = \sum_{k'=1}^{N_k} P(i', j', k')$, which are necessary to compute conditional probabilities, like $P(j'|i') = P(i', j')/P(i')$.

Therefore, by restricting the 3D-integrals of (26a,b) over the bounded domain, we get:

$$T(X_j \to X_i) \approx - \int_{L_i}^{U_i} d X_{i\prime} \int_{L_j}^{U_j} d X_{j\prime} \frac{\partial}{\partial X_i} \left[ \int_{L_j}^{X_{j\prime}} \rho(X_{j\prime\prime}|X_{i\prime}) dX_{j\prime\prime} \right] \int_{L_k}^{U_k} \rho(X_{i\prime}, X_{k\prime}) \frac{\partial R_i}{\partial X_j} d X_{k\prime} \quad (44)$$

where $\rho$ stands for pdf. In (44), the derivatives $\frac{\partial A}{\partial X_i}$ of a generic quantity $A$ at the voxel $(i', j', k')$ are estimated by centred differences as $\frac{\delta_i'[A]}{\Delta_i}$ where one uses the discretized operator $\delta_{i\prime}[A] \equiv \frac{1}{2}[A(i'+1, j', k') - A(i'-1, j', k')]$; while at the interval bounds: $i' = 1, N_i$ one takes the values at $i' = 2, N_i - 1$ respectively.

After the discretization of (44), one obtains:

$$T(X_j \to X_i) \approx - \frac{1}{\Delta_i} \sum_{i'=1}^{N_i} \sum_{j'=1}^{N_j} \sum_{k'=1}^{N_k} \delta_{i\prime} \left[ \sum_{j\prime\prime=1}^{j'} \frac{P(i', j'')}{P(i')} \right] P(i', k') \frac{\partial R_i}{\partial X_j}. \quad (45)$$

By using the prescribed analytical expressions values of $\frac{\partial F_i}{\partial X_j}$, $\frac{\partial g_{i,i}}{\partial X_j}$ and $\frac{\partial^2 g_{i,i}}{\partial X_j \partial X_i}$ to form $\frac{\partial R_i}{\partial X_j}$ in (45) at each voxel $(i', j', k')$, the rightmost factor of the above sum is:

$$P(i', k') \frac{\partial R_i}{\partial X_j} = P(i', k') \frac{\partial F_i}{\partial X_j} - \frac{1}{2\Delta_i} \delta_{i\prime}[P(i', k')] \frac{\partial g_{i,i}}{\partial X_j} - \frac{1}{2} P(i', k') \frac{\partial^2 g_{i,i}}{\partial X_j \partial X_i} \quad (46)$$

The truncation and discretization errors (due to finite bounds $L_i, U_i$ and non-infinitesimal $\Delta_i$ respectively) committed in the estimated causality, depends much on the model equations and pdf. For instance, if $[L_i, U_i]$ spans $n$ standard deviations around the average $X_i$, the estimation error $\varepsilon_A$ in the generic average of $A(X_i)$ within the probabilistic integral, satisfies, after using the Tchebitchev inequality, $|\varepsilon_A| \leq \frac{|A_{ave} - A_{ext}|}{n^2}$, where $A_{ave}$ is the estimated average in the bounded interval and $A_{ext}$ is the average of $A$ outside the interval $[L_i, U_i]$. In the problem of computing $T(X_j \to X_i)$, the above $A$ quantities are terms of $\frac{\partial R_i}{\partial X_j}$. Therefore, to minimize $|\varepsilon_A|$, high enough $n$ must be taken, especially if $A(X_i)$ is monotonic increasing with a positive power of $|X_i|$. Moreover, $|L_i|, |U_i|$ increase in the case of fatter pdf tails. As a rule of thumb, at least $n = 3$ standard deviations must be included in the interval used in the numerical integral. In practice, a stopping criterium for increasing $|L_i|, |U_i|$ and increasing resolution $N_i$ is that the estimated averages of the highest-order products of variables (e.g. $X_i^2 X_j^2$) have corrections smaller than 1% of that average. The deterministic and noise contributions $T(X_j \to X_i)_{F,AN}$ and $T(X_j \to X_i)_{g,AN}$ by AN, are computed and used as benchmarks for the other methods. The AN method is considered as a 'brute-force' method, since it requires the direct computation of the integrals. This cab be a very heavy computation, since the number of cells in domain integrals increases as a power of the system dimension. Moreover, the effective implementation increases in complexity with the dimension.

**4.2. Causal Sensitivity Method (CSM)**

In the second approach, which we developed in the context of this analysis, we identify the parts $F_{i,j}$ and $g_{i,i,j}$ (27) of the deterministic term $F_i$ and diffusivity $g_{i,i}$ that depends on $X_j$. Then, in the case, $F_{i,j}$ and $g_{i,i,j}$ are composed by additive terms with a separation of the causal variable, we apply the formulas (36a,b) or (37a,b), which dependent on the particular form of model equations.

In the application of the CSM, the expectations are computed as averages over a certain ensemble of different realizations of the model states. The ensemble can be: 1) a transient ensemble of a large number $N_r$ of time-evolving states driven by the equations of the model or 2) a long enough time-series of length $L_t$ with sampling-time $\Delta_t$, tunned to uncover the ergodic pdf of the model, with $N_r = L_t/\Delta_t$ realizations. The RET coming from a very large ensemble (long time-series) or the average of RETs obtained from shorter ensembles (time-series) can be taken. In that case, realizations of the estimated RET are obtained by generating a number of $N_s$ (30 in practice) time-series i.e. obtained with different noise generator seeds and initial conditions, leading to $N_s$ RET estimations (ensemble members) whose mean and standard deviation over the sampled realizations are computed. The sensitivity to the time series length is also studied.

In practice, the values of $F_{i,j}$ and $g_{i,i,j}$ are calculated for each realization (ensemble member or instant). One of the assumptions for applying the CSM formulas (36a,b) or (37a,b) is that the different additive terms of $F_{i,j}$ and $g_{i,i,j}$ can be factorized by factors depending on the different variables. That happens, for instance, when the deterministic terms are products of monomials of the different variables $X_i, X_j, X_k$. For each term, the single dependent functions of $X_i, X_j, X_k$ are identified, after which one computes the conditional expectations for a given value of $X_i$. Thanks to the product separability of forcings, the regressions are univariate. There, in general nonlinear regressions are performed in which we consider a certain set of basis functions (e.g. monomials $X_i^k, i = 0, ..., N_d$, up to a certain degree $N_d$ or a set of orthogonal functions, wavelets, trigonometric functions, polynomial splines over a set of subintervals of $X_i$) ranged by decreasing scale (e.g. polynomial order, wave number). The conditional expectation function must be differentiable to apply CSM with algorithmic expressions for the first and second derivatives. The most efficient and robust regression depends on the freedom choice of the basis functions or of the regression formula as a whole. In this respect, some strategies could be followed: 1) mixing different basis functions (e.g. polynomials, trigonometric functions, wavelets); 2) use of machine learning techniques such as Relevance Vector Machines [31] and 3) use of evolving basis functions towards an optimal set by using Symbolic Regression [32]. In general, the regression formula may depend nonlinearly on the fitted parameters.

For the sake of algorithmic simplicity and didactic presentation, we will restrict our analysis to polynomial basis functions. In the case of small ensembles or short time-series, the regressions must be accompanied by statistical significance tests in order to obtain robust regressions (e.g. forward step-regression or computation of the Akaike Information Criterium).

The regression coefficients are fixed for an ergodic pdf or time-evolving for transient ensembles of the model. For a good performance of CSM, the conditional averages like $E_k[f(X_k|X_i)]$ must be accurately estimated as well as their derivatives. Therefore, depending on the particular RET, $N_d$ must be high enough, such that regressions do not differ much from the probabilistic averages $\int \rho(X_k|X_i)f(X_k)\,dX_k$ along the interval $[L_i, U_i]$. Too small $N_d$ can lead to biases.

After, calculating the regressions, their first and second derivatives with respect to $X_i$ are computed as well as the required products and averages. Then ensemble (or time-averages) averages taken. If the regressions are computed using a basis of unbounded functions like monomials, the regressions as well as the values of the specific RETs $\mathcal{F}_{i,j,F}(X_i)$ and $\mathcal{F}_{i,j,F}(X_i)$ can reach (in absolute value) much higher values when $X_i$ is extrapolated beyond the bounds of the dataset. In those cases, a certain from of trimmed average (cutting the extremes) can be applied.

The computational complexity of CSM is much lower than in AN, being thus an execelent alternative to the highly-demanding AN especially in high dimensionsional systems.

### 4.3. Model fitting (MF) followed by CSM

The CSM approach assumes known model equations. However, in practice, both the deterministic terms and diffusivities $F_i, g_{ii} (i = 1, \dots, D)$ may be partially or totally unknown. In this case, an additional preliminar operation to apply CSM afterwards. This procedure, hereafter referred as 'Model fitting followed by CSM' (MF) is described bellow.

In order to apply the CSM, we suppose that, $F_i, g_{ii} (i = 1, \dots, D)$ are written as linear combinations of variable-separable basis functions (including the unity), $\Psi_l, \Phi_l$ belonging to certain prefixed classes (e.g. multivariate polynomials). Let us consider the deterministic term decomposition

$$F_i(\mathbf{X}) = \sum_l \theta_{F,l,i} \Psi_l(\mathbf{X}) \tag{47a}$$

$$\Psi_l(\mathbf{X}) = \prod_{r=1}^{D} \Psi_{l,r}(X_r) \tag{47b}$$

and the diffusivity decomposition

$$g_{i,i}(\mathbf{X}) = \sum_l \theta_{g,l,i} \Phi_l(\mathbf{X}) \geq 0 \tag{48a}$$

$$\Phi_l(\mathbf{X}) = \prod_{r=1}^{D} \Phi_{l,r}(X_r) \;\; ; \;\; \theta_{g,l,i}, \Phi_{l,r} \geq 0 \tag{48b}$$

where $\theta_{F,l,i}$ (merged in the matrix $\boldsymbol{\theta}_F$) and $\theta_{g,l,i}$ (merged in the matrix $\boldsymbol{\theta}_g$) are constant parameters. The corresponding fitted values from data are denoted respectively as $\hat{\theta}_{F,l,i}$ and $\hat{\theta}_{g,l,i}$, merged in matrices $\hat{\boldsymbol{\theta}}_F$ and $\hat{\boldsymbol{\theta}}_g$.

Like in the case of conditional expectancies, other regression techniques may be applied (e.g. machine learning techniques), paying attention to the product separability of forcings and algorithmic expressions of their derivatives. Moreover, robust regressions must be accompanied by statistical significance tests when the available time-series are short.

The estimation method is presented in Appendix B1. For instance in order to fit a linear model, we consider the class of basis-functions $\{\Psi_l(\mathbf{X})\} = \{1, X_1, \dots, X_D\}$.

Since the RETs of a linear combination of deterministic terms or diffusivities (47a, 48a) is the linear combination of the RETs driven by each term, we can write the total RET obtained by MF, as a sum over the basis-functions as:

$$T(X_j \to X_i)_{F,MF} = \sum_l T(X_j \to X_i)_{F,MF,l} \tag{49a}$$

$$T(X_j \to X_i)_{g,MF} = \sum_l T(X_j \to X_i)_{g,MF,l} \tag{49b}$$

By applying the CSM to each contribution (49a,b) and using the estimated parameters, one has the RETs associated to each basis function:

$$T(X_j \to X_i)_{F,MF,l} = \hat{\theta}_{F,l,i} \, E\left[\Psi_{l,i} \; \frac{d\, E(\Psi_{l,j}|X_i)}{dX_i} \; E(\prod_{r \neq i,j} \Psi_{l,r}|X_i)\right] \tag{50a}$$

$$T(X_j \to X_i)_{F,MF,l} = \frac{\hat{\theta}_{g,l,i}}{2} E\left[\Phi_{l,i} \quad \frac{d^2 E(\Phi_{l,j}|X_i)}{dX_i^2} \quad E(\prod_{r \neq i,j} \Phi_{l,r}|X_i)\right] \qquad (50b)$$

Therefore, looking in detail (50a,b), and for each pair $(i,j)$, and basis function $\Psi_l, \Phi_l$, one evaluates first the conditional expectations of the outer-variable factors. Then, conditional expectations $E(\Psi_{l,j}|X_i)$ and $E(\Phi_{l,j}|X_i)$ are assessed as well as their first and second derivatives, respectively. Then, the average products are taken multiplied by the fitted model coefficients. Finally the sum (49a,b) is calculated. We must note that the estimation of $T(X_j \to X_i)$, calls uniquely for the fitting of the evolution equation of the consequential variable $X_i$ as a linear combination of functions of type $f_1(X_i)f_2(X_j)f_3(X_{\sim(i,j)})$, separating causal, consequential and contextual variables. This fitting depends on the extension of space in which $X_i, X_j$ are embedded.

The RET estimations may have biases and sampling variances, coming from: 1) the model fitting through the estimation of time-derivatives (time-step, difference scheme used), 2) the extent and type of basis functions used in the model fitting, 3) overfitting of the model equations, 4) the ensemble size or time-series length, 5) the accuracy of the nonlinear regressions used to estimate the conditional expectations.

Despite the above constraints, it happens that the CSM and MF approaches are computationally cheaper and technically easier than the AN approach, even ih high system dimensions. Test of MF will be restricted to the Lorenz model here (see Sec. 6).

### 4.4. Multivariate Linear Approach

The multivariate linear (ML) approach uses the formulas obtained by Liang [18] and covered in (34) where, a multivariate linear model with additive Gaussian noise is fitted to data, using the maximum likelihood estimator of the model coefficients, and also minimizing the mean squared error of the time derivative of $X_i$. Then, the corresponding RET, denoted as $T(X_j \to X_i)_{F,ML}$ is obtained using the linear fitted model. Since the fitted model has no multiplicative noise, $T(X_j \to X_i)_{g,ML} = 0$.

The ML approach is equivalent to the CSM approach, when the fitted model is linear and the regressions used for the conditional expectations are linear, i.e. the order of the fitting poynomials is $N_d = 1$. This method is used to emphasize the limitations of such a linear approach which is not expected to work as accurately as a nonlinear estimator in strongly nonlinear situations. Therefore, it is expected that ML works under weak nonlinearity as will be shown below in examples.

### 4.5. Summary of the tested methods

Table 1 summarizes the methods of RET estimation, the used models in this study, the drawbacks, advantages and the computational cost (raw estimation of the number of simple algebraic operations) of each method in terms of size parameters: $D$ = state-space dimension, $N_v$ = Number of voxels (bins) taken in each variable, $N_r$ = Number of ensemble realizations or time-series values, $N_b$ = Number of basis-functions or free regression parameters. According to Table 1, computational complexity grows exponentialy with $D$ in the AN approach essentially due to the pdf estimation whereas the one of the CSM grows as a power of $D$, essentially due to the solution of regression problems. This shows clearly the much higher efficiency of CSM compared to AN.

| Method | Models used | Drawbacks or limitations | | Advantages | Comp. Cost |
|---|---|---|---|---|---|
| AN with exact model and pdf | Potential | 1: state-space discretization | | Exact model and pdf | $O(N_v^D)$ |

| | | 2: cut of pdf tails | | | |
|---|---|---|---|---|---|
| AN with exact model and estimated pdf | Lorenz | 3: Cost growing exponentially with dimension | Inaccuracy of estimated pdf from ensembles or time-series | Exact model | $O(N_v^D N_r)$. |
| CSM | Potential and Lorenz | 1: Factorization of forcings and diffusivities<br><br>2: Accuracy dependent on the time-series length and sampling | | 1: No explicit pdf needed,<br><br>2: Low comp. cost.<br><br>3: Nonlinear models | $O(D^2 N_b^3 N_r)$ |
| MF | Lorenz | | fitting of model equations | 4: Only time-series necessary in MF | $O(D^3 N_b^3 N_r)$ |
| ML | Potential and Lorenz | 1: Linear model assumption.<br><br>2:Inaccurate on highly nonlinear models.<br><br>2: Assumption of Gaussian errors. | | Very low comp. cost. | $O(D^3 N_r)$ |

Table 1: Summary of tested methods with the applied models, drawbacks, advantages and computational cost of the RET estimation.

## 5. Potential Model

### 5.1. Model description

The AN computation of RETs was performed by [16,17] on a stochastic potential model. Here, we will make independent estimations on the same model, adding some generalizations for the validation of three of the different methods of computing RETs, proposed in Sec. 4: AN, CSM and ML methods. The MF will be tested with the Lorenz model in Sec. 6.

The generalized potential model used here is described by a three-dimensional $(D = 3)$ vector $\mathbf{X} = (X_1, X_2, X_3)^T \in \mathbb{R}^3$ whose stochastic differential equations are derived from a potencial function $V(\mathbf{X})$ in the form:

$$dX_i = -\frac{\partial V(\mathbf{X})}{\partial x_i} dt + B(\mathbf{X}) dW_i, \quad i = 1,2,3 \tag{51a}$$

$$V(\mathbf{X}) = \frac{1}{2}[\alpha_{nl}(X_1^2 X_2^2 + X_3^2 X_2^2) + (X_1 - \alpha_l X_2)^2 + (X_2 - \alpha_l X_3)^2 + (X_3 - \alpha_l X_1)^2)] \geq 0 \tag{51b}$$

$$B(\mathbf{X}) = [\alpha_m V(\mathbf{X}) + b^2]^{1/2} \geq 0 \tag{51c}$$

where $\alpha_{nl} \geq 0, \alpha_l \geq 0$ are parameters weighting the nonlinear and linear terms respectively of the deterministic terms, which are given by:

$$F_1(X_1, X_2, X_3) = -\frac{\partial V}{\partial X_1} = -X_1[\alpha_{nl}X_2^2 + (1 + \alpha_l^2)] + \alpha_l(X_2 + X_3) \qquad (52a)$$

$$F_2(X_1, X_2, X_3) = -\frac{\partial V}{\partial X_2} = -X_2[\alpha_{nl}(X_1^2 + X_3^2) + (1 + \alpha_l^2)] + \alpha_l(X_1 + X_3) \qquad (52b)$$

$$F_3(X_1, X_2, X_3) = -\frac{\partial V}{\partial X_3} = -X_3[\alpha_{nl}X_2^2 + (1 + \alpha_l^2)] + \alpha_l(X_2 + X_1) \qquad (52c)$$

The stochastic forcing is driven by identical diffusivities $g_{i,i}(\mathbf{X}) = \alpha_m V + b^2 = B^2, i = 1,2,3$. The parameters $\alpha_m \geq 0$, $b \geq 0$, give the weight of the multiplicative and additive noises, respectively. The noises follow the law: $dW_i \sim \sqrt{dt}\, N(0,1)$.

The values $\alpha_{nl} = 1, \alpha_m = \alpha_l = 0$, reproduce the same situation studied by Liang [17]. The parameter $\alpha_l \neq 0$ produces a cross-linear symmetric term across the variables whereas $\alpha_{nl} = \alpha_l = 0$ leads to pure Ornestein-Ulenbeck processes (the equivalent to continuous red-noise processes).

The stationary solution of the Fokker-Planck equation (5) governing the model pdf is $\rho_{\mathbf{X}}(X_1, X_2, X_3) \equiv \hat{\rho}_{\mathbf{X}}/Z$, with

$$\hat{\rho}_{\mathbf{X}}(X_1, X_2, X_3) = \begin{cases} \exp\left(\frac{-2V}{b^2}\right), \alpha_m = 0 \\ \left(V + \frac{b^2}{\alpha_m}\right)^{-\left(1+\frac{2}{\alpha_m}\right)}, \alpha_m \neq 0, \end{cases} \qquad (53)$$

where $Z$ is the normalization constant such that the integral of $\rho_{\mathbf{X}}$ over $\mathbb{R}^3$ equals one.

### 5.2. Formulas of the rates of entropy transfer

From the 6 possible RETs of type $T(X_j \to X_i), (i, j = 1,2,3, i \neq j)$, only three of them are independent due to the functional symmetries of $F_i(X_1, X_2, X_3)$ and $g_{i,i}(X_1, X_2, X_3)$ written below:

$$[F_1(X_1, X_2, X_3) = F_3(X_3, X_2, X_1) \text{ and } g_{1,1}(X_1, X_2, X_3) = g_{3,3}(X_3, X_2, X_1)]$$

$$\Rightarrow [T(X_2 \to X_1) = T(X_2 \to X_3) \text{ and } T(X_3 \to X_1) = T(X_1 \to X_3)] \qquad (54a)$$

$$[F_2(X_1, X_2, X_3) = F_2(X_3, X_2, X_1) \text{ and } g_{2,2}(X_1, X_2, X_3) = g_{2,2}(X_3, X_2, X_1)]$$

$$\Rightarrow [T(X_1 \to X_2) = T(X_3 \to X_2)] \qquad (54b)$$

Therefore, in practice, we compute only $T(X_2 \to X_1), T(X_1 \to X_2)$ and $T(X_1 \to X_3)$.

For the implementation of the AN approach and computation of (45), one requires the sensitivities of the generalized speed $R_i$ with respect to $X_j$ (25) which can be written for the potential model as:

$$\frac{\partial R_i}{\partial X_j} = \frac{\partial F_i}{\partial X_j}\left(1 - \frac{\alpha_m}{2}\right) + \frac{1}{2}\alpha_m F_j \frac{\partial \log(\rho_{\sim j})}{\partial X_i}, \qquad (55a)$$

in which we have used:

$$\frac{\partial g_{i,i}}{\partial X_j} = -\alpha_m F_j \qquad (55b)$$

$$\frac{\partial^2 g_{i,i}}{\partial X_i X_j} = -\alpha_m \frac{\partial F_i}{\partial X_j} \qquad (55c)$$

The RETs based on the CSM approach are obtained by using the procedure described in Sec. 4.2, whose particular forms are given below.

### 5.2.1. Formula of $T(X_2 \to X_1)$

Here, the causal variable $X_j$ is $X_2$, whereas the consequential variable $X_i$ is $X_1$ and the single outer variable is $X_k = X_3$ or $i = 1, j = 2, k = 3$. The terms used in the computation are:

$$F_{1,2} = -\alpha_{nl} X_1 X_2^2 + \alpha_l X_2 \tag{56a}$$

$$g_{1,1,2} = \frac{\alpha_m}{2}\left[\alpha_{nl} X_2^2 (X_1^2 + X_3^2) + (1 + \alpha_l^2) X_2^2 - 2\alpha_l X_2 (X_1 + X_3)\right] \tag{56b}$$

Application of (37a,b) to each term of $F_{1,2}$ and $g_{1,1,2}$ leads to:

$$T(X_2 \to X_1)_F = -\alpha_{nl} E_1 \left[X_1 \frac{dE_2(X_2^2|X_1)}{dX_1}\right] + \alpha_l E_1 \left[\frac{dE_2(X_2|X_1)}{dX_1}\right] \tag{57a}$$

$$T(X_2 \to X_1)_g =$$

$$= \frac{\alpha_m}{4} E_1 \left\{[\alpha_{nl}[X_1^2 + E_3(X_3^2|X_1)] + (1 + \alpha_l^2)] \frac{d^2 E_2(X_2^2|X_1)}{dX_1^2} - 2\alpha_l [X_1 + E_3(X_3|X_1)] \frac{d^2 E_2(X_2|X_1)}{dX_1^2}\right\} \tag{57b}$$

The deterministic part (57a) is decomposed into a nonlinear contribution, proportional to $\alpha_{nl}$ and a linear one, proportional to $\alpha_l$ whereas the noise part (57b) is proportional to the multiplicative-noise parameter $\alpha_m$.

For the computation of (57a,b) by CSM, we need to assess four conditional expectations on $X_1$: $E_2(X_2|X_1), E_2(X_2^2|X_1), E_3(X_3|X_1), E_3(X_3^2|X_1)$ and their first and second derivatives.

### 5.2.2. Formula of $T(X_1 \to X_2)$

Applying the same procedure as before with $i = 2, j = 1, k = 3$, leads to:

$$F_{2,1} = -\alpha_{nl} X_2 X_1^2 + \alpha_l X_1 \tag{58a}$$

$$g_{2,2,1} = \frac{\alpha_m}{2}\left[\alpha_{nl} X_2^2 X_1^2 + (1 + \alpha_l^2) X_1^2 - 2\alpha_l X_1 (X_2 + X_3)\right] \tag{58b}$$

and:

$$T(X_1 \to X_2)_F = -\alpha_{nl} E_2 \left[X_2 \frac{dE_1(X_1^2|X_2)}{dX_2}\right] + \alpha_l E_2 \left[\frac{dE_1(X_1|X_2)}{dX_2}\right] \tag{59a}$$

$$T(X_1 \to X_2)_g = \frac{\alpha_m}{4} E_2 \left\{[\alpha_{nl} X_2^2 + (1 + \alpha_l^2)] \frac{d^2 E_1(X_1^2|X_2)}{dX_2^2} - 2\alpha_l [X_2 + E_3(X_3|X_2)] \frac{d^2 E_1(X_1|X_2)}{dX_2^2}\right\} \tag{59b}$$

For this specific computation we need three conditional expectations on $X_2$ namely: $E_1(X_1|X_2), E_1(X_1^2|X_2), E_3(X_3|X_2)$ and their first and second derivatives.

### 5.2.3. Formula of $T(X_1 \to X_3)$

In a similar way with $i = 3, j = 1, k = 2$ and

$$F_{3,1} = \alpha_l X_1 \tag{60a}$$

$$g_{3,3,1} = g_{2,2,1} = \frac{\alpha_m}{2}\left[\alpha_{nl} X_2^2 X_1^2 + (1 + \alpha_l^2) X_1^2 - 2\alpha_l X_1 (X_2 + X_3)\right] \tag{60b}$$

one gets:

$$T(X_1 \rightarrow X_3)_F = \alpha_l E_3 \left[\frac{dE_1(X_1|X_3)}{dX_3}\right] \quad (61a)$$

$$T(X_1 \rightarrow X_3)_g = \frac{\alpha_m}{4} E_3 \left\{[\alpha_{nl} E_2(X_2^2|X_3) + (1+\alpha_l^2)]\frac{d^2 E_1(X_1^2|X_3)}{dX_3^2} - 2\alpha_l[X_3 + E_2(X_2|X_3)]\frac{d^2 E_1(X_1|X_3)}{dX_3^2}\right\} \quad (61b)$$

In this case four dependent conditional expectations on $X_3$ are necessary: $E_1(X_1|X_3), E_1(X_1^2|X_3), E_2(X_2|X_3), E_2(X_2^2|X_3)$ and their first and second derivatives.

### 5.3. Results of the experiments

We have computed the RETs for different values of $\alpha_{nl}, \alpha_l, \alpha_m$. Three scenarios were considered:

1) An additive noise scenario with increasing non-linearity controlled by $\alpha_{nl}$.
2) A linear scenario with increasing multiplicative noise controlled by $\alpha_m$.
3) A multiplicative noise scenario with increasing non-linearity controlled by $\alpha_{nl}$.

In all validation cases, the pdf is ergodic and the CSM and ML estimates of the RET are obtained using model runs of size $L_t = 4000$ (and 400) time units corresponding to $N_r = 10^5 (10^4)$ time-steps $dt = \Delta t = 0.04$. Ergodicity was assessed by verifying the similarity of the statistical properties of independent model runs. The degree of polynomial regressions of the conditional expectations is fixed to $N_d = 4$ with some tests using $N_d = 2,8$. Model runs are started at $X_1 = X_2 = X_3 = 0.01$, using $N_s = 30$ random seeds for the noise generator. The AN integrals are estimated in a cube with edges $[-U, U], U = 10$. This will obviously limit the pdf tails, but to a relatively low extent. A discretization of $N_p = 200$ points along every edge of the cube is used, leading to a number of $200^3$ cubic voxels.

### 5.3.1. Additive noise with increasing non-linearity

Here, wet set $\alpha_l = 0., b = 1, \alpha_m = 0$ and $\alpha_{nl} \in [0,1]$, starting with a linear model ($\alpha_{nl} = 0$), and then increasing the amplitude of the non-linearity. The stochastic part of the RET vanishes since the noise is purely additive and additionally $T(X_1 \rightarrow X_3)_F = 0$ (61a). The value $\alpha_{nl} = 1$, reproduces the case studied by Liang [17].

Figure 1 shows the values of $T(X_2 \rightarrow X_1)$ and $T(X_1 \rightarrow X_2)$ computed by the AN, CSM and ML approaches as a function of $\alpha_{nl}$. The RETs increase with the strength of nonlinearity due to nonlinear cross-dependencies. In this case $T(X_2 \rightarrow X_1)_F = -\alpha_{nl} E_1\left[X_1 \frac{dE_2(X_2^2|X_1)}{dX_1}\right]$. Despite its explicit linear dependence on $\alpha_{nl}$, it increases in a nonlinear way due to the modifications of the pdf as a function of $\alpha_{nl}$. One obtains a good estimate with a quadratic regression of the conditional expectation as shown below. Therefore,

$$E_2(X_2^2|X_1) \approx E_2(X_2^2) + \frac{cov(X_2^2, X_1^2)}{var(X_1^2)}\left(X_1^2 - E_1(X_1^2)\right) \quad (62a)$$

where $cov$ and $var$ stand for covariance and variance, respectively. Taking the derivative, one gets:

$$\frac{dE_2(X_2^2|X_1)}{dX_1} = 2X_1 \frac{cov(X_2^2, X_1^2)}{var(X_1^2)}, \quad (62b)$$

leading to an estimate of the specific RET:

$$\mathcal{F}_{1,2,F}(X_1) = -\alpha_{nl} X_1 \frac{dE_2(X_2^2|X_1)}{dX_1} = -2\alpha_{nl} \frac{cov(X_2^2, X_1^2)}{var(X_1^2)} X_1^2 \quad (63)$$

varying with $X_1$ as expected in nonlinear systems. Its average is

$$T(X_2 \to X_1)_F = E_1[\mathcal{F}_{1,2,F}(X_1)] \approx -2\alpha_{nl} \frac{cov(X_2^2, X_1^2)}{var(X_1^2)} E_1(X_1^2) \qquad (64)$$

For instance for $\alpha_{nl} = 1$, we have, $cov(X_2^2, X_1^2) = -0.038$, $var(X_1^2) = 0.348$, $E_1(X_1^2) = 0.402$, the computation provides $T(X_2 \to X_1)_F \approx 0.088$, which is fairly close to $T(X_2 \to X_1)_{F,AN} = 0.094$, which was obtained independently by [15].

Figure. 1 show the AN value of RETs (dotted line) and the average (solid lines) and standard deviation (error bar length) over the 30 seeds of the RETs obtained by CSM and ML. The AN and CSM values are relatively close to each other, with a negative slight bias of CSM, lower than 7-10%, valid both for the long time series (Fig. 1a) ($L_t = 4000, N_r = 10^5$) and the short time series (Fig. 1b) ($L_t = 400, N_r = 10^4$). The difference is due to estimation errors both in the AN, due to the specific choices of the interval size $U$, and its resolution $N_p$ and in CSM due to the regression accuracy. By comparing Fig. 1a-b, we see that the standard deviation of RETs is approximately proportional to $1/\sqrt{N_r}$, being 5% of the RET by AN for the long time series. Therefore to achieve an acceptable error of less than 20%, we must have $L_t > 160$. The ML estimates of $T(X_1 \to X_2)$ and $T(X_2 \to X_1)$ are close to zero for the tested nonlinearity weights, which shows that the ML estimates are not able to detect RETs with those nonlinear terms.

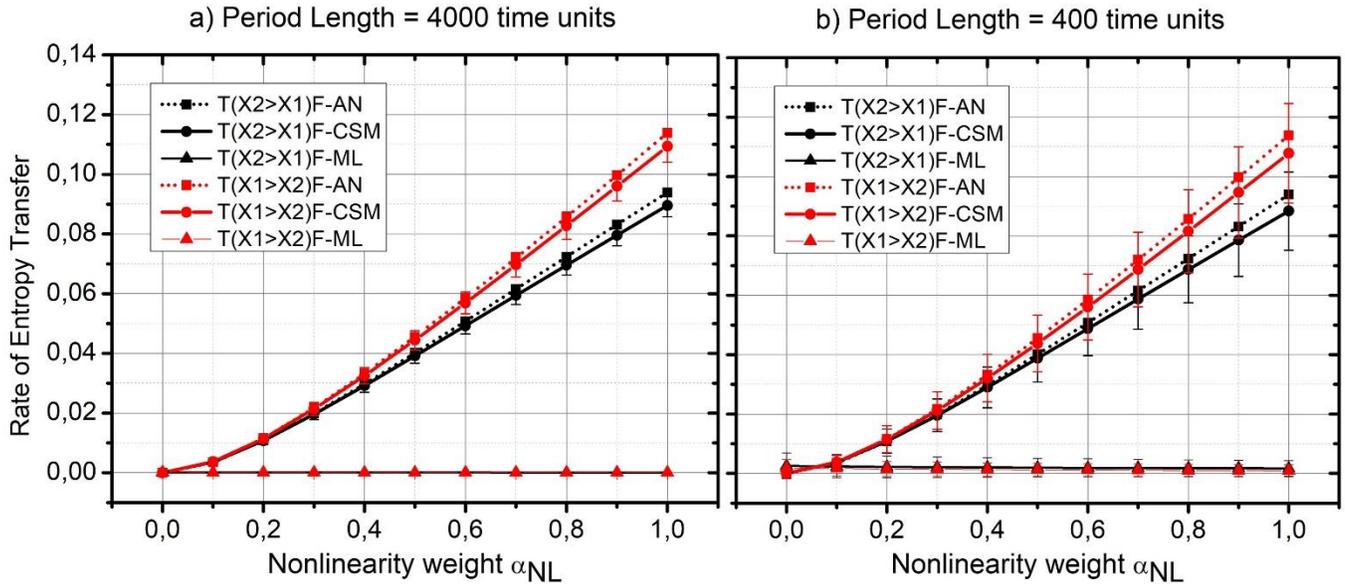

Figure 1. RETs for the scenario $\alpha_l = 0., b = 1, \alpha_m = 0$, with $\alpha_{nl} \in [0,1]$ of the potential model. Mean over 30 noise seeds of $T(X_1 \to X_2)$ (red curves) and $T(X_2 \to X_1)$ (black curves) by the AN (dotted line, squares), CSM (solid line, circles) and ML (solid line, triangles). Standard deviations over the ensemble of seeds are marked by the error bar length. Pannels a) and b) refer respectively to period (time series) length of 4000 and 400 time units. Note that curves for $T(X_1 \to X_2)$ and $T(X_2 \to X_1)$ for the ML approach superpose to each other.

### 5.3.2. Linear system with increasing multiplicative noise

Here we set $\alpha_{nl} = 0, \alpha_l = 0.5, b = 1$ and $\alpha_m \in [0,1]$, starting with a situation of a pure additive noise ($\alpha_m = 0$). Then, the multiplicative noise strength increases, producing non-Gaussian statistics, despite the fact the system is linear [34]

Thanks to the model symmetrization when $\alpha_{nl} = 0$, all the RETs are equal, and thus we only evaluate $T(X_2 \to X_1)_F$ and $T(X_2 \to X_1)_g$ for the diverse approaches (AN,CSM,ML). Note that in this scenario, both the deterministic term and noise scales as $\|\mathbf{X}\|$, leading to extreme behaviors when $\mathbf{X}$ is large. In fact, for the range $\alpha_m \in [0,1]$, the standard-deviation of $X_1$ increases from 0.92 to 1.94 and its kurtosis-excess increases from 0.04 to 34, which is an extreme scenario. This is easily verified through the pdf formula (53). For instance for $\alpha_m = 0.25$, the pdf scales as $\hat{\rho}_\mathbf{X} \sim V^{-9}$ while for $\alpha_m = 1$, the pdf scales as $\hat{\rho}_\mathbf{X} \sim V^{-3}$, producing a fat-tailled distribution.

In this case, the RET is approximated using a simple linear regression:

$$T(X_2 \to X_1)_F = \alpha_l E_1\left[\frac{dE_2(X_2|X_1)}{dX_1}\right] \approx \alpha_l \frac{cov(X_1,X_2)}{var(X_1^2)} = \alpha_l cor(X_1, X_2) \tag{65}$$

For all the range of $\alpha_m$, we have the correlation value $cor(X_1, X_2) \approx 0.66$ and thus $T(X_2 \to X_1)_F \approx 0.33 = 0.66 \times 0.5$, which perfectly agrees with the AN and CSM-based RET estimations shown in Fig. 2 through their average over noise seed values.

The ML estimation, also shown in Fig. 2, is able to detect the deterministic RET, though affected by a bias, which becomes large for highly leptokurtic (high kurtosis) probability distributions, associated with high $\alpha_m$.

The stochastic RET is:

$$T(X_2 \to X_1)_g = \frac{\alpha_m}{4} E_2\left\{(1+\alpha_l^2)\frac{d^2 E_2(X_2^2|X_1)}{dX_1^2} - 2\alpha_l[X_1 + E_3(X_3|X_1)]\frac{d^2 E_2(X_2|X_1)}{dX_1^2}\right\} \tag{66}$$

For symmetry reasons all the cross statistics are symmetric and thus $E_1(X_1|X_2)$ and $E_2(X_2|X_1)$ must have the same linear functional dependency, leading to $\frac{d^2 E_2(X_2|X_1)}{dX_1^2} = 0$. Moreover, by using a quadratic regression for $X_2^2$, we get $\frac{d^2 E_2(X_2|X_1)}{dX_1^2} \approx 2\, cor(X_1^2, X_2^2) \approx 0.52$ in all the range. Therefore, one obtains:

$$T(X_2 \to X_1)_g = \frac{\alpha_m}{2}(1+\alpha_l^2)cor(X_1^2, X_2^2) \approx \alpha_m \times 0.325 \tag{67}$$

Figure 2 also shows the values of the S-RET $T(X_2 \to X_1)_g$, i.e. the stochastic RET, obtained with the AN and CSM approaches. Both approaches agree quite well for $\alpha_m \leq 0.6$, after which, the excess-kurtosis is larger than 3 and the AN value becomes highly biased if too short interval bounds are used. Here the interval $[-10,10]$ was shown to be sufficient for accounting for the pdf tails. The CSM value grows almost linearly with $\alpha_m$ in agreement with (67). This suggests that the CSM estimation can even provide a more accurate and outlier-resistant estimate than the AN method in certain circumstances. The standard deviations of RETs obtained by CSM grows with the intensity $\alpha_m$ of the multiplicative noise and scale approximately as $1/\sqrt{N_r}$ (as in scenario of Sec. 5.3.1) by comparing the values for the long period (Fig. 2a) and short period (Fig. 2b).

Another interesting aspect is that, in certain extreme noise conditions, the stochastic RET can be larger than the deterministic RET (e.g. $\alpha_m > 1$), which means that information flow occurs mostly trough the noise than through the deterministic dynamics.

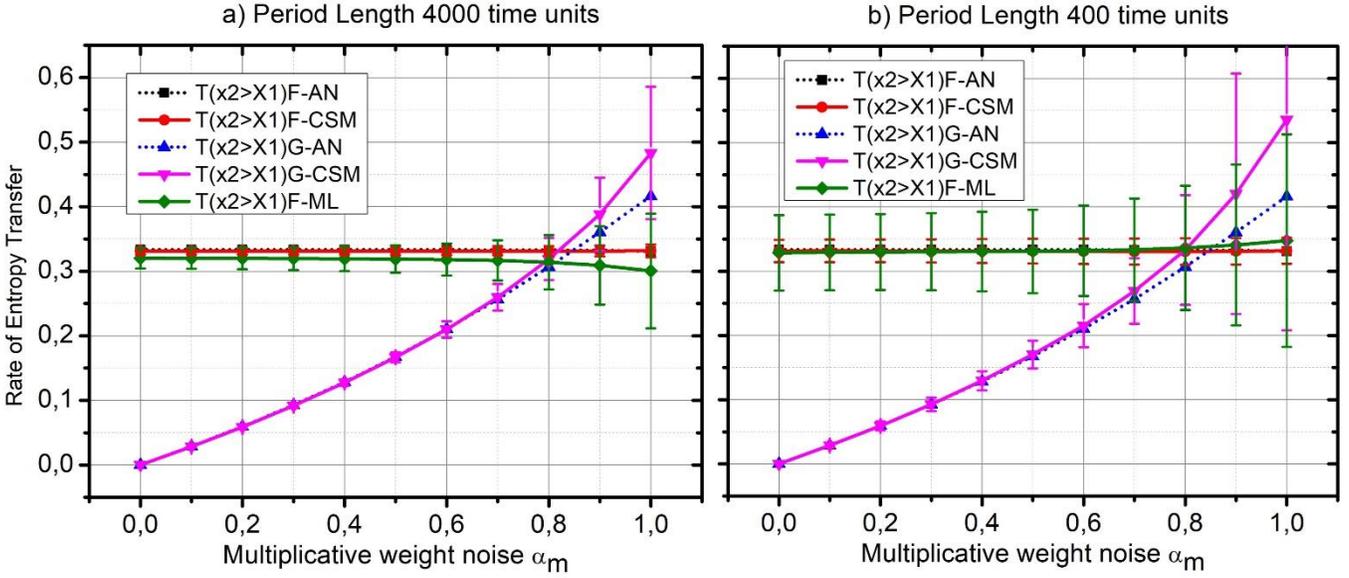

Figure 2. RETs for the scenario $\alpha_{nl} = 0, \alpha_l = 0.5, b = 1$ and $\alpha_m \in [0,1]$ of the potential model. Mean over 30 noise seeds of $T(X_2 \to X_1)_F$ by CSM (red curve, squares), by ML (green solid, diamonds) and of $T(X_2 \to X_1)_g$ by CSM (magenta solid, triangles). Values of $T(X_2 \to X_1)_F$ by AN (black dotted) and $T(X_2 \to X_1)_g$ by AN (blue dotted). Standard deviations over the ensemble of seeds are marked by the error bar length. Pannels a) and b) refer respectively to period (time series) length of 4000 and 400 time units.

### 5.3.3. Multiplicative noise with increasing nonlinearity

Here we set $\alpha_l = 0.5, \alpha_m = 0.3, b = 1$ and $\alpha_{nl} \in [0,1]$, simulating a mixture of non-linearity and multiplicative noise. The symmetry between variables is broken through a non-vanishing $\alpha_{nl}$. Moreover, according to (51a-c), the deterministic terms are scaled as $\|\mathbf{X}\|^3$ whereas the noise scales as $\|\mathbf{X}\|^2$, and therefore that contributes for smaller kurtosis of the system variables than in the previous scenario.

As concerns $T(X_2 \to X_1)_F$ (57a), and using the linear regression for $E_2(X_2|X_1)$ and the quadratic regression for $E_2(X_2^2|X_1)$, one gets:

$$T(X_2 \to X_1)_F \approx -2\alpha_{nl} \frac{cov(X_2^2, X_1^2)}{var(X_1^2)} E_1(X_1^2) + \alpha_l \frac{cov(X_1, X_2)}{var(X_1)} \tag{68}$$

Figure 3 shows the values of $T(X_2 \to X_1)_F$ using the AN and CSM approaches, using the regression degree $N_d = 8$, as well as the minimal regression approach given by (68) (acronym MN) and the ML method. Figure 3 shows that all AN, CSM and MN estimations are quite close to each other. The ML approach is reasonable only for small values of the nonlinearity (small $\alpha_{nl}$), i.e. weak nonlinearity.

The specific RET is not constant, as expected from nonlinear systems and has higher amplitude for larger values of $X_1^2$ according to:

$$\mathcal{F}_{1,2,F}(X_1) \approx -2\alpha_{nl} X_1^2 \frac{cov(X_2^2, X_1^2)}{var(X_1^2)} + \alpha_l \frac{cov(X_1, X_2)}{var(X_1)} \tag{69}$$

The minimal regression of $T(X_2 \to X_1)_g$ (59b), leads to

$$T(X_2 \to X_1)_g \approx \frac{\alpha_m}{2} \{\alpha_{nl}[E_1(X_1^2) + E_3(X_3^2)] + (1 + \alpha_l^2)\} \frac{cov(X_2^2, X_1^2)}{var(X_1^2)} \tag{70}$$

Figure 3 also shows the values of $T(X_2 \to X_1)_g$ (Fig3a: long period, Fig 3b: short period), obtained with the AN, CSM and MN approximations. The AN and CSM agree quite well, with quite short standard deviation over noise-generated samples, as seen in error bars of Fig.3a,b, while MN presents lower values of RET. Despite this bias, the dependence on $\alpha_{nl}$ is well detected by the minimal regression of MN. The standard deviation of sampled RETs by CSM are also approximately scaled as $1/\sqrt{N_r}$ as in previous scenarios.

The graphics for $T(X_1 \to X_2)$ and $T(X_1 \to X_3)$ are similar to those of Fig. 3 (not shown).

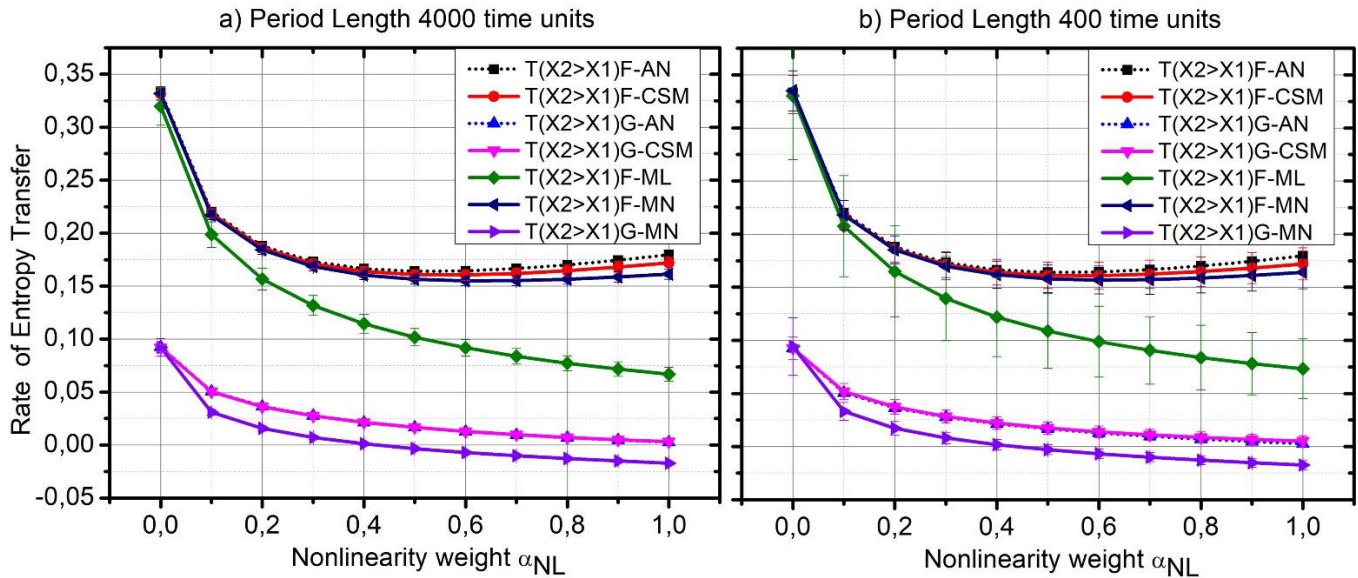

Figure 3. RETs for the scenario $\alpha_l = 0.5, \alpha_m = 0.3, b = 1$ and $\alpha_{nl} \in [0,1]$ of the potential model. Mean over 30 noise seeds of $T(X_2 \to X_1)_F$ by CSM (red, solid), by ML (gree, solid), by MN (deep blue, solid) and of $T(X_2 \to X_1)_g$ by CSM (magenta, solid) and by MN (purple, solid). Values of $T(X_2 \to X_1)_F$ by AN (black, dotted) and $T(X_2 \to X_1)_g$ by AN (blue, dotted). Standard deviations over the ensemble of seeds are marked by the error bar length. Pannels a) and b) refer respectively to period (time series) length of 4000 and 400 time units.

## 6. Lorenz Model

In this section we compute the RETs without the explicit knowledge of the model pdf as in the previous section. For that extension of the analysis to more classical models, the Lorenz's 1963 model [26] driven by a noise term, is used [35]. The Lorenz model and its extensions have been used in numerous theoretical and methodological studies, such as predictability [36,37], data assimilation [38], bifurcation theory [39], ergodic theory [40], thermodynamics [41], signal processing [42] among many others. The model is a minimal Galerkin truncation of the Boussinesq equations of the two-dimensional fluid convection, with heating at the bottom and cooling at the top, where the model variables are $\hat{X}_1$, proportional to the rate of convection and $\hat{X}_2, \hat{X}_3$ which are proportional to the temperature variation in the horizontal and vertical directions, respectively. The usual parameter values: $\sigma = 10$ (Prandtl number), $R_a = 27$ (rescaled Rayleigh number) and $\beta = \frac{8}{3}$ (aspect ratio), are used. The model exhibits a chaotic attractor and an ergodic pdf with expectations $\mu_1 = \mu_2 = 0, \mu_3 \approx 22.61$ and standard deviations $\sigma_1 \approx 7.74, \sigma_2 \approx 8.76$ and $\sigma_3 \approx 8.35$ for $\hat{X}_1, \hat{X}_2, \hat{X}_3$, respectively. The model is slightly transformed to describe the evolution of the standardized variables $X_i \equiv \frac{\hat{X}_i - \mu_i}{\sigma_i}, i = 1,2,3$. We run the stochastic Lorenz models through a Predictor-Corrector scheme (see Appendix B2), and setting $\mu_i, \sigma_i. i = 1,2,3$ to the above values in all experiments. The model equations are:

$$dX_1 = \left[\left(\frac{\sigma\sigma_2}{\sigma_1}\right)X_2 - \sigma X_1\right]dt + B_1 dW_1 \tag{71a}$$

$$dX_2 = \left\{\left[\frac{\sigma_1}{\sigma_2}(R_a - \mu_3)\right]X_1 - \left(\frac{\sigma_1\sigma_3}{\sigma_2}\right)X_1 X_3 - X_2\right\}dt + B_2 dW_2 \tag{71b}$$

$$dX_3 = \left[\left(\frac{\sigma_1\sigma_2}{\sigma_3}\right)X_1 X_2 - \beta X_3 - \left(\frac{\beta\mu_3}{\sigma_3}\right)\right]dt + B_3 dW_3, \tag{71c}$$

where we set equal noise coefficients:

$$B_1 = B_2 = B_3 = [b^2 + \alpha_m \exp(-X_1^2)]^{1/2} = g^{1/2}, \tag{71d}$$

with $b^2 \geq 0$ and $\alpha_m \geq 0$ being parameters associated with the amplitudes of the additive and multiplicative noises respectively. The noise-free model is set to $b = \alpha_m = 0$. The noises follow the law: $dW_i \sim \sqrt{dt}\ N(0,1)$. The model converges to pullback attractors whose properties depend on the parameter values [34].

### 6.1. Rates of entropy transfer and budget of entropy

The deterministic terms in (71a-c) are composed of polynomials, satisfying the separability assumption for the application of CSM formulas (37a,b). The deterministic and stochastic RETs are easily computed as:

$$T(X_2 \to X_1)_F = \frac{\sigma\sigma_2}{\sigma_1} E_1 \left[\frac{dE_2(X_2|X_1)}{dX_1}\right]; \tag{72a}$$

$$T(X_3 \to X_1)_F = 0; \tag{72b}$$

$$T(X_1 \to X_2)_F = E_2\left[\left[\frac{\sigma_1}{\sigma_2}(R_a - \mu_3) - \left(\frac{\sigma_1\sigma_3}{\sigma_2}\right)E_3(X_3|X_2)\right]\frac{dE_1(X_1|X_2)}{dX_2}\right]; \tag{72c}$$

$$T(X_1 \to X_2)_g = \frac{\alpha_m}{2} E_2\left[\frac{d^2 E_1(\exp(-X_1^2)|X_2)}{dX_2^2}\right]; \tag{72d}$$

$$T(X_3 \to X_2)_F = -\left(\frac{\sigma_1\sigma_3}{\sigma_2}\right) E_2\left[E_1(X_1|X_2)\frac{dE_3(X_3|X_2)}{dX_2}\right]; \tag{72e}$$

$$T(X_1 \to X_3)_F = \frac{\sigma_1\sigma_2}{\sigma_3} E_3\left[E_2(X_2|X_3)\frac{dE_1(X_1|X_3)}{dX_3}\right]; \tag{72f}$$

$$T(X_1 \to X_3)_g = \frac{\alpha_m}{2} E_3\left[\frac{d^2 E_1(\exp(-X_1^2)|X_3)}{dX_3^2}\right]: \tag{72g}$$

$$T(X_2 \to X_3)_F = \frac{\sigma_1\sigma_2}{\sigma_3} E_3\left[E_1(X_1|X_3)\frac{dE_2(X_2|X_3)}{dX_3}\right]; \tag{72h}$$

$$T(X_2 \to X_1)_g = T(X_3 \to X_1)_g = T(X_3 \to X_2)_g = T(X_2 \to X_3)_g = 0 \tag{72i}$$

Since the normalization is an affine variable change, the RETs are the same as those from the non-normalized variables according to theorem 6 (43b).

In what concerns the entropy budget, the terms of (20) can be easily estimated using the CSM formulas of Secs. 2 and 3. The generic entropy balance for the three variables is:

$$E\left(\frac{\partial F_i}{\partial X_i}\right) + T(X_{\sim i} \to X_i)_{F,sing} + T(X_{\sim i} \to X_i)_{F,syn} + \left(\frac{dH_{X_i}}{dt}\right)_{g,a} + \left(\frac{dH_{X_i}}{dt}\right)_{g,m} = \frac{dH_{X_i}}{dt} = 0, i = 1,2,3 \tag{73}$$

where the first term is the self-determinist generation (SEG), the second term is the sum of single RETs (42a) the third one is the synergetic RET (42b), the fourth and fifth terms are the additive-noise and multiplicative-noise terms respectively, which have the following expressions:

$$\left(\frac{dH_{X_i}}{dt}\right)_{g,a} = E\left[\left(\frac{\alpha_m}{2}\exp(-X_1^2) + \frac{b^2}{2}\right)\left(\frac{d\log\rho_i}{dX_i}\right)^2\right] \tag{74a}$$

$$\left(\frac{dH_{X_i}}{dt}\right)_{g,m} = -\frac{\alpha_m}{2}(1+\delta_{i,1})E\left[\frac{d^2E(\exp(-X_1^2)|X_i)}{dX_i^2}\right] \tag{74b}$$

where $\delta_{i,1}$ is the Kronecker symbol. For the Lorenz system, we still have:

$$E\left(\frac{\partial F_1}{\partial X_1}\right) = -\sigma;\; E\left(\frac{\partial F_2}{\partial X_2}\right) = -1, E\left(\frac{\partial F_3}{\partial X_3}\right) = -\beta \tag{75a}$$

$$T(X_{\sim 1} \to X_1)_F = T(X_{2,3} \to X_1)_F = \frac{\sigma\sigma_2}{\sigma_1}E\left[\frac{dE(X_2|X_1)}{dX_1}\right] \tag{75b}$$

$$T(X_{\sim 2} \to X_2)_F = T(X_{1,3} \to X_2)_F = \frac{\sigma_1}{\sigma_2}(R_a - \mu_3)E_2\left[\frac{dE(X_1|X_2)}{dX_2}\right] - \left(\frac{\sigma_1\sigma_3}{\sigma_2}\right)E\left[\frac{dE(X_1X_3|X_2)}{dX_2}\right] \tag{75c}$$

$$T(X_{\sim 3} \to X_3)_F = T(X_{1,2} \to X_3)_F = \left(\frac{\sigma_1\sigma_2}{\sigma_3}\right)E\left[\frac{dE(X_1X_2|X_3)}{dX_3}\right] \tag{75d}$$

We must note that, in the Lorenz model, there are two quadratic terms responsible for the nonlinearity: $X_1 X_3$ in $\frac{dX_2}{dt}$ (71b) and $X_1 X_2$ in $\frac{dX_3}{dt}$ (71c), yielding synergetic RETs in $T(X_{1,3} \to X_2)_F$ and $T(X_{1,2} \to X_3)_F$:

$$T(X_{1,3} \to X_2)_{F,syn} = -\left(\frac{\sigma_1\sigma_3}{\sigma_2}\right)E\left[\frac{d\,cov(X_1,X_3|X_2)}{dX_2}\right] \tag{76a}$$

$$T(X_{1,2} \to X_3)_{F,syn} = \left(\frac{\sigma_1\sigma_2}{\sigma_3}\right)E\left[\frac{d\,cov(X_1,X_2|X_3)}{dX_3}\right] \tag{76b}$$

These terms are relevant for the entropy budget as we will show below and are also important in the turbulent chaotic regime simulated by the Lorenz model. The conditional covariance terms $cov(X_1,X_3|X_2)$ and $cov(X_1,X_2|X_3)$ are essentially the vertical and horizontal heat fluxes and the derivatives in (76a), (76b) are their parametrizations with respect to temperature gradients. Therefore, the synergetic terms (76a), (76b) are simply associated with the horizontal and vertical turbulent heat flux diffusivities as a function of temperature gradients. This example suggests that formulas provided by the CSM can be good tools to identify physical factors influencing the entropy budget and the inferred causality.

All the remaining terms in the RETs come from the cross-dependent linear terms in equations: $\left(\frac{\sigma\sigma_2}{\sigma_1}\right)X_2$ in $\frac{dX_1}{dt}$ (71a) and $\left[\frac{\sigma_1}{\sigma_2}(R_a - \mu_3)\right]X_1$ in $\frac{dX_2}{dt}$ (71c), which are detected by the ML approach.

## 6.2. Setup of experiments

The RET values are calculated by the AN, CSM, MF and ML approaches for three situations: noise free ($b = \alpha_m = 0$); additive noise ($b = 1., \alpha_m = 0$) and multiplicative noise ($b = 0, \alpha_m = 1.$). The variance of different variables is 1 for the free-noise case and within the interval [0.9,1.1] for the noisy scenarios which are particularly extreme because the signal to noise ratio is small, and the noise can considerably affect the S-RETs.

The numerical 3D integration, used in AN is made over the cube $[-L, L]^3, L = 4.$, regularly discretized into cubic voxels of side $\Delta = L/100$. Two integration periods are tested, after a relaxation of 100 time units and with lengths $L_t = 400$ time units (long run), and $L_t = 80$ time units (short run), simulating real situations of low data availability. Used time-steps are of size $dt = \Delta t = 0.02$. An extra time-step $dt = \Delta t = 0.01$ is used in the MF where time-derivatives are computed by forward differences. The sampling mean and variability of RETs are assessed by generating $N_s = 30$ independent integration realizations by varying the seed of the noise generator and initial condition.

The ergodic pdf of the model is estimated by a kernel-based approach, using the probabilities $P(\mathbf{X})$ at every state-space voxel centroid $\mathbf{X} \in [-L, L]^3$ estimated as:

$$P(\mathbf{X}) = \frac{1}{Z}\sum_{t=1}^{N_r} \exp[-d^2]\,\mathbf{1}[d \leq \sqrt{3}\Delta] \quad ; \quad d^2 = \|\mathbf{X} - \mathbf{X}(t)\|^2 \tag{77}$$

The instantaneous state $\mathbf{X}(t)$ has an effect only at points within a Cartesian distance $\sqrt{3}\Delta$, which is set by using the indicator function $\mathbf{1}$ (1 or 0 if the condition is true or false respectively). $Z$ is the normalizing constant. Accuracy tests were performed, by comparing probabilistic and temporal averages of powers (up to degree 4) of the variables. After the $P(\mathbf{X})$ computation, the AN-based RETs are assessed at the voxel centroids.

In the CSM approach, we use polynomial regressions of degree $N_d = 10$ to obtain accurate representations of the conditional expectations and their derivatives, though lower degrees are sufficient for certain terms. This will be further discussed below in the validation section.

In the MF approach, the deterministic terms and diffusivities $F_i, g_{i,i}$ are estimated independently and used afterwards in the CSM formulas. The fitting of time-derivatives $\frac{dX_i}{dt}$ is performed using the class of 10 mixed monomials (base-functions) $\Psi_l(\mathbf{X})$ (47b) of maximum degree 2, i.e. $\{\Psi_l(\mathbf{X})\} = \{1, X_1, X_2, X_3, X_1^2, X_1X_2, X_1X_3, X_2^2, X_2X_3, X_3^2\}$ to which the model (71a-c) belongs to, allowing to get an unbiased fitting. For the fitting of noise, we chose the constrained class $\{\Phi_l(\mathbf{X})\} = \{1, \exp(-X_1^2)\}$. The model fitted coefficients are then used to compute D-RETs and S-RETs, using the sums over the basis functions (49a,b) and (50a,b).

### 6.3. Minimal CSM results (MN)

We provide here approximations of the D-RETs, by doing minimal (MN) regressions appearing in the CSM formulas for the noise-free situation. The conditional expectations, valid for the standardized variables are approximated by:

$$E_2(X_2|X_1) \approx cor(X_1, X_2)X_1 = 0.88\,X_1 \tag{78a}$$

$$E_3(X_3|X_1) \approx \frac{cor(X_3, X_1^2)}{std(X_1^2)}(X_1^2 - 1) = 0.563(X_2^2 - 1) \tag{78b}$$

$$E_1(X_1|X_2) \approx cor(X_1, X_2)X_2 = 0.88\,X_2 \tag{78c}$$

$$E_3(X_3|X_2) \approx \frac{cor(X_3, X_2^2)}{std(X_2^2)}(X_2^2 - 1) = 0.068(X_2^2 - 1) \tag{78d}$$

$$E_1(X_1|X_3) = 0 \tag{78e}$$

$$E_1(X_2|X_3) = 0 \tag{78f}$$

Those regressions are quite obvious from Fig. 4 showing the two-dimensional pdfs $\rho_{1,2}(X_1, X_2)$, $\rho_{1,3}(X_1, X_3)$, and $\rho_{2,3}(X_2, X_3)$. In fact, the linear correlation $cor(X_1, X_2) > 0$ appears from the straight positive slope of $\rho_{1,2}$ contours (Fig. 4a) whereas the positive quadratic correlations $cor(X_3, X_1^2)$ and $cor(X_3, X_2^2)$ come from the convex parabolic shapes of $\rho_{1,3}$ (Fig. 4b) and $\rho_{2,3}$ (Fig. 4c) respectively. The pdfs were estimated by the kernel-based method, with a run of 10000 time units.

By substituting the minimal regressions into the D-RET formulas, we get:

$$T(X_2 \rightarrow X_1)_F \approx \frac{\sigma \sigma_2}{\sigma_1} \times 0.88 = 9.96 \tag{79a}$$

$$T(X_1 \rightarrow X_2)_F = E_2\left[\left[\frac{\sigma_1}{\sigma_2}(R_a - \mu_3) - \left(\frac{\sigma_1 \sigma_3}{\sigma_2}\right)E_3(X_3|X_2)\right]\frac{dE_1(X_1|X_2)}{dX_2}\right] \approx \frac{\sigma_1}{\sigma_2}(R_a - \mu_3) \times 0.88 = 3.88 \tag{79b}$$

$$T(X_3 \rightarrow X_2)_F = -\left(\frac{\sigma_1 \sigma_3}{\sigma_2}\right)E_2\left[E_1(X_1|X_2)\frac{dE_3(X_3|X_2)}{dX_2}\right] \approx -\left(\frac{\sigma_1 \sigma_3}{\sigma_2}\right)E(0.88X_2 \times 2.\times 0.068X_2) = -0.883 \tag{79c}$$

The values of $T(X_1 \rightarrow X_2)_g$ and $T(X_1 \rightarrow X_3)_g$ calls for the Taylor expansion of $\exp(-X_1^2)$ and $X_3$ regression of at least 4$^{th}$ degree. All the remaining single D-RETs are zero.

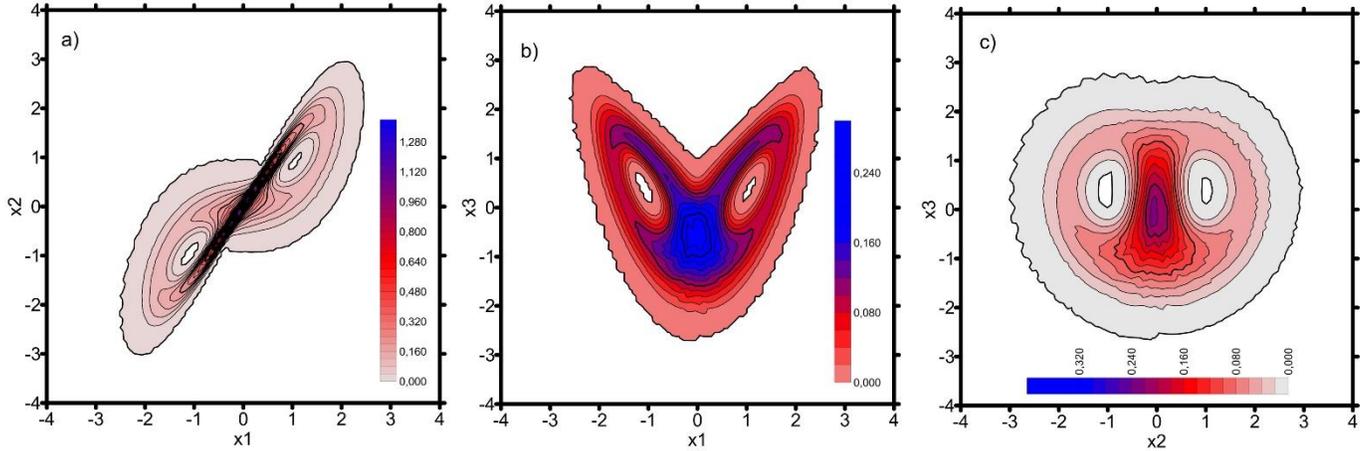

Figure 4. Two-dimensional pdfs $\rho_{1,2}(X_1, X_2)$ (a) $\rho_{1,3}(X_1, X_3)$ (b) and $\rho_{2,3}(X_2, X_3)$ (c) of the standardized Lorenz model (unit variance of each variable) (71a-c) in the noise-free scenario in the domain $[-4,4]^2$.

### 6.4. Validation of the RET estimation

Table 2 presents the overall results of the different deterministic and stochastic single RETs, computed with the two period lengths ($L_t = 400$ and $L_t = 80$ with $\Delta t = 0.02$ and $L_t = 80$ with $\Delta t = 0.01$) for the different scenarios: noise free, additive noise, and multiplicative noise and using the four tested approaches: AN, CSM, MF and ML. The sample averages and sample standard deviations of RETs were computed from an ensemble of 30 independent realizations of model runs. Polynomials of order 10 have been applied for the nonlinear regressions in CSM and MF. Model fitting uses polynomials of second order. The ensemble average AN values obtained with the long run are taken as reference values (in bold in Table 2). The sources of estimation errors are discussed in Sec. 4 and summarized in Table 1.

From the analysis of the table 2, the main conclusions about the RETs estimations are:

1) The deterministic RETs obtained by minimal CSM, i.e. MN, compare quite well with the AN-based values showing that relatively simple regressions can recover the RETs. The biases are associated to the representation of the conditional expectations by polynomials of very low order.
2) There is a good agreement between the AN and CSM estimations with biases ranging within 1-5%, for both run lengths and noise scenarios. The CSM variability, measured by the standard deviation among samples range within 1-14% for the long-runs being roughly proportional to $\frac{1}{\sqrt{L_t}}$. (inverse of the root square of the run length), i.e. ~2.2 larger in the short-run samples. The CSM variability is in most cases less than the AN variability, justifying the robustness of CSM.

3) The MF estimations have comparable biases to those of CSM, except in $T(X_2 \rightarrow X_1)_F$. That bias in MF can be decreased by taking smaller time-steps (see comparison between dt=0.02 and 0.01). The sampling standard deviation of MF ranges within 2-10% in the long runs and 4-25% in the short runs, being generally larger than that of CSM due to model fitting errors and being also roughly proportional to $\frac{1}{\sqrt{L_t}}$. It shows that even with very short time series without the knowledge of model equations, it is possible to obtain relatively accurate causality diagnostics.
4) In what concerns the ML values, $T(X_2 \rightarrow X_1)_F$ agrees with those from AN, due to the absence of nonlinear terms in (71a). All the remaining ML values are close to zero, except $T(X_1 \rightarrow X_2)_F$ because the product $-\left(\frac{\sigma_1 \sigma_3}{\sigma_2}\right) X_1 X_3$ in (71b) correlates with $-X_1$, leading to a negative (highly biased) D-RET (see Table 2). The zero ML values agree with AN in the cases $T(X_1 \rightarrow X_3)_F, T(X_2 \rightarrow X_3)_F$ because of two coincidences: the absence of cross-linear terms in the equation governing $X_3$ (71c) and the symmetry of pdfs $\rho_{1,3}(X_1, X_3)$ and $\rho_{2,3}(X_2, X_3)$ (see Fig. 4).
5) The AN values computed with short runs exhibit high variability (sampling standard deviation), reaching twice to three times larger than that of CSM, especially due the inaccuracy of estimated pdfs. Therefore, , the time-series length and model dimension are severe limitations to the applicability of AN.
6) The deterministic RETs (D-RETs), which are mostly associated to nonlinear terms in $T(X_1 \rightarrow X_2)_F, T(X_3 \rightarrow X_2)_F$ decrease in amplitude in the noisy scenarios because the nonlinear dependencies weaken with respect to the noise-free situation. In particular, in the additive-noise scenario, the model becomes closer to Ornstein-Uhlenbeck models.

The sensitivity of the CSM and MF values with respect to the degree $N_d$ of polynomial regression used in the conditional expectations is illustrated in Fig. 5, showing the average and standard deviation (over 30 noise seeds) of two RETs, essentially of nonlinear nature: $T(X_1 \rightarrow X_2)_F, T(X_3 \rightarrow X_2)_F$ for the noise free and additive noise scenarios, short run ($L_t = 80$) and dt=0.02. We conclude from the figure that the speed of convergence as Nd grows, depends on the considered RET. For the noise-free scenario, $T(X_1 \rightarrow X_2)_F$ stabilizes for $N_d \geq 4$ whereas $T(X_3 \rightarrow X_2)_F$ stabilizes for $N_d \geq 10$, justifying that choice in the CSM validation. In the noisy scenario, since the pdf is less non-Gaussian than in the noise-free situation, and non-linearity decreases, the RETs have smaller absolute value and the convergence is faster, showing that in both RETs stabilize for $N_d \geq 4$. The final bias (difference with respect to the analytical value) is still nonzero indicating other sources of errors, either from the analytical or from the CSM and MF algorithms.

The remaining RETs converge rather quickly at $N_d = 2$ (not shown). This makes clear that the representativness of the conditional expectations call in some cases for higher-order polynomial regressions or eventually for more appropriate regression functions, namely with finite L2 norms in the real domain.

The convergence speed of the polynomial regressions of the conditional expectations generally depends on the nonlinearities of the studied model, the shape of the corresponding pdf and also of the set of basis functions (e.g. polynomials, trigonometric functions, orthogonal eigen-functions, wavelets). Two factors weakening the convergence of regressions are the abruptness of pdf variations along state-space, the existence of topologically complex pdf shape and the 'strangeness' and fractal nature of chaotic attractors, especially if the Hausedorff dimension is small. In the 3D noise-free Lorenz model, the Hausedorff dimension is ~2.06 [43] and the atractor projects quite well on each of the three variables, thus leading to smooth univariate and bivariate pdfs (Fig. 4). The strength of the Gaussian white noise forcings tends to make pdfs more, Gaussian and smoothed, leading to faster convergence regressions as shown in Fig. 5.

| RET | Noise type | Run | NA | CSM | MF | ML |
|---|---|---|---|---|---|---|
| $T(X_2 \to X_1)_F$ | Free | 400<br>80<br>80(a) | **9.92±0.023**<br>9.96±0.109 | 9.96±0.002<br>9.96±0.013 | 8.87±0.002<br>8.87±0.011<br>9.44±0.009 | 9.98±0.003<br>9.98±0.014 |
| | Add. | 400<br>80<br>80(a) | **9.30±0.068**<br>9.24±0.217 | 9.30±0.059<br>9.37±0.113 | 8.29±0.127<br>8.35±0.270<br>8.89±0.285 | 9.21±0.085<br>9.24±0.172 |
| | Mult. | 400<br>80<br>80(a) | **9.82±0.040**<br>9.70±0.152 | 9.86±0.020<br>9.85±0.050 | 8.77±0.094<br>8.73±0.284<br>9.36±0.220 | 9.53±0.065<br>9.51±0.144 |
| $T(X_1 \to X_2)_F$ | Free | 400<br>80 | **3.62±0.026**<br>3.63±0.098 | 3.58±0.014<br>3.57±0.052 | 3.83±0.078<br>3.76±0.168 | -9.18±0.010<br>-9.18±0.032 |
| | Add. | 400<br>80 | **2.28±0.115**<br>2.29±0.238 | 2.24±0.117<br>2.26±0.187 | 2.30±0.205<br>2.42±0.375 | -6.47±0.082<br>-6.42±0.185 |
| | Mult. | 400<br>80 | **2.47±0.086**<br>2.46±0.234 | 2.43±0.082<br>2.45±0.233 | 2.87±0.278<br>2.97±0.573 | -7.31±0.066<br>-7.21±0.149 |
| $T(X_3 \to X_2)_F$ | Noise | 400<br>80 | **-0.76±0.046**<br>-0.85±0.166 | -0.73±0.022<br>-0.72±0.047 | -0.80±0.022<br>-0.77±0.049 | 0.0003±0.0005<br>0.0004±0.002 |
| | Add. | 400<br>80 | **-0.55±0.089**<br>-0.77±0.303 | -0.52±0.075<br>-0.57±0.159 | -0.56±0.078<br>-0.597±0.162 | 0.001±0.0022<br>0.008±0.016 |
| | Mult. | 400<br>80 | **-0.44±0.070**<br>-0.58±0.200 | -0.41±0.057<br>-0.45±0.138 | -0.43±0.057<br>-0.48±0.143 | 0.001±0.0019<br>0.005±0.0087 |
| $T(X_1 \to X_3)_F$ | Noise | 400<br>80 | **0.03±0.028**<br>0.005±0.107 | 0.01±0.035<br>0.06±0.091 | 0.008±0.025<br>0.04±0.061 | -0.002±0.0036<br>-0.009±0.010 |
| | Add. | 400<br>80 | **0.01±0.039**<br>0.03±0.178 | 0.005±0.013<br>0.045±0.053 | 0.003±0.013<br>0.003±0.054 | -0.003±0.0048<br>-0.004±0.016 |
| | Mult. | 400<br>80 | **0.02±0.037**<br>0.07±0.138 | 0.02±0.023<br>0.07±0.052 | 0.02±0.018<br>0.05±0.056 | -0.005±0.0077<br>-0.017±0.033 |
| $T(X_2 \to X_3)_F$ | Noise | 400<br>80 | **0.008±0.028**<br>-0.02±0.145 | 0.005±0.013<br>0.041±0.057 | -0.002±0.045<br>0.009±0.094 | 0.003±0.0046<br>0.01±0.012 |
| | Add. | 400<br>80 | **0.002±0.053**<br>0.06±0.247 | 0.003±0.011<br>0.047±0.059 | 0.001±0.019<br>0.030±0.070 | -0.003±0.0055<br>-0.018±0.037 |
| | Mult. | 400<br>80 | **0.007±0.054**<br>-0.007±0.218 | 0.007±0.009<br>2.53±0.051 | 0.001±0.019<br>0.02±0.067 | -0.002±0.0089<br>-0.01±0.029 |
| $T(X_1 \to X_2)_g$ | Mult. | 400<br>80 | **-0.11±0.010**<br>-0.10±0.056 | -0.11±0.0068<br>-0.11±0.018 | -0.09±0.0059<br>-0.09±0.016 | na<br>na |
| $T(X_1 \to X_3)_g$ | Mult. | 400<br>80 | **-0.03±0.012**<br>-0.06±0.056 | -0.021±0.002<br>-0.018±0.076 | -0.020±0.0016<br>-0.015±0.0062 | na<br>na |

Table 2 Values of D-RETs and S-RETs (indicated in the first column), for the 3 scenarios (noise-free, additive and multiplicative noise), estimated by the 4 approaches (AN, CSM, MF and ML). The values are given in the format (μ±σ) where μ and σ are sampling average and sampling standard deviation, obtained from an ensemble 30 independent model run realizations, using runs of 400 and 80 time units (indicated in third column) and $\Delta t = 0.02$.

The value 80(a) refers to MF values using runs of 80 time units and $\Delta t = 0.01$. The reference values (AN with long run) is given in boldface type. The RET value of $T(X_3 \to X_1)_F$ is not included since it is theoretically null.

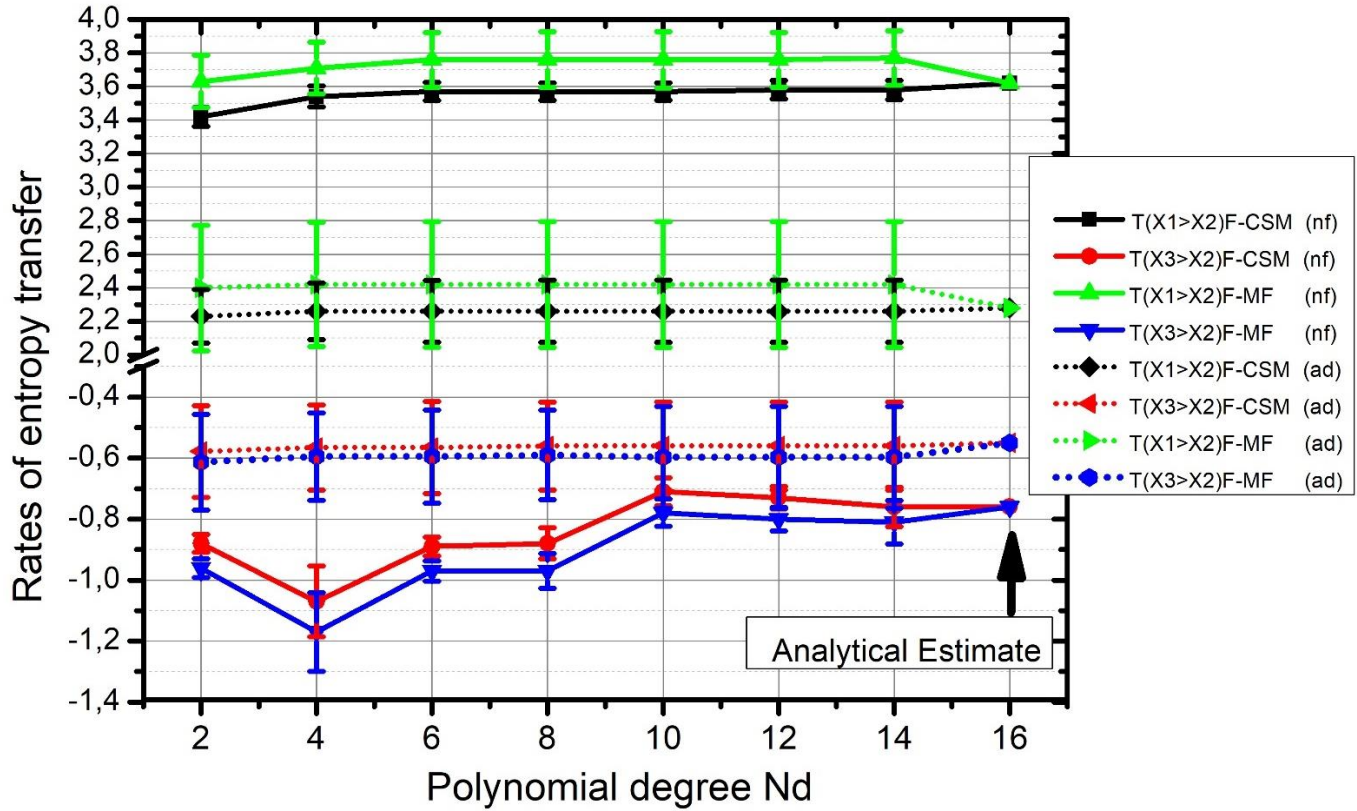

Figure 5. Values (sample averages and standard deviations in error bars) of $T(X_1 \to X_2)_F$ by CSM (black) and by MF (green) and $T(X_3 \to X_2)_F$ by CSM (red) and by MF (blue) for polynomial regressions of degrees Nd=2,4,6,8,10,14. Soild and dotted lines refer to noise-free and additive noise scenarios respectively. The analytical value is marked at Nd=16.

### 6.5. Discussion of the entropy budget

The computation of the entropy budget terms and specific contributions is important to identify the leading dynamical mechanisms and provide key information about predictability, causality and synergies. We present in Table 3 the different terms (rows) of the budget (73) for the three scenarios (columns) and the three variables (sub-columns) using the CSM values based on a long run of 400 time units. As expected, the sum of terms (last row) is very close to zero, being of the order of the sampling variability (in Table 2). The entropy budget points out the very important role of synergetic RETs coming from the nonlinearities of the Lorenz system. In fact, while, for $X_1$, the deterministic terms are linear, leading to vanishing synergetic RETs, the synergetic RETs for $X_2$ (joint effect of $X_1, X_3$) and $X_3$ (joint effect of $X_1, X_2$) are different from 0. In the case of $X_3$, the synergetic RET gets the total value of the RET (Table 3). It shows, that, despite the presence of $X_1$ and $X_2$ in the dynamics of $X_3$, it is rather their joint effect, through the product $X_1 X_2$ in (71c) that contribute to the entropy change. This must partially be a result of the high interrelation between $X_1$ and $X_2$, coming from their sharp pdf (Fig. 4a) and high linear correlation (~0.88 in the noise free situation).

The computation of single RETs (one to one entropy transfers or more appropriately, entropy generation of one variable under the effect of another one), as well their sum, hereby denoted as $T(X_{\sim i} \to X_i)_{F,sing}$, both in the Lorenz system and probably in most nonlinear and complex network systems (like the atmospheric-oceanic system) appears to be an incomplete picture of the causality due to the effect of synergies. Therefore, the set of one-to-one causal links, not only of the LK type but also issued from other techniques (e.g., conditional mutual information (CMI) [44], Peter and Clark Momentary Conditional Independence (PCMCI) [45], Granger causality [2,6]) may not reproduce the 'full story' about the system causality.

Moreover, the importance of synergetic terms reveals the usefulness (possibly holding on more general models), of merging highly interrelated variables and concentrating them onto low-order spaces or mixed variables (e.g. through principal component analysis, independent component and subspace analysis [46,47], nonlinear principal component analysis [42] and projection pursuit [48]), followed by computation of RETs between those mixed variables and the remaining variables.

In order to illustrate that rationale in the noise-free scenario, we consider the pair of variables $(X_3, X_4)$ where $X_4 \equiv X_1^2 - E(X_1^2) = X_1^2 - 1$ and the linear regression: $X_2 X_1 = \alpha_3 X_3 + \alpha_4 X_4 + E(X_2 X_1) + \varepsilon_{reg} = -0.56 X_3 + 1.20 X_4 + 0.88 + \varepsilon_{reg}$, where the regression error $\varepsilon_{reg}$ has zero average and variance 0.14 along the attractor. The evolution equations (71c) and (71a multiplied by $2X_1$), of $X_3$ and $X_4$ respectively become:

$$\frac{dX_3}{dt} = \left[\left(\frac{\sigma_1 \sigma_2}{\sigma_3}\alpha_3 - \beta\right)X_3 + \left(\frac{\sigma_1 \sigma_2}{\sigma_3}\alpha_4\right)X_4 + \left(\frac{\sigma_1 \sigma_2}{\sigma_3}E(X_1 X_2) - \frac{\beta \mu_3}{\sigma_3}\right)\right] + \frac{\sigma_1 \sigma_2}{\sigma_3}\varepsilon_{reg} \approx -7.21 X_3 + 9.74 X_4 \quad (80a)$$

$$\frac{dX_4}{dt} = \left[\left(\frac{2\sigma \sigma_2}{\sigma_1}\alpha_3\right)X_3 + \left(\frac{2\sigma \sigma_2}{\sigma_1}\alpha_4\right)X_4 + \left(\frac{2\sigma \sigma_2}{\sigma_1}E(X_1 X_2) - 1\right)\right] + \frac{2\sigma \sigma_2}{\sigma_1}\varepsilon_{reg} \approx -12.67 X_3 + 7.16 X_4 \quad (80b)$$

where the neglected terms are about 10-20 times less than the retained ones. The two-dimensional system (80a,80b) appears as approximately linear and quasi Hamiltonian, due the near vanishing of the sum of forcing divergences: $-7.21 + 7.16 \sim 0$, (i.e. the preservation of area elements with the dynamical system flow) and a nearly conserving positive-defined Hamiltonian $6.34 X_3^2 + 4.87 X_4^2 - 7.18 X_3 X_4$. The driving matrix $\begin{bmatrix} -7.21 & 9.74 \\ -12.67 & 7.61 \end{bmatrix}$ of (80a,80b) has two imaginary conjugate eigenvalues $\pm i\omega$ leading to oscillations of $(X_3, X_4)$ with a period $\frac{2\pi}{\sqrt{12.67 \times 9.74 - 7.21 \times 7.61}} \sim 0.76$, which is close to that of Lorenz system oscillations. The time evolution of $(X_3, X_4)$ is shown in Fig. 6, in which the oscillations are evident. The system also satisfy closed boundary conditions since the state-space speed is tangent to the axis $X_4 = 0$, thus being the validy conditions of the presented theory (Secs. 2 and 3)

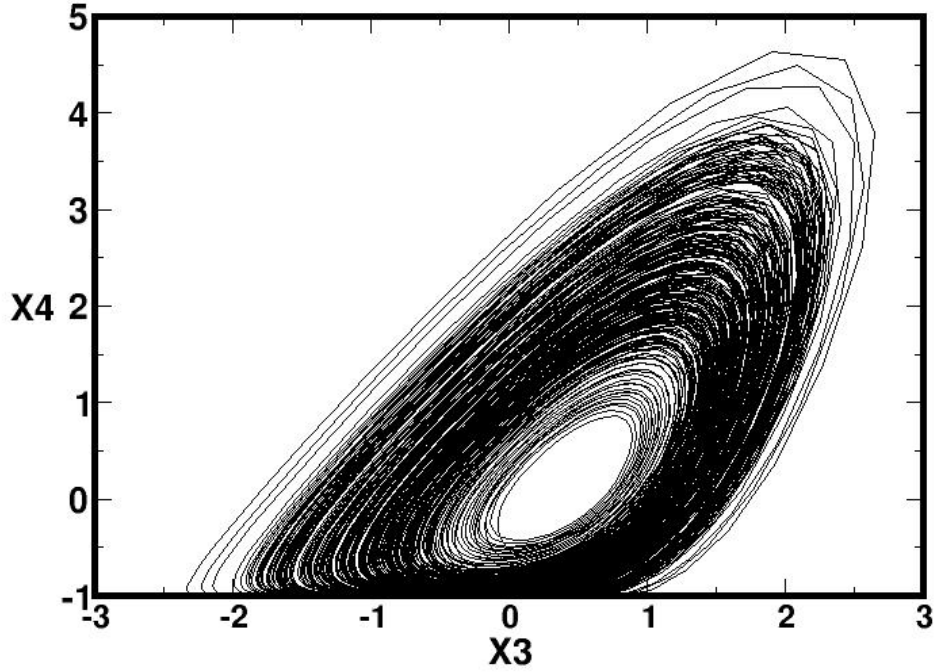

Figure 6. Time evolution of $X_3$ and $X_4 \equiv X_1^2 - 1$ in the noise-free scenario of the Lorenz 63 system.

In what concerns the LK causality, the system (80a, 80b) is nearly linear and thus the RET formula (34), valid for linear systems can be applied leading to the approximation:

$$T(X_4 \to X_3) \sim 9.74 \frac{cov(X_3,X_4)}{var(X_3)} = 7.11 \tag{81a}$$

$$T(X_3 \to X_4) \sim -12.67 \frac{cov(X_3,X_4)}{var(X_4)} = -7.09 \tag{81b}$$

The above RET values compensate very well the deterministic entropy self-generations, $-7.21$ and $7.16$, respectively for $X_3$ and $X_4$ (80a,80b), confirming the equilibrium of the entropy budget in both variables. Those near symmetric values of the entropy transfer appear as a characteristic of two-dimensional linear Hamiltonian systems. This has been verified independently by empirical estimations of $T(X_1^2 \to X_3)$ and $T(X_3 \to X_1^2)$, respectively 6.99 and $-7.03$ as well as in [49]. Therefore, the entropy appears as being self-generated by $X_1^2$ (i.e. the square of convection intensity), then transferred to $X_3$ (intensity of the vertical temperature gradient), being self-dissipated afterwards by $X_3$. The variable $X_2^2$, (square of the intensity of the horizontal temperature gradient) is nonlinearly driven by $(X_3, X_4)$, whose approximate evolution equation is obtained by (71b) multiplied by $2X_2$ after using the regression for $X_2 X_1$ with a positive RET $T(X_3, X_4 \to X_2)$ compensating the dissipation. In a similar way, an approximately Hamiltonian system will be obtained by considering the pair $(X_2 X_1, X_3)$.

The Liang-Kleeman entropy diagnostics can thus suggest groups of variables which, when mixed together, can provide a simplified picture of the entropy transfers.

For stochastic systems, the presence of noise induces a decrease of the amplitude of the deterministic RETs (singles and synergetic), since the noise contributes positively to the entropy balance. In the case of additive noise and as far its amplitude increase, every single variable becomes more approximately governed by Ornstein-Uhlenbeck processes producing RETs of smaller amplitude as far as noise variance increases

An important step forward is to make a systematic study of the RETs in stochastic systems and how they evolve along the space of bifurcation parameters, noise intensity and its characteristics, which is out of the scope for the present study. Another possible step is the merging of signal-processing techniques and blind source separation methods with the Liang-Kleeman entropy diagnostics, using the nonlinear approach devided in this study.

|  | Noise-free | | | Add. Noise | | | Mult. Noise | | |
|---|---|---|---|---|---|---|---|---|---|
|  | $X_1$ | $X_2$ | $X_3$ | $X_1$ | $X_2$ | $X_3$ | $X_1$ | $X_2$ | $X_3$ |
| $T(X_{\sim i} \to X_i)_{F,sing}$ | 9.97 | 2.85 | 0.02 | 9.30 | 1.72 | 0.01 | 9.86 | 2.02 | 0.01 |
| $T(X_{\sim i} \to X_i)_{F,syn}$ | 0 | -1.82 | 2.61 | 0 | -1.17 | 2.06 | 0 | -1.19 | 2.33 |
| $\left(\dfrac{dH_{X_i}}{dt}\right)_{g,m}$ | 0 | 0 | 0 | 0 | 0 | 0 | 0.05 | 0.11 | 0.02 |
| $\left(\dfrac{dH_{X_i}}{dt}\right)_{g,a}$ | 0 | 0 | 0 | 0.65 | 0.49 | 0.56 | 0.08 | 0.09 | 0.28 |
| $E\left(\dfrac{\partial F_i}{\partial X_i}\right)$ | -10 | -1 | -2.66 | -10 | -1 | -2.66 | -10 | -1 | -2.66 |
| Budget Sum | -0.03 | 0.03 | -0.03 | -0.05 | -0.04 | -0.03 | -0.01 | 0.03 | -0.03 |

Table 3. Terms of the budget of entropy for the 3 variables of the stochastic Lorenz system and the three noise scenarios.

For the specific RETs, it allows determining the regions of state-space that most contribute to the RET. By focusing on the noise-free situation, the deterministic contributions (12b,31b) for $X_1$ and $X_3$ are nearly constant along the range of the consequential variable, being retained by the single and synergetic terms, respectively for $X_1$ and $X_3$, i.e.:

$$\mathcal{F}_{1,\sim 1,F}(X_1) = \frac{\sigma\sigma_2}{\sigma_1} \frac{dE(X_2|X_1)}{dX_1} \sim \mathcal{F}_{1,\sim 1,F,sing}(X_1) \sim T(X_{\sim 1} \to X_1)_{F,sing} = 9.974 \quad (82a)$$

$$\mathcal{F}_{2,\sim 2,F}(X_2) = \frac{\sigma_1}{\sigma_2}(R_a - \mu_3)\frac{dE(X_1|X_2)}{dX_2} - \left(\frac{\sigma_1\sigma_3}{\sigma_2}\right)\frac{dE(X_1 X_3|X_2)}{dX_2} \sim T(X_{\sim 2} \to X_2)_F = 1.03 \quad (82b)$$

$$\mathcal{F}_{3,\sim 3,F}(X_3) = \left(\frac{\sigma_1\sigma_2}{\sigma_3}\right)\frac{dE(X_1 X_2|X_3)}{dX_3} \sim \mathcal{F}_{3,\sim 3,F,syn}(X_3) \sim T(X_{\sim 3} \to X_3)_{F,syn} = 2.61 \quad (82c)$$

Concerning $X_2$, there are really strong variations and changes of the sign of the specific deterministic RETs $\mathcal{F}_{2,1,F}(X_2), \mathcal{F}_{2,3,F}(X_2)$ and $\mathcal{F}_{2,\sim 2,F,syn}(X_2)$ along the range of $X_2$ (see Fig. 7) which is a clear mark of nonlinearity, since on linear systems the specific RETs are constant. In fact, while, the specific value of $T(X_1 \to X_2)_F$, i.e. $\mathcal{F}_{2,1,F}(X_2) = $ is mostly positive and constant $\sim 3.7$ (black curve in Fig. 7), the specific $T(X_3 \to X_2)_F$ i.e. $\mathcal{F}_{2,3,F}(X_2)$ (red curve in Fig. 6) is positive near the $X_2 = 0$ but reaches large negative values at the lobes of the attractor while the synergetic effect of $(X_1, X_3)$ through $\mathcal{F}_{2,\sim 2,F,syn}(X_2)$ (blue curve in Fig. 7) is mostly negative, displaying an opposite behavior. This means that, both $X_1$ and $X_3$ separately contribute to increase the entropy of $X_2$, whereas their joint combination contributes to decrease entropy of $X_2$.

Most of the RETs take place near the point of maximum joint pdf at $(X_1 = 0, X_2 = 0, X_3 \sim -0.3)$ where most of sign changes of the convective cells take place, when the system orbit approaches the unstable fixed point of no convection: $\left(X_1 = 0, X_2 = 0, X_3 = \frac{-\mu_3}{\sigma_3} = -2.71\right)$. Here, the average and specific RETs are evaluated along the ergodic pdf. However, they can be evaluated on forecast ensembles, particulary on short-to-long range weather forecasting. There, entropy is interpreted as the entropy of forecast error, coming mostly from the ensemble spread. That is useful to identify the drivers of predictability or unpredictability on specific situations.

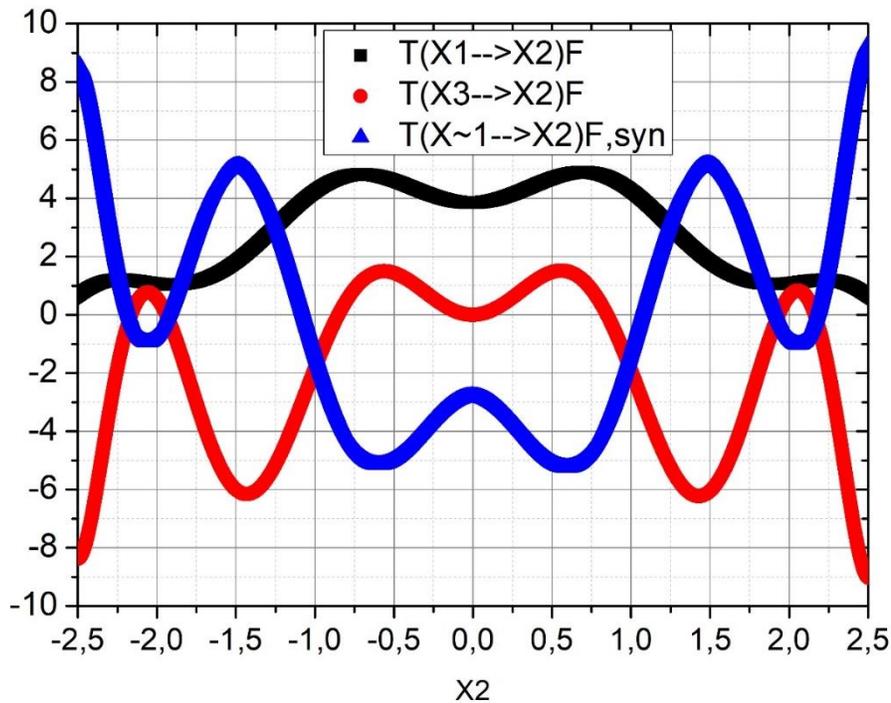

Figure 7. Specific D-RET contributions for $T(X_{\sim 2} \to X_2)_F$, in the noise-free scenario, through $T(X_1 \to X_2)_F$, (black curve), $T(X_3 \to X_2)_F$ (red curve) and $T(X_{\sim 2} \to X_2)_{F,syn}$ (blue curve).

## 7. Discussion and Conclusions

A general theory is derived for computing Shannon entropy transfers in nonlinear (deterministic and/or stochastic) systems, holding under quite generic state-space boundary conditions, namely: 1) unbounded pdf support set; 2) closed boundaries, i.e. null deterministic and diffusive fluxes at the pdf support set borders; 3) periodic conditions in state-space. The theory extends the Liang-Kleeman (LK) framework [16] of causality inference, by decomposing the evolution of the Shannon entropy of a single state vector component (taken as the consequential variable), into a self entropy generation (SEG) and a global rate of entropy transfer (RET) issued from the complementary space of the consequential variable, with contributions assigned either, to the deterministic or the stochastic forcings. In the case of open conditions, some extra terms must be added to the entropy balance equation, related to externally driven fluxes of entropy to the state-space, imposed by nun-null speeds or diffusivities at the borders of the pdf support set in the space-space.

The RET comes from the existence of cross variable dependencies, either in the deterministic components or in probabilistic diffusivities in the multiplicative-noise case. The global RET is further decomposed into the sum of RETs from the single variables spanning the complementary space plus a synergetic RET accounting for synergetic causal effects due to groups (e.g. triplets, quartets) of variables [28]. The synergetic term vanishes in linear systems, being thus of purely nonlinear origin.

Synergies due to nonlinearities (e.g. products of variables in the forcings), as well as redundancies between variables suggest the extended RET computation between a generic causal subspace and a generic consequential subspace (or more generally manifolds), and for given contextual or outer subspace. As a climatic example of the causal, consequential and contextual subspaces, we can choose sub-spaces spaning the low-frequency variability of separate oceanic basins and the overlying atmosphere where indirect oceanic links may take place via an 'atmospheric dynamical bridge' [50]. Consistent expressions of the space-to-space RETs, are being devised (author's future manuscript), following the same CSM technique presented here.

The diagnostics of causality between two particular variables (inferred by RETs) can strongly depend on the set of contextual or remaining variables, spaning the so-called outer space acting as room of 'hidden' indirect interactions between that pair of variables. This effect can be amplified when the allowed interactions jump from linear to nonlinear and the dimension of the 'outer' space increases, thus calling for a proper analysis of the extension of the 'causality outer space'.

An effective method of estimating RETs, the 'Causal Sensitivity Method' (CSM) is then introduced, as an alternative to the 'brute-force' and more expensive estimation approaches calling for the explicit knowledge of the pdf of the system, which may have severe pitfals. In fact, the pdf estimation can be unreliable even for moderate system dimensions if not enough data are available, and being biased by effect of the 'curse of dimensionality' (fact that multivariate outliers become more probable than near average values).

The CSM relies on RET formulas depending explicitly on derivatives (sensitivities) of conditional expectations of the deterministic terms and noise diffusivities, conditioned on the consequential variable. The CSM is fully appropriate to cases where an ensemble of realizations is available, from which the conditional expectations are estimated using nonlinear differentiable regressions based upon an extended set of basis functions (e.g. monomials, orthogonal functions, wavelets) which can be fixed or optimized through signal processing techniques and machine learning, as relevant vector machines and symbolic regression. Those ensembles can be time-evolving or obtained from a sampled long time-series covering the attractor of the system. In particular in ensemble forecasting (e.g. medium-to-long ensemble weather forecasting), entropy of forecast errors grows with the ensemble spread, and the RETs become important to diagnose drivers of predictability or unpredictability of chosen variables.

Formulas for the CSM-based specific contributions to RETs, along the state-space are available, being uniform (varying) on linear (nonlinear) systems, and thus providing information where the RETs are intense or not. The CSM-based RET formulas could also help to propose certain parametrizations of the effects of the outer variables on the causal-consequential pair, as for instance in parameterizing sub-grid scale processes in turbulence or in meteorology.

To exemplify the usefulness of the CSM formulas, four RET estimation approaches are compared: The 'brute-force' computationally expensive analytical-numerical (AN) method, relying upon numerical integrals on the state-space depending on the system pdf and taken as the reference in our study; the CSM, when the dynamical equations are known; the MF method based on the model fitting followed by the CSM; and the multivariate linear (ML) approach of Liang [18]. The methods are validated in two three-dimensional models: a model derived from a potential function and the classical chaotic Lorenz model [26], both subject to a variety of noises and nonlinearities. The CSM values appear to be robust and close to the reference AN values, being thus appropriate for application with real data. The ML approach only works under weak nonlinearity, being potentially strongly biased when strong nonlinearities and synergies across variables are present. This justifies the need for using a nonlinear estimate for nonlinear problems. Diagnostics of RETs can suggested ways of separating system variables leading to more simplified pictures of entropy fluxes.

The application of the CSM, with or without model fitting, is susceptible of generalization to differentiable maps [12] and estimation of RETs between subspaces of variability, thus opening a straightfull, and computationally cheap methodology to infer information flows across variables or groups of variables of a system, for instance spanning sub-spaces of variability of the climatic system. Furthermore, this study has raised the importance of a systematic study of RETs on chaotic and stochastic system as a function of bifurcation and noise parameters.

The application of the CSM, as well as other causality diagnostics to real systems of moderate dimension $D$ (10-20) and high dimension represents a great endeavor, both in the cases of known and unknown evolution equations, with the computational cost of many problems growing exponentially with $D$ and the 'curse of dimensionality' effect. Furthermore, when available time series are short, for selected spatial and temporal scales, we add the need of making robust model fittings, avoiding overfitting. In high-dimension systems, governed by complex networks (like the climatic system), there are possible high levels of linear or nonlinear redundancies (measured for instance by mutual information). Therefore, it is advisable to first compute low-dimensional spaces (by linear and/or nonlinear principal component analysis and independent and/or subspace component analysis), and then diagnose the causality between the spaning variables or between subspaces, following the same CSM strategy. Those variables can be substituted by other relevant common indices (e.g. climatic indices in the climatic system, like El-Niño index), acting as drivers (causal variables) of the raw system variables, for instance, helping to locate which regions are more or less influenced.

The implementation of the CSM in real cases studies without available model equations has important prerequisites related to efficient and robust regressions both in the fitting of model equations and also in the expression of CSM conditional expectations. Robust regressions should preferentially be accompanied by statistical significance tests. The fitting may be performed by standard linear or non-linear regression techniques or more sophisticated machine-learning-based and genetic programming (e.g. relevance vector machines), keeping in mind that basis-functions must be multiplicatively separable in terms of the causal/consequential/outer variables. The basis-functions should preferentially be non-recurrent (issued from sets of orthogonal functions), and ranging from large to small scales.

Further applications to real-world class of problems like the interaction between different components of the climate system [50] with presence of feedbacks is certainly a way forward that is taken up and will be reported in the near future. Moreover, the CSM technique can be applied to ensemble forecasting (e.g. short-to-long weather range forecasting) where the Shannon entropy mostly comes from the ensemble spread and forecast error variance. There, the identification of causes (drivers) of the ensemble forecast spread through the RETs can be useful to determine predictability conditions of selected variables, thus opening a 'world' of applications.

**CRediT authorship contribution statement**

Carlos Pires: Conceptualization, Methodology, Formal Analysis, Validation, Investigation, Writing – original draft, Writing – review & editing, Visualization. David Docquier: Methodology, Validation. Stéphane Vannitsem: Methodology, Formal analysis ,Writing – review & editing.

**Declaration of competing interest**

The authors declare that they have no known competing financial interests or personal relationships that could have appeared to influence the work reported in this paper.

**Declaration of Generative AI and AI assisted technologies in the writing process**

Generative AI/AI assisted technologies were not used in the preparation of this work.

**Data availability**

Data will be made available on request.

**Acknowledgements**

We thank Reik Donner and Giorgia Di Capua for very fruitful discussions during the preparation of this work and also the constructive comments of two anonymous reviewers. It is funded by the Portuguese ´Fundação para a Ciência e a Tecnologia' (FCT) I.P./MCTES through national funds (PIDDAC) – UIDB/50019/2020- IDL and the project JPIOCEANS/0001/2019 (ROADMAP: 'The Role of ocean dynamics and Ocean–Atmosphere interactions in Driving Climate variations and future Projections of impact–relevant extreme events'). D. Docquier and S. Vannitsem also acknowledge the partial financial support from the Belgian Federal Scientific Policy Office (Belspo) under contract B2/20E/P1/ROADMAP.

**Appendix A Proofs of the theoretical results**

**A.1. Proof of (4)**

Let us denote: $X_i(t + dt) = X_i(t) + Y_i(dt)$; $Y_i(dt) \equiv F_i(\mathbf{X}, \boldsymbol{\theta}, t)dt + \sum_{k=1}^{D_n} B_{i,k}(\mathbf{X}, \boldsymbol{\theta}, t)dW_k$, $i = 1, \ldots, D$ and $\rho_{i,self}(X_i)$ the pdf of $X_i$ in the stochastic process where $X_{\sim i}$ is frozen in the infinitesimal period $[t, t + dt]$. The entropy of that process, needs the writing of the Frobenius-Perron (FP) operator $\wp[\rho_{i,self}(X_i)]$ evolving $\rho_{i,self}$ from $t$ to $t$+dt. For that, we use a similar procedure to that of Liang in 2016 [16] (his Proposition VI.3) in which the FP operator of the pdf of $X_1$ (denoted as $\rho_{1,\backslash 2}$) under frozen $X_2$ is written as:

$$\wp[\rho_{1,\backslash 2}(X_1)] = \rho_{1,\backslash 2}(X_1 + Y_1) - dt \int_{\mathbb{R}^{D-2}} \frac{\partial F_1 \rho_{\sim 2}}{\partial X_1} dX_3 \ldots dX_D + \frac{dt}{2} \int_{\mathbb{R}^{D-2}} \frac{\partial^2 g_{1,1} \rho_{\sim 2}}{\partial^2 X_1} dX_3 \ldots dX_D + o(dt)$$

(A.1)

where $\rho_{\sim 2} = \int_{\mathbb{R}} \rho(\mathbf{X}) dX_2$. Now, in a similar way, the transformed pdf is $\wp[\rho_{i,self}(X_i)]$, which is obtained by changing, in the previous equation, the sub-index 1 into $i$, the sub-index 2 into $\sim i$ (corresponding to the set of frozen variables), and the integration $\int_{\mathbb{R}^{D-2}}(..)dX_3 \ldots dX_D$ into the integrand function $(..)$. WE thus have the FP:

$$\wp[\rho_{i,self}(X_i)] = \wp[\rho_{i,\backslash \sim i}(X_i)] = \rho_{i,self}(X_i + Y_i) - dt \frac{\partial F_i \rho_i}{\partial X_i} + \frac{dt}{2} \frac{\partial^2 g_{i,i} \rho_i}{\partial^2 X_i} + o(dt) \qquad (A.2)$$

where the pdf $\rho_{\sim 2}$ was changed into $\rho_{\sim\sim i} = \rho_i$. The corresponding entropy $H_{X_i,self}(t + dt)$ is obtained by taking minus logarithm of the above pdf, followed by the expectation operator $E$ and the Taylor expansion of $\rho_{i,self}(X_i + Y_i)$ around $X_i$ (see the similar procedure in Proposition VI.2 and VI.3 of [16]). Then, after subtracting $H_{X_i,self}(t) = H_{X_1}(t)$, dividing by $dt$ and taking the limit $dt \to 0$, we obtain the SEG given by the time derivative:

$$\frac{d H_{X_i,self}}{dt} = -E\left[F_i \frac{\partial \log \rho_i}{\partial X_i}\right] - \frac{1}{2} E\left[g_{i,i} \frac{\partial^2 \log \rho_i}{\partial^2 X_i}\right] + E\left[\frac{1}{\rho_i} \frac{\partial F_i \rho_i}{\partial X_i}\right] - \frac{1}{2} E\left[\frac{1}{\rho_i} \frac{\partial^2 g_{i,i} \rho_i}{\partial^2 X_i}\right] \qquad (A.3)$$

According to Proposition VI.2 of [16], the time derivative $\frac{d H_{X_i}}{dt}$ is the sum of the third and fourth terms of the r.h.s. of the above equation. Consequently, the RET is given by

$$T(X_{\sim i} \to X_i) = \frac{d H_{X_i}}{dt} - \frac{d H_{X_i,self}}{dt} = -E\left(\frac{1}{\rho_i} \frac{\partial F_i \rho_i}{\partial X_i}\right) + \frac{1}{2} E\left(\frac{1}{\rho_i} \frac{\partial^2 g_{i,i} \rho_i}{\partial X_i^2}\right) \qquad (A.4)$$

Providing the result (4).

## A.2. Proof of (10a)

We have $T(X_{\sim i} \to X_i)_F = -E\left(\frac{1}{\rho_i}\frac{\partial F_i \rho_i}{\partial X_i}\right) = -\int_{\sim i}\int_i \frac{\rho_X}{\rho_i}\frac{\partial F_i \rho_i}{\partial X_i} d\mathbf{X} = -\int_{\sim i}\int_i \rho_{\sim i|i}\frac{\partial F_i \rho_i}{\partial X_i} d\mathbf{X} = \int_{\sim i}\int_i F_i \rho_i \frac{\partial \rho_{\sim i|i}}{\partial X_i} d\mathbf{X} = \int_{\sim i}\int_i F_i \frac{\rho_X}{\rho_{\sim i|i}}\frac{\partial \rho_{\sim i|i}}{\partial X_i} d\mathbf{X} = E\left(F_i \frac{\partial \log \rho_{\sim i|i}}{\partial X_i}\right)$, where $\int_i$ and $\int_{\sim i}$ stand for integration on $X_i$ and $X_{\sim i}$ respectively. In the forth equality we apply integration by parts and $\lim_{\|\mathbf{X}\|\to\infty} \rho_\mathbf{X} F_i = 0$. Now, let us take (9a): $F_i = \hat{F}_i(X_i) + F_{i,\sim i}$. The part depending on $\hat{F}_i$ is $\int_{\sim i}\int_i \hat{F}_i \rho_i \frac{\partial \rho_{\sim i|i}}{\partial X_i} d\mathbf{X} = \int_i \hat{F}_i \rho_i \frac{d}{dX_i}\int_{\sim i}\rho_{\sim i|i} d\mathbf{X} = 0$, since $\int_{\sim i}\rho_{\sim i|i} dX_{\sim i} = 1$, and the the second equality of (10a) follows.

## A.3. Proof of (11)

(11) comes from: $E\left[F_{i,\sim i}\frac{\partial \log \rho_{\sim i|i}}{\partial X_i}\right] = \int_{\sim i}\int_i \rho_\mathbf{X} F_{i,\sim i}\frac{\partial \log \rho_{\sim i|i}}{\partial X_i} d\mathbf{X} = \int_{\sim i}\int_i \rho_i F_{i,\sim i}\frac{\partial \rho_{\sim i|i}}{\partial X_i} d\mathbf{X} = \int_{\sim i}\int_i \rho_i \frac{\partial F_{i,\sim i}\rho_{\sim i|i}}{\partial X_i} d\mathbf{X} - \int_{\sim i}\int_i \rho_i \rho_{\sim i|i}\frac{\partial F_{i,\sim i}}{\partial X_i} d\mathbf{X} = \int_i \rho_i \frac{d}{dX_i}\int_{\sim i}\rho_{\sim i|i} F_{i,\sim i} d\mathbf{X} - E\left(\frac{\partial F_{i,\sim i}}{\partial X_i}\right) = E_i\left[\frac{d}{dX_i}E_{\sim i}(F_{i,\sim i}|X_i)\right] - E\left(\frac{\partial F_{i,\sim i}}{\partial X_i}\right)$.

## A.4. Proof of (16a)

(16a) is obtained through: $\left(\frac{dH_{X_i,self}}{dt}\right)_{g,m} = \left(\frac{dH_{X_i}}{dt}\right)_{g,m} - T(X_{\sim i} \to X_i)_g = -\frac{1}{2}\int_{\sim i}\int_i g_{i,i}\rho_{\sim i|i}\frac{d^2\rho_i}{dX_i^2} d\mathbf{X} - \frac{1}{2}\int_{\sim i}\int_i \rho_{\sim i|i}\frac{\partial^2 g_{i,i}\rho_i}{\partial X_i^2} d\mathbf{X} = -\frac{1}{2}\int_{\sim i}\int_i \left(\rho_i \frac{\partial^2 g_{i,i}\rho_{\sim i|i}}{\partial X_i^2} + \rho_{\sim i|i}\frac{\partial^2 g_{i,i}\rho_i}{\partial X_i^2}\right) d\mathbf{X} = -\frac{1}{2}\int_{\sim i}\int_i \left(\frac{d^2 g_{i,i}\rho_{\sim i|i}\rho_i}{dX_i^2} - 2\frac{\partial g_{i,i}\rho_{\sim i|i}}{\partial X_i}\frac{d\rho_i}{dX_i}\right) d\mathbf{X} = \int_{\sim i}\int_i \frac{\partial g_{i,i}\rho_{\sim i|i}}{\partial X_i}\frac{d\rho_i}{dX_i} d\mathbf{X} = -\int_i \rho_i\frac{d^2}{dX_i^2}\int_{\sim i} g_{i,i}\rho_{\sim i|i} d\mathbf{X} = E_i\left(-\frac{d^2 E(g_{i,i}|X_i)}{dX_i^2}\right)$. In the third equality we apply integration by parts twice with respect to $X_i$ and once in the fifth and sixth equalities.

## A.5. Proof of (17)

(17) is obtained from: $T(X_{\sim i} \to X_i)_g = \frac{1}{2}E\left[\frac{1}{\rho_i}\frac{\partial^2 g_{i,i\sim i,\rho_i}}{\partial X_i^2}\right] = \frac{1}{2}\int_{\sim i}\int_i \rho_{\sim i|i}\frac{\partial^2 g_{i,i\sim i}\rho_i}{\partial X_i^2} d\mathbf{X} = \frac{1}{2}\int_{\sim i}\int_i \rho_{\sim i|i}\left(\rho_i\frac{\partial^2 g_{i,i\sim i}}{\partial X_i^2} + g_{i,i\sim i}\frac{d^2\rho_i}{dX_i^2} + 2\frac{\partial g_{i,i\sim i}}{\partial X_i}\frac{d\rho_i}{dX_i}\right) d\mathbf{X} = \int_{\sim i}\int_i \left[\frac{1}{2}\rho_{\sim i|i}\rho_i\frac{\partial^2 g_{i,i\sim i}}{\partial X_i^2} + \frac{1}{2}\rho_i\frac{d^2 \rho_{\sim i|i} g_{i,i\sim i}}{dX_i^2} - \rho_i\frac{d}{dX_i}\left(\rho_{\sim i|i}\frac{\partial g_{i,i\sim i}}{\partial X_i}\right)\right] d\mathbf{X} = \frac{1}{2}E\left(\frac{\partial^2 g_{i,i\sim i}}{\partial X_i^2}\right) + \frac{1}{2}E_i\left[\frac{d^2}{dX_i^2}E_{\sim i}(g_{i,i\sim i}|X_i)\right] - E_i\left[\frac{d}{dX_i}E_{\sim i}\left(\frac{\partial g_{i,i\sim i}}{\partial X_i}\Big|X_i\right)\right]$, thus yielding the result.

## A.6. Proof of (22a-c, 23a,b)

According to formula (2) of Liang [16], the rate of entropy transfer to $X_i$ due to the influence of $X_j$ is:

$$T(X_j \to X_i) = -\int_{\mathbb{R}^D} \rho_{j|i}\frac{\partial(F_i \rho_{\sim j})}{\partial X_i} d\mathbf{X} + \frac{1}{2}\int_{\mathbb{R}^D} \rho_{j|i}\frac{\partial^2(g_{i,i} \rho_{\sim j})}{\partial X_i^2} d\mathbf{X} \tag{A.5}$$

Integration by parts with respect to $X_i$ and assuming that $\lim_{X_i \to \pm\infty} \rho_{j|i} F_i \rho_{\sim j} = 0$, leads the deterministic term (first of the r.h.s.) of (A.5) to be written as:

$$\int_{\mathbb{R}^D} F_i \rho_{\sim j}\frac{\partial(\rho_{j|i})}{\partial X_i} d\mathbf{X} = \int_{\mathbb{R}^D} \rho_{j,\sim j} F_i \frac{\rho_{\sim j}\rho_{j|i}}{\rho_{j,\sim j}}\frac{\partial \log \rho_{j|i}}{\partial X_i} d\mathbf{X} = E\left[F_i \frac{\rho_{j|i}}{\rho_{j|\sim j}}\frac{\partial \log(\rho_{j|i})}{\partial X_i}\right], \tag{A.6}$$

where the full pdf writes as $\rho_{j,\sim j}$ leading to the expectation $E[(...)] = \int_{\mathbb{R}^D} \rho_{j,\sim j} (...) d\mathbf{X}$ and we use the definition of conditional pdf $\frac{\rho_{j,\sim j}}{\rho_{\sim j}} = \rho_{j|\sim j}$. Similar procedure applied to the noise term (second of the r.h.s.) of (A.5) leads it to be written in a similar form of (A.6):

$$-\frac{1}{2}\int_{\mathbb{R}^D}\frac{\partial(g_{i,i}\rho_{\sim j})}{\partial X_i}\frac{\partial(\rho_{j|i})}{\partial X_i}d\mathbf{X} = \int_{\mathbb{R}^D}\rho_{j,\sim j}\frac{\rho_{\sim j}\rho_{j|i}}{\rho_{j,\sim j}}\left(-\frac{1}{2\rho_{\sim j}}\frac{\partial(g_{i,i}\rho_{\sim j})}{\partial X_i}\right)\frac{\partial\log\rho_{j|i}}{\partial X_i}d\mathbf{X} =$$
$$E\left[\left(-\frac{1}{2\rho_{\sim j}}\frac{\partial(g_{i,i}\rho_{\sim j})}{\partial X_i}\right)\frac{\rho_{j|i}}{\rho_{j|\sim j}}\frac{\partial\log(\rho_{j|i})}{\partial X_i}\right]. \tag{A.7}$$

The sum of deterministic and noise terms leads to the result by defining the generalized speed $R_i = F_i - \frac{1}{2\rho_{\sim j}}\frac{\partial(g_{i,i}\rho_{\sim j})}{\partial X_i}$.

### A.7. Proof of (24, 25,26a-b)

The derivative of the cumulated distribution function (cdf) is the pdf, and thus the term appearing in the expected value (1a) writes as $\frac{\partial\log(\rho_{j|i})}{\partial X_i} = \frac{1}{\rho_{j|i}}\frac{\partial^2 Pr(j|i)}{\partial X_i \partial X_j} = \frac{-1}{\rho_{j|i}}\frac{\partial^2 Pr_{ex}(j|i)}{\partial X_i \partial X_j}$. The manipulation of the full and conditional pdfs leads to

$$E\left[R_i \frac{\rho_{j|i}}{\rho_{j|\sim j}}\frac{\partial\log(\rho_{j|i})}{\partial X_i}\right] = -\int_{\sim j}\rho_{\sim j}\,dX_{\sim j}\int_j R_i \frac{\partial^2 Pr_{ex}(j|i)}{\partial X_i \partial X_j}dX_j \tag{A.8}$$

where the $\sim j$-integration means integration with respect to all variables different from $X_j$ (i.e $X_{\sim j}$) and $\rho_{\sim j}$ is the pdf of $X_{\sim j}$. Applying integration by parts with respect to $X_j$, we get $\int_{-\infty}^{\infty}R_i\frac{\partial^2 Pr_{ex}(j|i)}{\partial X_i \partial X_j}dX_j = R_i\frac{\partial Pr_{ex}(j|i)}{\partial X_i}|_{X_j=-\infty}^{X_j=\infty} - \int_{-\infty}^{\infty}\frac{\partial Pr_{ex}(j|i)}{\partial X_i}\frac{\partial R_i}{\partial X_j}dX_j$. The first term vanishes because it depends on the derivatives of a probability: zero or one, at $X_j = -\infty$ and $X_j = \infty$ respectively. After its substitution and recalling that $\rho_{\sim j} = \frac{\rho_{j,\sim j}}{\rho_{j|\sim j}}$ and applying the expectation operator on the state-space, the result follows.

### A.8. Proof of (29a)

After expanding the expectation in terms of pdfs with $R_i$ substituted by $R_{i,j}$ we get:

$$T(X_j \to X_i) = E\left[R_{i,j}\frac{\rho_{j|i}}{\rho_{j|i,k}}\frac{\partial\log(\rho_{j|i})}{\partial X_i}\right] = \int_i \rho_i\,dX_i \int_j \frac{\partial\rho_{j|i}}{\partial X_i}dX_j \int_k \rho_{k|i}R_{i,j}\,dX_k = \int_i \rho_i\,dX_i \int_j \bar{R}_{i,j}\frac{\partial\rho_{j|i}}{\partial X_i}dX_j \tag{A.9}$$

where $\bar{R}_{i,j} \equiv \int_k \rho_{k|i}R_{i,j}\,dX_k = \bar{F}_{i,j} - \frac{1}{2}\frac{\partial\bar{g}_{i,i,j}}{\partial X_i} - \frac{1}{2}\frac{d\log\rho_i}{dX_i}\bar{g}_{i,i,j} = \bar{F}_{i,j} - \frac{1}{2\rho_i}\frac{\partial\rho_i\bar{g}_{i,i,j}}{\partial X_i}$ with conditional averages over the outer variables given by $\bar{F}_{i,j} \equiv \int_k \rho_{k|i}F_{i,j}\,dX_k$ and $\bar{g}_{i,i,j} \equiv \int_k \rho_{k|i}g_{i,i,j}\,dX_k$.

Integration by parts with respect to $X_i$, leads to: $T(X_j \to X_i) = -\int_j dX_j \int_i dX_i\,\rho_{j|i}\frac{\partial\rho_i\bar{R}_{i,j}}{\partial X_i} = T(X_j \to X_i)_F + T(X_j \to X_i)_g$. The term due to the deterministic term develops as:

$$T(X_j \to X_i)_F = -\int_j dX_j \int_i dX_i\,\rho_{j|i}\rho_i\left(\frac{\partial\bar{F}_{i,j}}{\partial X_i}\right) - \int_j dX_j \int_i dX_i\,\rho_{j|i}\bar{F}_{i,j}\left(\frac{\partial\rho_i}{\partial X_i}\right) \tag{A.10}$$

The first term of the r.h.s is $-E_{i,j}\left(\frac{\partial\bar{F}_{i,j}}{\partial X_i}\right)$, whereas in the second term we apply integration by parts with respect to $X_i$ yielding: $-\int_j dX_j \int_i dX_i\,\rho_{j|i}\bar{F}_{i,j}\left(\frac{\partial\rho_i}{\partial X_i}\right) = \int_i \rho_i dX_i\frac{\partial}{\partial X_i}\int_j \rho_{j|i}\bar{F}_{i,j}dX_i = E_i\left[\frac{d}{dX_i}E_j(\bar{F}_{i,j}|X_i)\right]$, providing the formula of $T(X_j \to X_i)_F$.

## A.9. Proof of (29b)

The term due to the diffusivity is:

$$T(X_j \to X_i)_g = \frac{1}{2}\int_j dX_j \int_i dX_i\, \rho_{j|i} \frac{\partial^2 \rho_i \bar{g}_{i,i,j}}{\partial X_i^2} = \frac{1}{2}\int_j dX_j \int_i dX_i\, \rho_{j|i} \left(\rho_i \frac{\partial^2 \bar{g}_{i,i,j}}{\partial X_i^2} + \bar{g}_{i,i,j} \frac{\partial^2 \rho_i}{\partial X_i^2} + 2\frac{\partial \bar{g}_{i,i,j}}{\partial X_i}\frac{\partial \rho_i}{\partial X_i}\right). \quad (A.11)$$

Application of integration by parts with respect to $X_i$, twice and once, to the second and third terms, respectively, leads to the result:

$$T(X_j \to X_i)_g = \frac{1}{2}E_{i,j}\left(\frac{\partial^2 \bar{g}_{i,i,j}}{\partial X_i^2}\right) + \frac{1}{2}E_i\left[\frac{d^2}{dX_i^2}E_j(\bar{g}_{i,i,j}|X_i)\right] - E_i\left[\frac{d}{dX_i}E_j\left(\frac{\partial \bar{g}_{i,i,j}}{\partial X_i}\Big|X_i\right)\right] \quad (A.12)$$

## A.10. Proof of (36a)

Let us consider w.l.g. $A_F = f_1(X_j)f_2(X_i, X_k) = F_{i,j}$. Let us build the terms appearing in the l.h.s. of Theorem 3. Then $\bar{F}_{i,j} = f_1 E_k(f_2|X_i)$ and $\frac{\partial \bar{F}_{i,j}}{\partial X_i} = f_1 E_k(f_2|X_i)'$ where the prime (double prime) stands for the first (second) derivative with respect to $X_i$. Then $E_{i,j}\left(\frac{\partial \bar{R}_{i,j}}{\partial X_i}\right) = E_i\big(E_j(f_1|X_i)E_k(f_2|X_i)'\big)$. Now, $E_j(\bar{F}_{i,j}|X_i) = E_k(f_2|X_i)E_j(f_1|X_i)$, its derivative is $\frac{d}{dX_i}E_j(\bar{R}_{i,j}|X_i) = E_k(f_2|X_i)'E_j(f_1|X_i) + E_k(f_2|X_i)E_j(f_1|X_i)'$ and the $X_i$ expectation comes as $E_i\left[\frac{d}{dX_i}E_j(\bar{R}_{i,j}|X_i)\right] = E_i(E_k(f_2|X_i)'E_j(f_1|X_i) + E_k(f_2|X_i)E_j(f_1|X_i)')$. Subtracting $E_{i,j}\left(\frac{\partial \bar{R}_{i,j}}{\partial X_i}\right)$, we get $T(X_j \to X_i)_{A_F} = E_i\left[E_k(f_2|X_i)\frac{d}{dX_i}E_j(f_1|X_i)\right]$.

## A.11. Proof of (36b)

Considering $A_g = f_3(X_j)f_4(X_i, X_k) = g_{i,i,j}$, one obtains $\bar{g}_{i,i,j} = f_3 E_k(f_4|X_i)$. Now, playing with the terms whose $X_i$ average is taken in $T(X_j \to X_i)_g$, we get:

$$\frac{1}{2}E_j\left(\frac{\partial^2 \bar{g}_{i,i,j}}{\partial X_i^2}\Big|X_i\right) = \frac{1}{2}E_j(f_3|X_i)E_k(f_4|X_i)'' \quad (A.13)$$

$$\frac{1}{2}\frac{d^2}{dX_i^2}E_j(\bar{g}_{i,i,j}|X_i) = \frac{1}{2}E_j(f_3|X_i)E_k(f_4|X_i)'' + E_j(f_3|X_i)'E_k(f_4|X_i)' + \frac{1}{2}E_j(f_3|X_i)''E_k(f_4|X_i) \quad (A.14)$$

$$-\frac{d}{dX_i}E_j\left(\frac{\partial \bar{g}_{i,i,j}}{\partial X_i}\Big|X_i\right) = -E_j(f_3|X_i)E_k(f_4|X_i)'' - E_j(f_3|X_i)'E_k(f_4|X_i)' \quad (A.15)$$

where primes are derivatives with respect $X_i$. By summing up, and taking $X_i$ averaging, we obtain

$$T(X_j \to X_i)_{A_g} = \frac{1}{2}E_i\left[E_k(f_4|X_i)\frac{d^2}{dX_i^2}E_j(f_3|X_i)\right] \quad (A.16)$$

## A.12. Proof of (43b)

The equation governing $\hat{X}_i$ writes:

$$d\hat{X}_i = \hat{F}_i(\hat{\mathbf{X}}, \boldsymbol{\theta})dt + \sum_{k=1}^{D_n} \hat{B}_{i,k}(\hat{\mathbf{X}}, \boldsymbol{\theta})dW_k, \quad (A.17)$$

where $\hat{\mathbf{X}}$ is the changed vector, $\hat{F}_i = \mathfrak{I}_i F_i$ and $\hat{B}_{i,k} = \mathfrak{I}_i B_{i,k}$. The changed diffusion term comes as $\hat{g}_{ii} = \mathfrak{I}_i^2 g_{ii}$. By using the expression of Theorem 2 we have:

$$T(\hat{X}_j \to \hat{X}_i) = E\left[\frac{1}{\hat{\rho}_{j|\sim j}} \frac{\partial \widehat{Pr}_{ex}(j|i)}{\partial \hat{X}_i} \frac{\partial \hat{R}_i}{\partial \hat{X}_j}\right] \quad \text{(A.18)}$$

where hatted quantities $(\widehat{\ldots})$ refer to quantities obtained with the changed variables. Then we take into account that: $\frac{1}{\hat{\rho}_{j|\sim j}} = \frac{\Im_j}{\rho_{j|\sim j}}$, $\frac{\partial}{\partial \hat{X}_j} = \frac{1}{\Im_j} \frac{\partial}{\partial X_j}$, $\widehat{Pr}_{ex}(j|i) = \text{Prob}(\hat{u}_j \geq \hat{X}_j | \hat{X}_i) = \text{Prob}(u_j \geq X_j | X_i) = Pr_{ex}(j|i)$ and $\frac{\partial}{\partial \hat{X}_i} = \frac{1}{\Im_i} \frac{\partial}{\partial X_i}$. Furthermore, we have $\hat{R}_i = \hat{F}_i - \hat{g}_{i,i} \frac{1}{2} \frac{\partial \log(\hat{\rho}_{\sim j})}{\partial \hat{X}_i} - \frac{1}{2} \frac{\partial \hat{g}_{i,i}}{\partial \hat{X}_i}$. There, recall that $\frac{\partial \log(\hat{\rho}_{\sim j})}{\partial \hat{X}_i} = \frac{1}{\Im_i} \frac{\partial \log(\rho_{\sim j})}{\partial X_i} - \frac{1}{\Im_i} \frac{\partial \log(\Im_i)}{\partial X_i}$ and $\frac{\partial \hat{g}_{i,i}}{\partial \hat{X}_i} = \frac{1}{\Im_i} \frac{\partial \Im_i^2 g_{i,i}}{\partial X_i} = \Im_i \frac{\partial g_{i,i}}{\partial X_i} + 2\Im_i g_{i,i} \frac{\partial \log(\Im_i)}{\partial X_i}$, which yields $\hat{R}_i = \Im_i \left(R_i - \frac{1}{2} g_{i,i} \frac{\partial \log(\Im_i)}{\partial X_i}\right)$ and the derivative $\frac{\partial \hat{R}_i}{\partial \hat{X}_j} = \frac{\Im_i}{\Im_j} \left(\frac{\partial R_i}{\partial X_j} - \frac{1}{2} \frac{\partial g_{i,i}}{\partial X_j} \frac{\partial \log(\Im_i)}{\partial X_i}\right)$. Then, plugging all the terms into $T(\hat{X}_j \to \hat{X}_i)$, we get the result.

**Appendix B Technical Details about the RET estimation methods**

**B.1. The fitting of nonlinear models with multiplicative noise**

In order to fit the model, let us consider the empirical $X_i$ time-derivatives, $\dot{X}_i \equiv [X_i(t+dt) - X_i(t)]/dt$, obtained with, either successive ensemble realizations at a particular instant $t$ or taken from a single time-series in the ergodic case.

Referring to [33], one can find the minimum mean square difference between $\dot{X}_i$ and $F_i$ by applying ordinary multivariate linear regression yielding the fitted matrix of parameters through the matrix product:

$$\hat{\boldsymbol{\theta}}_F = \mathbf{C}_{\Psi,\Psi}^{-1} E(\boldsymbol{\Psi}\dot{\mathbf{X}}^T) \quad \text{(B.1)}$$

where $\mathbf{C}_{\Psi,\Psi}^{-1}$ is the inverse of the matrix of averaged function products $\mathbf{C}_{\Psi,\Psi(r,s)} = E(\Psi_r \Psi_s)$ and matrix $E(\boldsymbol{\Psi}\dot{\mathbf{X}}^T)$ contains the average product entries $E(\boldsymbol{\Psi}\dot{\mathbf{X}}^T)_{(r,s)} = E(\Psi_r \dot{X}_s)$, where $\boldsymbol{\Psi} = (\Psi_1, \ldots)^T$.

The fitting of the multiplicative noise is obtained from the residuals $\varepsilon_i = \dot{X}_i - \sum_l \hat{\theta}_{F,l,i} \Psi_l(\mathbf{X})$. By dealing with the stochastic equation, their squares are $d_i = \varepsilon_i^2 = \frac{g_{ii} W_i^2}{dt}$, filling the vector $\mathbf{d}$ and where $W_i \sim N(0,1)$ are independent white standard Gaussian noises. After multiplying $\varepsilon_i^2$ by the positive basis-functions $\Phi_l$ and taking average, we obtain a linear system of equations in terms of the parameters $\theta_{g,l,i}$. Therefore, the matrix of fitted parameters is obtained as:

$$\hat{\boldsymbol{\theta}}_g = dt\, \mathbf{C}_{\Phi,\Phi}^{-1} E[\boldsymbol{\Phi}\mathbf{d}^T] \quad \text{(B.2)}$$

where $\mathbf{C}_{\Phi,\Phi}^{-1}$ is the inverse of the matrix $\mathbf{C}_{\Phi,\Phi}$ of averaged function products used to fit the diffusivities and $(r,s)$−entries of matrix $E[\boldsymbol{\Phi}\mathbf{d}^T]$ are the expectation products $E(\Phi_r \varepsilon_s^2)$. The fitting is consistent only if $\hat{\theta}_{g,l,i}$.

The simplest example of 1D model with both a deterministic and a multiplicative noise part depending on one function each is

$$\dot{x}dt = x(t+dt) - x(t) = [\theta_{F,0} + \theta_{F,1}\Psi(x(t))]dt + \sqrt{dt}[\theta_{g,0} + \theta_{g,1}\Phi(x(t))]^{1/2} W(t) \quad \text{(B.3)}$$

where $W(t) \sim N(0,1)$ is a standard Gaussian white noise.

The fitted parameters are:

$$\hat{\theta}_{F,0} = E(\dot{x}) - \theta_{F,1} E(\Psi) \quad \text{(B.4a)}$$

$$\hat{\theta}_{F,1} = \frac{cov(\dot{x},\Psi)}{var(\Psi)} \tag{B.4b}$$

The square errors of the tendency are $\varepsilon^2 = \dot{x} - [\hat{\theta}_{F,0} + \theta_{F,1}\Psi(x)]$ from which one infers:

$$\hat{\theta}_{g,0} = dt\, E(\varepsilon^2) - \theta_{g,1}E(\Phi) \tag{B.5a}$$

$$\hat{\theta}_{g,1} = dt\, \frac{cov(\varepsilon^2,\Phi)}{var(\Phi)} \tag{B.5b}$$

**B.2. The Predictor-Corrector scheme**

The stochastic differential equations are integrated in time using the predictor-corrector scheme described by the steps below:

$$X^*_{i,t+dt} = X_{i,t} + dt\, F_i(\mathbf{X}_t) + \sqrt{2dt}\, B_i(\mathbf{X}_t)W_{1,t} \tag{B.6}$$

$$X^{**}_{i,t+dt} = X_{i,t} + dt\, F_i(\mathbf{X}^*_t) + \sqrt{2dt}\, B_i(\mathbf{X}^*_t)W_{2,t} \tag{B.7}$$

$$X_{i,t+dt} = \frac{1}{2}\left(X^*_{i,t+dt} + X^{**}_{i,t+dt}\right) \tag{B.8}$$

where $X^*_{i,t+dt}, X^{**}_{i,t+dt}$ are preliminar estimates of $X_i$ at time $t + dt$ and $W_{1,t}, W_{2,t}$ are independent realizations of standard Gaussian white noise processes.

**Appendix C List of Symbols and achronyms**

| | |
|---|---|
| $B_{i,k}$ | noise-diffusion coefficients merged in matrix $\mathbf{B}$ |
| $\frac{dH_{X_i}}{dt}, \left(\frac{dH_{X_i}}{dt}\right)_F, \left(\frac{dH_{X_i}}{dt}\right)_g$ | Rate of entropy change (REC) and its deterministic and stochastic parts. |
| $\left(\frac{dH_{X_i}}{dt}\right)_g, \left(\frac{dH_{X_i}}{dt}\right)_{g,a}, \left(\frac{dH_{X_i}}{dt}\right)_{g,m}$ | Stochastic part of the rate of entropy change (REC) and its additive and multiplicative noise parts |
| $\frac{dH_{X_i,self}}{dt}, \left(\frac{dH_{X_i,self}}{dt}\right)_F, \left(\frac{dH_{X_i,self}}{dt}\right)_g$ | Self entropy generation (SEG) and its deterministic (D-RET) and stochastic (S-RET) parts |
| $F_i, F_{i,j}$ | component $i$ of the vector $\mathbf{F}$ of deterministic terms and the part dependent on $X_j$ only |
| $\mathcal{F}_{i,\sim i,F}(X_i), \mathcal{F}_{i,\sim i,g}(X_i)$ | Specific functions of D-RET $T(X_{\sim i} \to X_i)\,_F$ and S-RET $T(X_{\sim i} \to X_i)\,_g$ |
| $\mathcal{F}_{i,j,F}(X_i), \mathcal{F}_{i,j,g}(X_i)$ | Specific functions of the single D-RET and S-RET |
| $g_{i,i}, g_{i,i,j}$ | component $i,i$ of the diffusivities, merged in matrix $\mathbf{G}$ and part dependent on $X_j$ |
| $H_{X_i}$ | Shannon entropy of variable $X_i$ |
| $\wp[\ldots]$ | Frobenius Perron operator |
| $R_i, R_{i,j}$ | Component $I$ of the generalized speed and the part dependent on $X_j$ |
| $T(X_{\sim i} \to X_i), T(X_{\sim i} \to X_i)\,_F, T(X_{\sim i} \to X_i)\,_g$ | Rate of entropy transfer (RET) between the complementary vector of $X_i$ and its deterministic (D-RET) and stochastic (S-RET) parts |
| $T(X_{\sim i} \to X_i)_{sing}, T(X_{\sim i} \to X_i)_{syn}$ | RETs associated to single RETs and synergies |
| $T(X_j \to X_i), T(X_j \to X_i)_F, T(X_j \to X_i)_g$ | Single rate of entropy transfer (RET) between the causal variable $X_j$ and the consequential variable $X_i$ and its deterministic (D-RET) and stochastic (S-RET) parts |
| $X_i, X_{\sim i}$ | component *i* of the state vector **X** and its complementary vector |
| AN | Analytical-Numerical Method |
| CSM | Causal Sensitivity Method |
| MF | Model Fitting followed by CSM |
| ML | Multivariate Linear Method |
| MN | Minimal CSM |